\documentclass[11pt]{article}

\RequirePackage[OT1]{fontenc}
\usepackage{graphicx}
\RequirePackage{amsthm,amsmath}
\RequirePackage{natbib}
\RequirePackage[colorlinks,citecolor=blue,urlcolor=blue]{hyperref}
\usepackage{amsfonts}

\usepackage{multirow}
\usepackage{verbatim}

\newcommand*{\graphFont}{\scriptsize{\fontfamily{pag}\selectfont}}

\newcommand{\B}{\boldsymbol}
\newcommand{\M}{\mathbf}
\newcommand{\sgn}{\operatorname{sgn}}

\newcommand{\TAGS}{TAGS } 
\newcommand{\TAGSs}{TAGS } 

\usepackage[capposition=bottom]{floatrow}
\usepackage[hmargin=3.8cm,vmargin=3.cm]{geometry}
\usepackage{authblk}
\usepackage{adjustbox}

\DeclareMathOperator*{\argmin}{arg\,min}

\DeclareMathOperator*{\mini}{minimize}

\title{Mining Events with Declassified Diplomatic Documents}

\author[1]{Yuanjun Gao\thanks{gaoyuanjun0430@gmail.com}}
\author[2]{Jack Goetz\thanks{jrgoetz@umich.edu}}
\author[3]{Rahul Mazumder\thanks{rahulmaz@mit.edu}}
\author[4]{Matthew Connelly \thanks{mjc96@columbia.edu}}
\affil[1]{Department of Statistics, Columbia University, New York, NY.}
\affil[2]{Department of Statistics, University of Michigan, Ann Arbor, MI.}
\affil[2]{Massachusetts Institute of Technology, Cambridge, MA.}
\affil[3]{Department of History, Columbia University, New York, NY.}

\begin{document}

\maketitle

\begin{abstract}
Since 1973 the State Department has been using electronic records systems to preserve classified communications. Recently,  approximately 1.9 million of these records from 1973-77 have been made available by the U.S. National Archives.  While some of these communication streams have periods witnessing an acceleration in the rate of transmission; others do not show any notable patterns in communication intensity.  Given the sheer volume of these communications -- far greater than what had been available until now -- scholars need automated statistical techniques to identify the communications that warrant closer study. We develop a statistical framework that can semi-automatically identify  from a large corpus of documents a handful 
that historians would consider more interesting electronic records. Our approach brings together related but distinct statistical concepts from nonparametric signal estimation and statistical hypothesis testing -- which when put together help us identify and analyze various geometrical aspects of the communication streams. 
Dominant periods of heightened and sustained activities aka \emph{bursts}, as identified through these methods, correspond well with historical events recognized by standard reference works on the 1970s.
\end{abstract}




\section{Introduction}


For more than forty years, social scientists have been developing datasets of events for the quantitative analysis of world politics. The last decade has witnessed a dramatic increase in activity in this area, much of it focused on automatic event detection for purposes of explaining and predicting political crises \cite{Beieler2016}. All of these efforts however, have used public information, such as newspaper or wire service reporting. Rather than directly measuring political activity, these systems can only count what reporters write about, which can vary over time and geography depending on many extraneous factors \cite{jenkins2016should}. Together with the intrinsic challenges in automatic extraction, this has produced datasets that purport to track the same kind of events, such as political protests, but that are completely uncorrelated \cite{Hanna2010}. Moreover, some of the most important political activity is not immediately reported, and may not become publicly known until decades later, when formerly secret records are declassified. Even then, the sheer volume of these records can make it difficult even for the diligent researcher to identify individual events and assess their relative importance. 

In this paper we study a new dataset of declassified documents and use statistical methods to identify and rank political events. Since 1973, the State Department has been using electronic records systems to preserve classified communications. The National Archives\footnote{Website:~\url{https://aad.archives.gov/aad/series-list.jsp?cat=WR43}}  
now makes available over 1.4 million declassified cables from 1973-77 and in addition, the metadata of more than 0.4 million other communications originally delivered by diplomatic pouch. They are all machine-readable and rich with metadata, creating many opportunities for statistical modeling.
\begin{figure}[h!]
    \centering
    		\scalebox{.95}{\begin{tabular}{rcc}
 					&{\scriptsize{\sf (a) \# of communications for \TAGS CY}} & {\scriptsize{\sf  (b) \# of communications for \TAGS VS}} \\
	 \rotatebox{90}{~~~~~~~~~~~~~~~{\graphFont Count}}		
	 &\includegraphics[width=.5\textwidth,height = 0.2\textheight,clip = true,trim = 30 0 0 0]{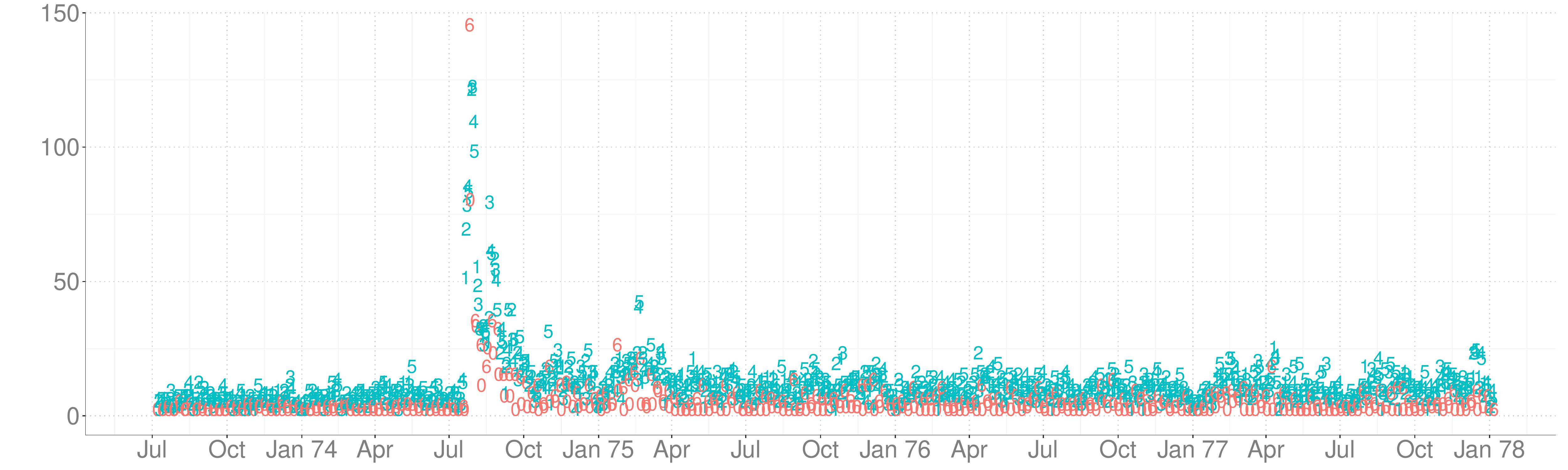} &
								 \includegraphics[width=.5\textwidth,height = 0.2\textheight,clip = true,trim = 30 0 0 0]{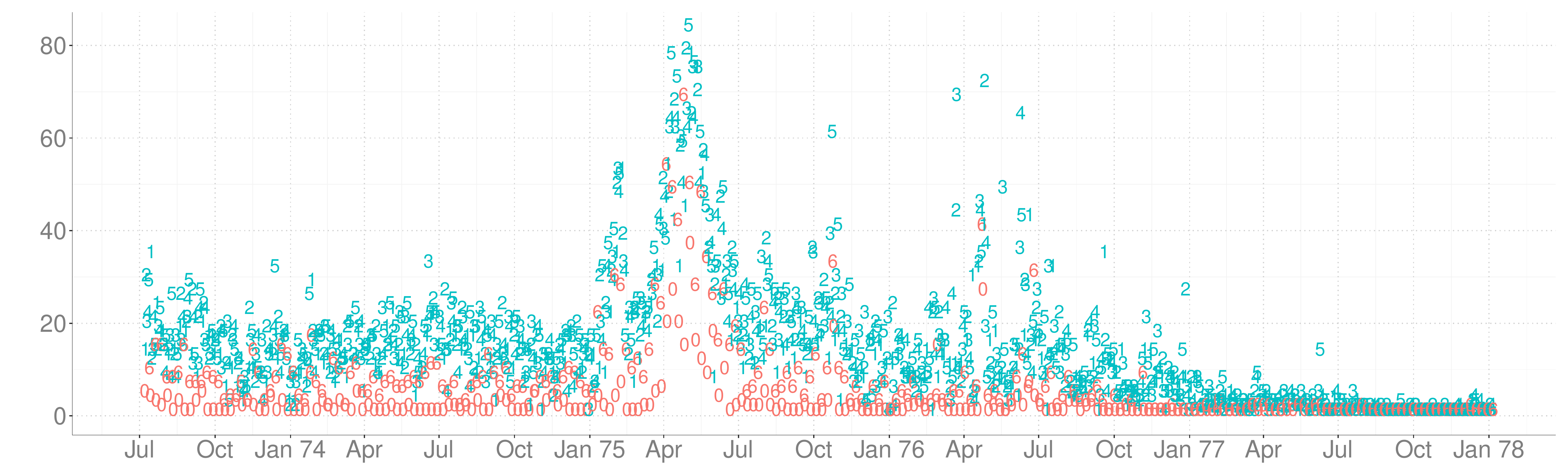}\\~\\ 

					& {\scriptsize{\sf (c) \# of communications for \TAGS UNGA }}& {\scriptsize{\sf (d) \# of communications for \TAGS FI}} \\
	 \rotatebox{90}{~~~~~~~~~~~~~~~{{\graphFont Count}}}		
	 & {\includegraphics[width=.5\textwidth,height = 0.2\textheight,clip = true,trim = 30 10 0 0]{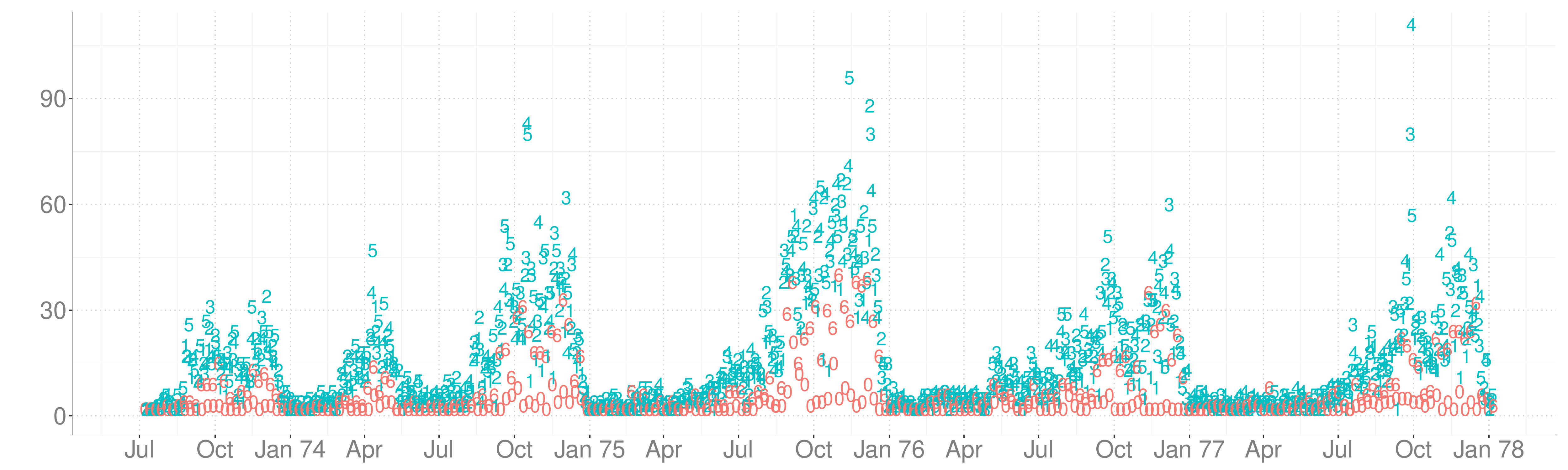}} & {\includegraphics[width=.5\textwidth,height = 0.2\textheight,clip = true,trim = 30 10 0 0]{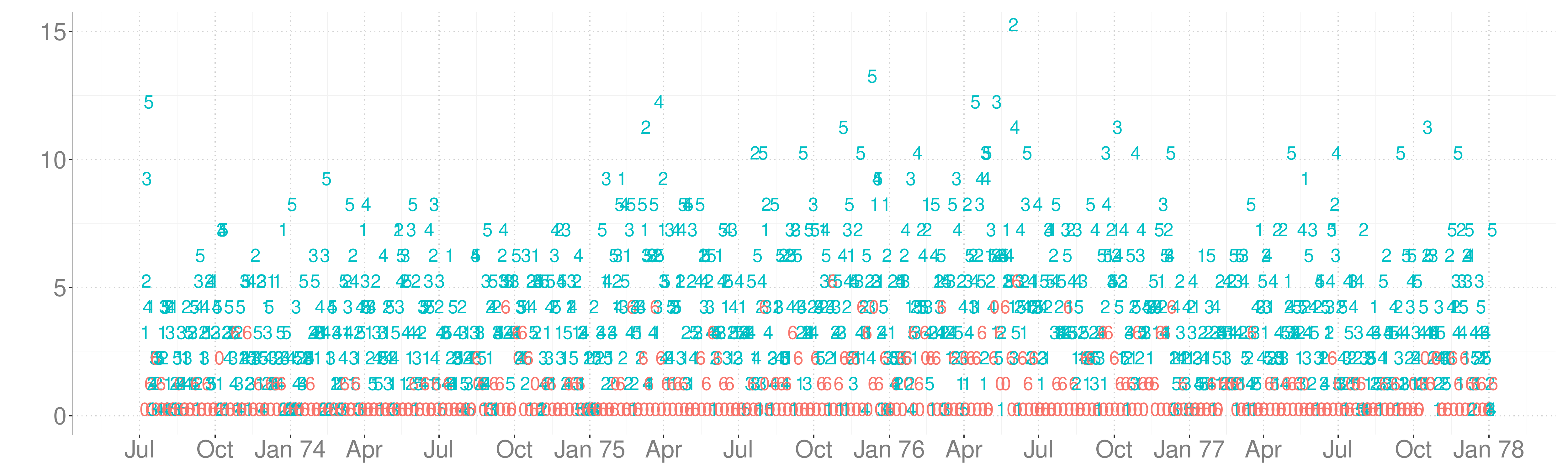}}   \\
		\end{tabular}}
				\caption{\small{Figures showing counts of communications sent on each day, in the period 1973-1978. The numbers in the plot represent day-of-week (0-Sunday, 1-Monday, 2-Tuesday, ..., 6-Saturday), with weekdays colored in blue and weekends in red. Figures (a)--(d) show the communications
			restricted to different TAGS. The apparent heightened activities in the communication streams correspond 
		to (a) Cyprus coup (b) Fall of Saigon (the most intense one) (c) the yearly United Nation General Assembly meetings. 
		There does not seem to be any interesting activity for panel (d), showing cables related to Finland. A goal of this paper is to create statistical methods to automatically identify series with heightened 
		diplomatic communications and further describe their structural patterns. 
		 }} \label{fig:raw}
\end{figure}
Our goal is to explore modern statistical techniques that can automatically identify statistically interesting events in an important corpus of historical documents, which will continue to grow year-by-year as millions more communications are declassified. 
For these communication streams, we are interested in studying and identifying ``interesting'' statistical patterns -- we contend that these patterns correspond to heightened diplomatic activity; and validate our findings with standard reference works on the 1970s.
A statistically interesting pattern can mean several things.
Loosely speaking, this can correspond to sudden localized changes or abrupt ``jumps'' in communication traffic, regardless of the overall series-specific baseline activity (a communication stream may be very active or have very low traffic intensity overall).
There can also be continuous periods in a communication stream, where the data lies consistently above a series-specific baseline that corresponds to a representative global activity-level of that stream -- these 
are ``bursts'' of activity in the temporal structure of the document streams that probably correspond with heightened diplomatic activity, such as the start or end of a war.
An interesting event can also correspond to a heightened traffic intensity that plays out over longer periods, such as an increase over time, whether or not there are jumps. 


When these communications were first entered in the State Department system, they were assigned one or more \TAGSs (Traffic Analysis by Geography and Subject), which indicate what countries or subjects each cable is related to. For example, ``VS''  signifies South  Vietnam, and ``SHUM'' concerns human rights. By using these content-based \TAGS as the feature, we avoid the complication of language processing 
and focus on identifying statistically relevant activity patterns in the communication streams.

\subsection{A brief exploratory description of the data}

A glimpse of  processed data in the form of communication streams is shown in Figure~\ref{fig:raw}. 
The data shows that there is less traffic on weekends and holidays (including the end of the year). In addition, the number of communications sent in 1973 seems to be smaller compared to later years, due to fewer measurements. Since the overall (aggregated across all TAGS) number of communications sent across time  did not have any systematic pattern indicative of events of historical importance, we decided to study the time series at a more granular level, by restricting to different types of  TAGS. In Figure~\ref{fig:raw}, panels (a)--(d) represent the communication streams, when restricted by \TAGS type. Panels 
(a)--(c) show noticeable forms of increased activities in portions of the series -- these are indicative of ``interesting'' historical events. For example, in panel (a) the increased activity in July 1974 corresponds to the Cyprus coup; in panel (b) the increase in number of diplomatic communications in April 1975 corresponds to the Fall of Saigon. Panel (c) shows multiple bursts in the number of cables, containing the particular \TAGS UNGA (which stands for United Nations General Assembly) -- interestingly, they all correspond to the annual United Nations General Assembly meetings. In addition to these visible bursts there seem to be some shorter periods of heightened activities,  such as the smaller peaks for VS (South Vietnam) a year after the fall of Saigon corresponding to the ensuing refugee crisis. 

In contrast to panels (a)--(c)  in Figure~\ref{fig:raw}, panel (d), for cables related to Finland (FI), does not seem to show any period of heightened activity during the time period under consideration. 
These prototypes are representative of the different TAGS-specific series: Exploratory analyses of the database of TAGS specific communication streams suggest that 
there are several series with some ``interesting event'' (as in Panels (a)--(c)), while others seem to be less active (as in panel (d)).
A first goal of our work is to quantitatively define traits that separate communication streams like the figure in panel (d) from those in panels (a)--(c). 
We develop statistical methods that can \emph{mine} these (\TAGS specific) time-series and identify communication streams that exhibit statistically 
interesting activities in them. 
Once we identify these interesting communication streams, we develop algorithms that perform a deeper investigation of each series and identify time intervals where the signal undergoes abrupt localized changes in communication traffic.
The informal ideas described above are made more precise in Section~\ref{sec:methods} of this paper.

Exploratory analysis suggests that changes in the proportion of a particular \TAGS appearing in a communication stream are better representatives of 
identifying whether a period is active or not, as compared to tracking the corresponding counts.
Due to the noticeable difference in the number of cables that were communicated over the weekdays and low-traffic days, the communication streams shown in Figure~\ref{fig:raw} seem to be a superposition of
a high traffic series and a low-traffic (weekend and holiday) series. 
As a pre-processing step, we filtered out the days where the total number of cables being communicated were very small -- they lead to more reliable estimates
of proportions.

\begin{figure}[h!]
	\caption{{\small{Communication streams with different significance scores in the spirit of Section~\ref{sec:interest-1}:
	(a): relating to Finland, corresponds to a null model (b): relating to scientific grants, shows a weak deviation from the null model -- perhaps due a slow decreasing trend in the series (c): relating to foreign policy -- this generally shows significant deviation from the null model, which is due to sudden changes/jumps after Oct '76 (d): relating to internal political affairs, shows deviation from the null model but not due to a jump as prominent as (c) -- this series seems to exhibit some systematic pattern of heightened activity after Oct '76, leading to a small p-value. The small p-values suggest the presence of a statistically interesting event in each series, and can be used to identify interesting communication streams -- the p-value however, does not provide additional insights into the finer structural patterns of the streams. Additional examples can be found in Figure~\ref{fig:all_sig}.}}}
	\centering
	\scalebox{.95}{\begin{tabular}{ccc}
	&{\scriptsize{\sf (a)\TAGS FI, p-value = 0.181}} & {\scriptsize{\sf (b)\TAGS OSCI, p-value = 0.005}} \\
\rotatebox{90}{~~~~~~~~~~~~{\graphFont Proportion}}	& 
\includegraphics[width=0.5\textwidth,height = .18\textheight, clip = true,trim = 35 0 0 60]{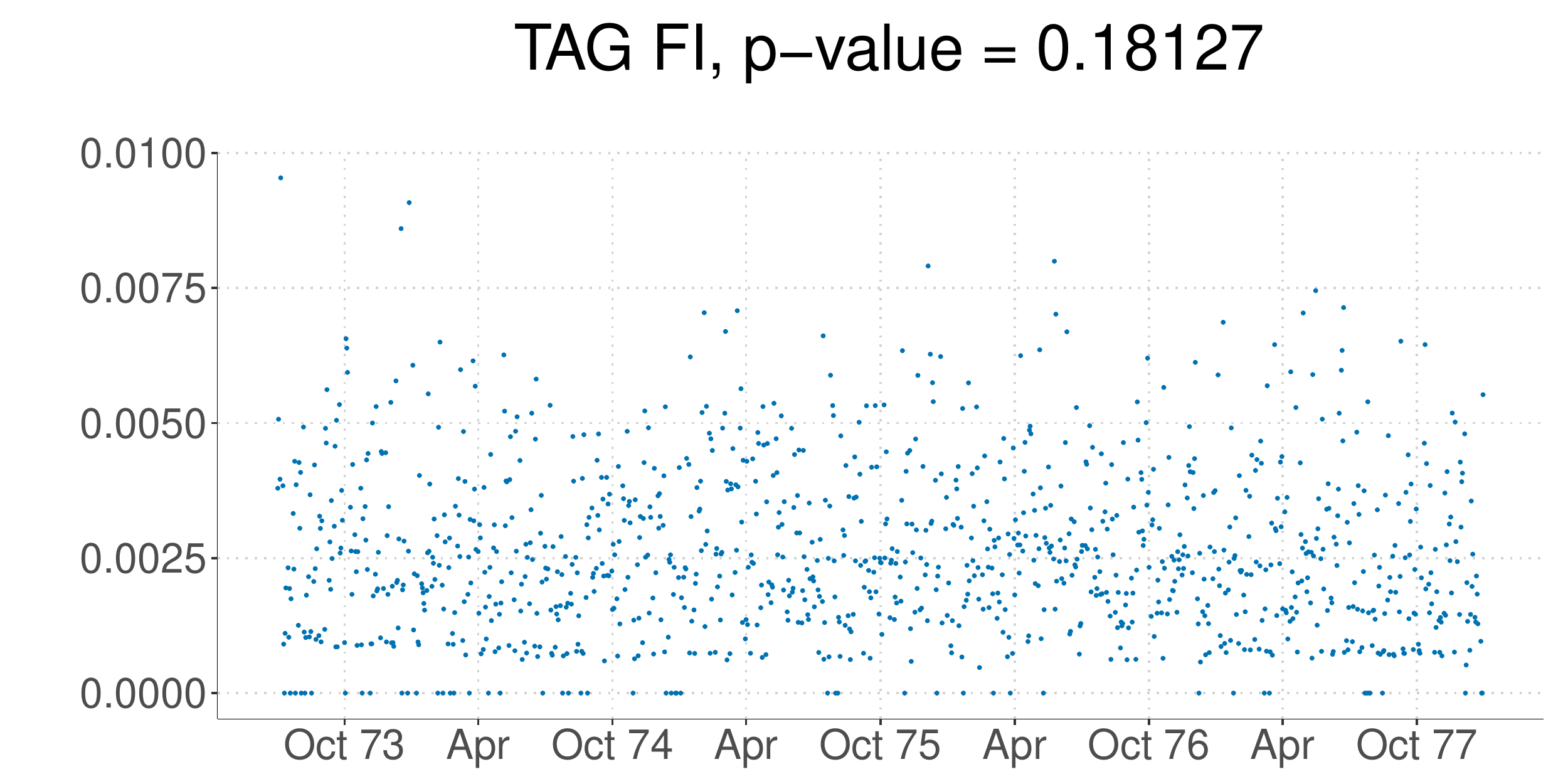} &
\includegraphics[width=0.5\textwidth,height = .18\textheight, clip = true,trim = 35 0 0 60]{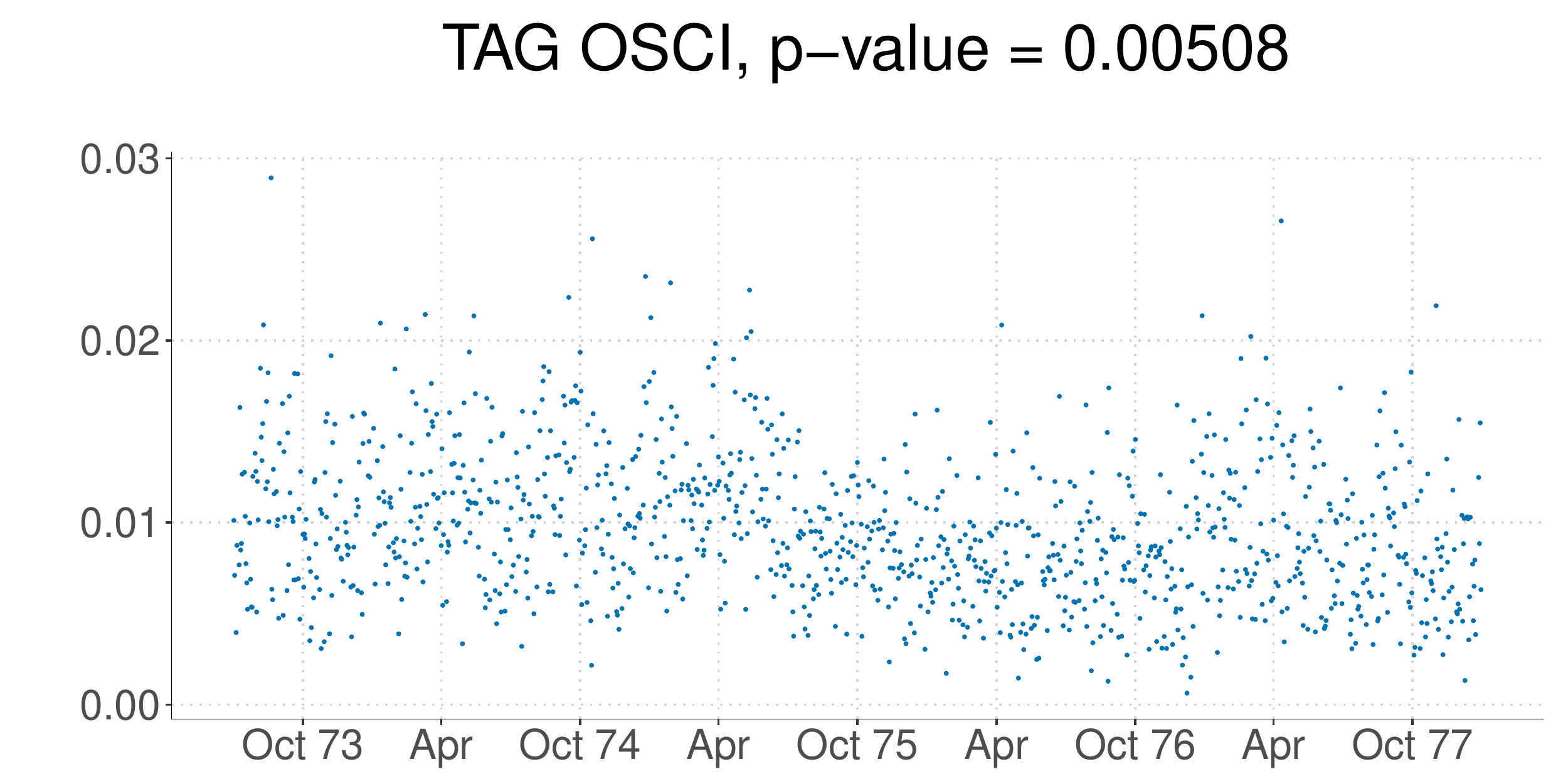} \\~\\
&{\scriptsize{\sf (c) \TAGS PFOR, p-value = 0}} & {\scriptsize{\sf (d) \TAGS PINT, p-value = 0}} \\
	\rotatebox{90}{~~~~~~~~~~~~{\graphFont Proportion}}				&
	\includegraphics[width=0.5\textwidth,,height = .18\textheight, clip = true,trim = 35 0 0 60]{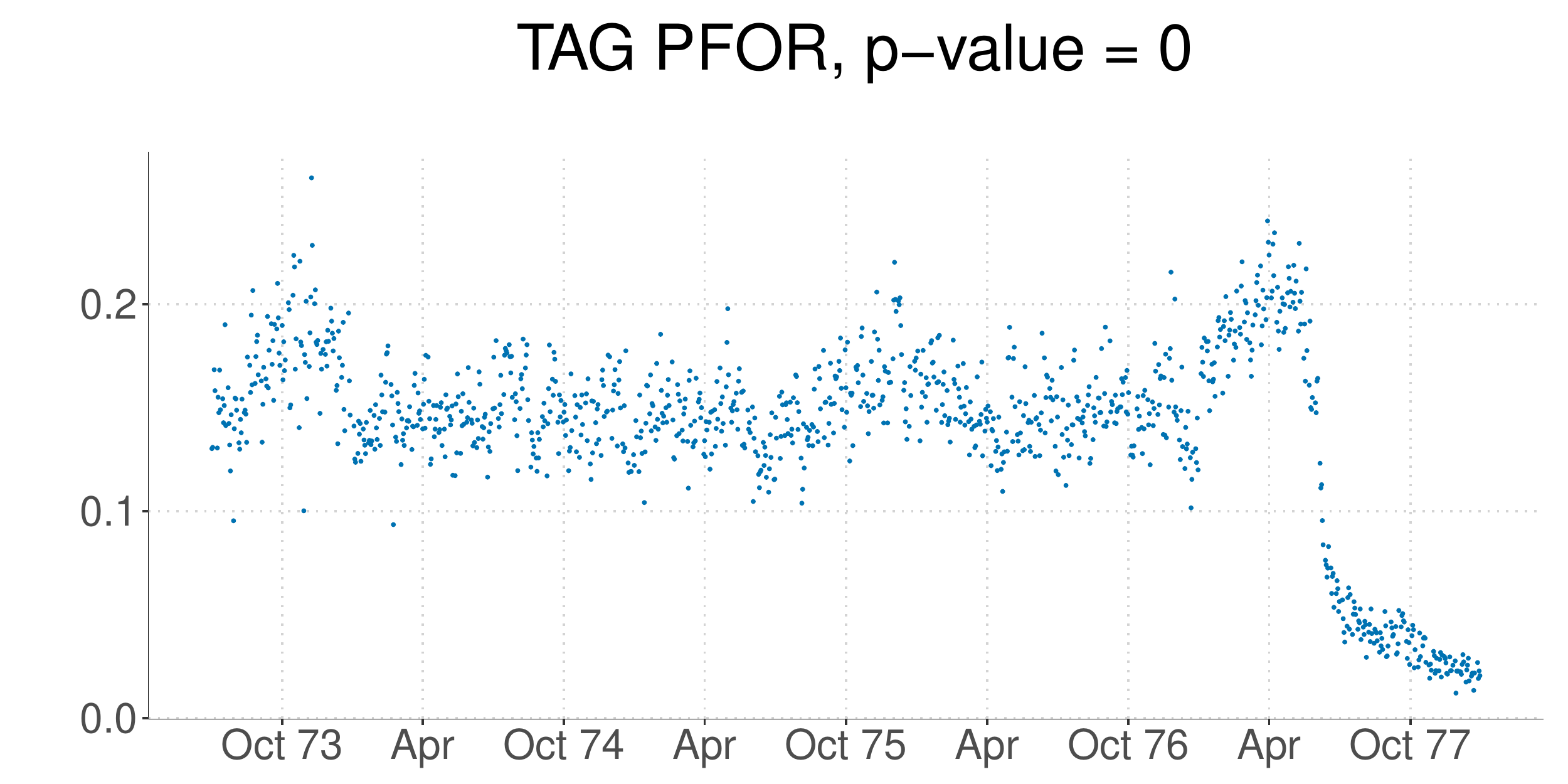} &				
\includegraphics[width=0.5\textwidth,,height = .18\textheight, clip = true,trim = 35 0 0 60]{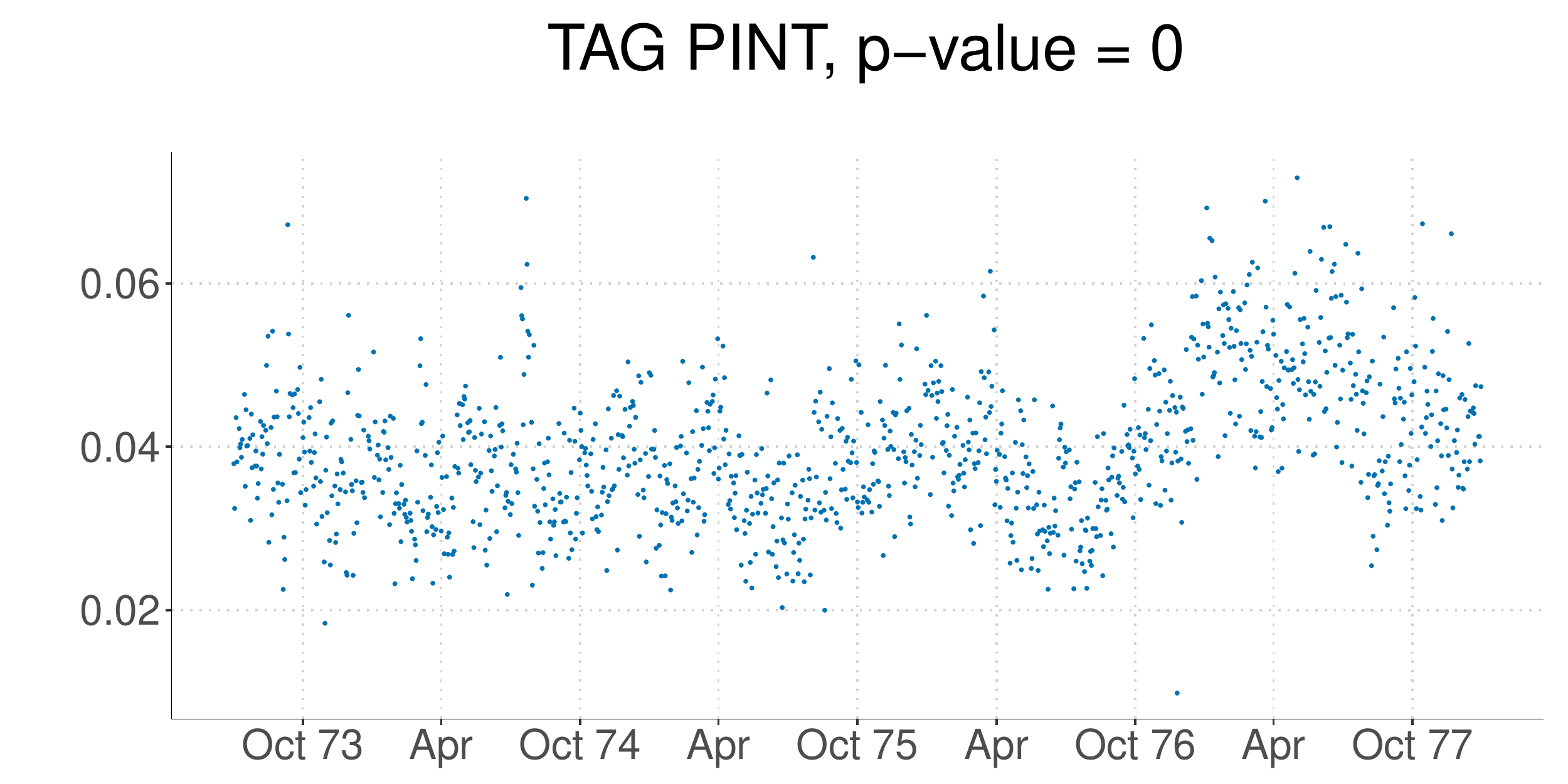}
	\end{tabular}}
	\label{fig:all-together1}
	\end{figure}


\section{Statistical Methodology}\label{sec:methods}

For the reader's convenience, we first present an outline of the main statistical approaches that we discuss in this paper. 
Section~\ref{sec:interest-1} addresses our first question: how do we determine whether a communication stream, among several hundreds, is  interesting or not? We address this as a statistical hypothesis testing problem, where we consider the
problem of whether a communication stream is generated from a homogeneous binomial process -- in other words: is the proportion of documents containing the particular \TAGS uniform across time?
We select a collection of \TAGSs for which there seems to be some form of a statistically interesting event, and explore their geometry in further detail.  
If $p_{t}$ denotes the proportion of cables containing a \TAGS at time $t$, we estimate the function $t \mapsto p_{t}$ with a piecewise constant signal -- this is performed by using a regularized negative log-likelihood criterion based on the fused Lasso penalty~\cite{TSRZ2005} or its $\ell_{0}$-counterpart~\citep{boysen2009consistencies,Johnson2013fldp}.
 For efficient computation, we develop new fast algorithms 
for these penalties by adapting existing work for the least squares loss function to the case of the generalized linear model likelihood, studied herein -- see Section~\ref{sec:model}.
The piecewise constant segments lead to  localized changes in the underlying signal $t \mapsto p_{t}$ -- we use hypothesis testing ideas based on sample splitting~\cite{wasserman2009high} to rank-order the strengths of the jumps
based on the associated p-values -- see Section~\ref{deeper-invest-1}.
The locations of discontinuities or jumps of the signal are aggregated to obtain locally contiguous subintervals of heightened diplomatic communications, which we call bursts, adopting the terminology 
coined by~\cite{kleinberg2003bursty} in a different context (Section~\ref{sec:signals-post-1}). Finally in Section~\ref{pwise-linear-segments}, we discuss preliminary results on estimating the underlying signal with more flexible local models, beyond piecewise constant segments. 

\subsection{Identifying Interesting Communication Streams}\label{sec:interest-1}


We develop a framework to identify communication streams that exhibit some form of a heightened or in other words, statistically interesting activity.
Consider a TAGS-specific series $(y_{t}, n_{t}), t =1, \ldots, N$, where, $y_{t}$ denotes the number of documents containing the specific TAGS among $n_{t}$ cables, with proportion $p_{t}$. We will assume that the conditional distributions of $(y_{t}|n_{t},p_{t})$'s are all independent across $t$.
We ask: 

\emph{Is there any evidence of (localized) heightened intensity of the proportions, compared to a baseline model, where all proportions are the same?} 

To measure a localized change (increase) in intensity, we fix a window of size $2\Delta$ and consider 
all the points in the $\Delta$ neighborhood of a time point $i$, given by: $\text{N}(\Delta; i) = \{ j : 1 \leq j \leq N,  | j - i | \leq \Delta \}$ (we pick $\Delta = 5$ days in our experiment).
The average proportion in this neighborhood: $p_{i}^\text{ave} := \sum_{ j \in  \text{N}(\Delta; i)} n_{j} p_{j}/\sum_{ j \in  \text{N}(\Delta; i)} n_{j}$, is a measure of 
communication traffic around the reference time point $i$. 
We say that a large value of  $p_{i}^\text{ave}$ compared to a baseline value $p$ (for example, the global proportion),
indicates the presence of an intense activity of some form; and declare such an occurrence to be statistically \emph{interesting}. We hypothesize such an 
event to be associated with an event of historical interest; and subsequently validate this belief by factoring in the insights of a historian or social scientist. 

Formally, we consider a simple testing problem 
with $H_0$: $p_t =p$ $\forall t$ versus $H_{1}:$ there exists an $i$ such that $p^{\text{ave}}_i$ is larger than the (global) average  proportion.
Inspired by popularly used scan statistics~\cite{glaz2001scan}, we propose to use the following test statistic: 
\begin{equation}\label{scan-stat-1}
\mathcal{T} = \max\limits_{t} ~T_{t}~~~\text{where,}~~~~ T_{t}:=   (\widehat{p_{t}^\text{ave}} - \hat{p}_{H_0})/ \hat{\sigma}_{t},
\end{equation}
where, $\hat{p}_{H_0}$ is the (estimated) global proportion of the signal estimated under the null hypothesis;
$\widehat{p_{t}^\text{ave}}$ is an estimate of $p_{t}^\text{ave}$;
and $\hat{\sigma}_{t}$ is the standard deviation of $\widehat{p_{t}^\text{ave}}$ evaluated under the
null ($H_0$). $\mathcal{T}$ measures the strength of a locally contiguous period of heightened activity--the larger this value, the more pronounced is the localized traffic. 
We use a permutation based approach to compute the distribution of $\mathcal{T}$ under the null. 
Figure~\ref{fig:all-together1} shows different communication streams with their associated p-values computed using the framework described above;
and varying levels of activity in the different representative communication streams -- a large p-value for panel (a) (corresponding to \TAGS FI) signifies a lack of 
interesting activity in this series -- this aligns with an expert's understanding of historical events.
A table providing a summary of how many cables survive different p-value thresholds are provided in Table~\ref{tab:cutoffs-prelim} (Appenfix).

%

We also used a test-statistic depending upon the likelihood ratio test: we replaced $T_{t}$ (in display~\eqref{scan-stat-1}) by the local likelihood ratio derived from the binomial distribution; assuming that 
$(y_{t}|n_{t},p_{t})$ are independently distributed $\text{Bin}(n_{t}, p_{t})$ for $t = 1, \ldots, N$.
 The results obtained via this likelihood based approach was quite similar to that obtained from the model-free testing procedure 
 described above; and are hence omitted for the sake of brevity. 


\subsection{Identifying Jumps in Communication Streams}
\label{sec:model}
The method in Section~\ref{sec:interest-1} simply identifies whether something statistically interesting, in the form of a 
heightened localized activity (for example), is occurring somewhere in a communication stream. It does not inform us about the structure or 
precise location of such an activity. 
We proceed to explore some finer structural properties of the series.  Inspired by popularly used signal segmentation/estimation methods~\cite{TSRZ2005,Johnson2013fldp,mammen1997locally} in statistical signal processing, we seek to identify breaks or jumps in a piecewise constant approximation of the signal $t \mapsto p_{t}$ -- these changes are  more localized than the (global) deviations studied in Section~\ref{sec:interest-1}.

We use the notation of Section~\ref{sec:interest-1}.
We assume herein that $(y_{t}|n_{t},p_{t}) \stackrel{\text{ind}}{\sim} \text{Bin}(n_{t}, p_{t})$, where, 
$p_{t}$ denotes the probability of success. 
Our model assumes that the parameters $p_{t}$'s are piecewise constant. Locations where the underlying signal $t \mapsto p_{t}$ exhibits a discontinuity will be called a ``jump'' in the communication stream.
Instead of working directly with the proportion $p_{t}$'s we will use the logistic-link function: $p_{t} = \exp(\theta_{t})/(1+ \exp(\theta_{t}))$ and will treat 
$\theta_{t}$ as a natural parameter.
 This leads to the following regularized estimation problem:
\begin{equation}\label{eqn-opt-1}
\mini_{\theta_{t}, 1 \leq t \leq N} \;\; \phi(\B\theta):= \sum_{t=1}^{N} \Big(-y_{t} \theta_{t} + n_{t}  \log(1 + \exp(\theta_{t})) \Big) +  \lambda H(\B{\theta}),
\end{equation}
where, 
${\mathcal L}(\B\theta):=\sum_{t=1}^{N} \left(-y_{t} \theta_{t} + n_{t}  \log(1 + \exp(\theta_{t})) \right)$, the negative logarithm of the likelihood is the data-fidelity term and $H(\B\theta)$ is the regularizer.
$H(\B\theta)$ encourages the estimated $\theta_{t}$'s (and hence the proportion $p_{t}$'s) to be piecewise constant and the regularization parameter $\lambda >0$ controls the amount of shrinkage. Two common examples of $H(\B\theta)$ that we use herein are~\cite{TSRZ2005,mammen1997locally,boysen2009consistencies,Johnson2013fldp}:
\begin{itemize}
\item $\ell_1$-segmentation (Fused Lasso): $H(\B\theta) = H_{\ell_{1}}(\B\theta) = \sum\limits_{t=1}^{N-1} |\theta_{t+1} - \theta_{t}|$, which penalizes the total variation of a signal, which may also be 
thought as a soft-version of the number of the number of jumps in $\theta_{t}, t \geq 1$.
\item $\ell_0$-segmentation: Here we take $H(\B\theta) = H_{\ell_{0}}(\B\theta) = \sum\limits_{t=1}^{N-1} \M{1} (\theta_{t+1} \neq \theta_{t})$, which penalizes the number of jumps in the signal $\theta_{t}$.
\end{itemize}

We assume above and in the discussion below, that the time points are equally spaced. 
If they are not equispaced, the penalty function needs to be adjusted in a straightforward fashion
as discussed in Section~\ref{admm-details-irreg}. 

For the $\ell_{1}$ penalty function $H_{\ell_{1}}(\B\theta)$, Problem~\eqref{eqn-opt-1} is a convex optimization problem. The $\ell_{1}$-penalty on the successive differences 
in $\B\theta$, is commonly referred to as the fused lasso or total-variation penalty~\cite{TSRZ2005,mammen1997locally}. This semi-norm induces shrinkage along with sparsity on the coefficient differences $\theta_{t+1} - \theta_{t}$. However, the $\ell_{1}$-based penalty often over-estimates the number of jumps when the tuning parameter is chosen so as to obtain a model with 
good data-fidelity -- this is especially true if the tuning parameter is chosen based on a held out test set to minimize test error. This is due to the shrinkage effect of the $\ell_{1}$-penalty, which severely penalizes large values of the 
jumps $\theta_{t+1} - \theta_{t}$. To obtain a model with fewer jumps, the regularization parameter needs to be made larger -- this however, may lead to a model where some of  the important jumps are missed.  Many of these limitations can be overcome by using a $\ell_{0}$-based penalty~\cite{boysen2009consistencies} which directly penalizes the number of jumps and is less agnostic to the precise value of the jump.
The caveat however, is that the resulting optimization problem becomes non-convex and discrete optimization 
methods are required to obtain the global minimum of such problems~\cite{boysen2009consistencies,Johnson2013fldp}. Developing global optimization algorithms for Problem~\eqref{eqn-opt-1} (for the penalty $H_{\ell_0}(\B\theta)$)
along the lines of~\cite{mazumder2017discrete,bertsimas2016best} is beyond the scope of the current paper. Herein, we describe fast and robust algorithms to obtain good quality solutions to Problem~\eqref{eqn-opt-1}. 
Our proposal is motivated by first-order optimization based algorithms~\cite{nesterov2004introductorynew} pioneered in the convex optimization community. Loosely speaking, these are iterative methods that can be used to obtain high quality approximate solutions for  convex optimization tasks, compared to off-the-shelf interior point methods that are difficult to scale to large problems.  
These methods apply to both $H_{\ell_1}(\B\theta)$ and $H_{\ell_0}(\B\theta)$ -- however, there are certain subtleties as we describe below. 
 
\subsubsection{Model Fitting: Optimization Algorithm}\label{opt-algo-1}
Problem~\eqref{eqn-opt-1} is of the composite form, i.e., the objective function can be written as the sum of a smooth convex function ${\mathcal L}(\B\theta)$ and a non-smooth penalty function $H(\B\theta)$. We will apply proximal gradient 
descent methods~\cite{fista-09} for this problem. The negative log-likelihood function ${\mathcal L}(\B\theta)$ is continuously differentiable and satisfies:
\begin{equation}\label{lip-one-1}
\| \nabla {\mathcal L}(\M{u})  -   \nabla {\mathcal L}(\M{v})  \| \leq \ell \| \M{u} - \M{v} \|,~~~~~\forall \M{u}, \M{v};
\end{equation}
with $\ell = \frac{1}{4} \max_{i=1} y_{i}$. This follows by noting that the $i$th coordinate of $\nabla {\mathcal L}(\M{u})$ is:
$\{\nabla {\mathcal L}(\M{u})\}_i = -y_{i} + n_{i}\exp(u_{i})/(1 + \exp(u_{i}))$
and $\nabla^2 {\mathcal L}(\M{u})$ is a diagonal matrix with the $i$th diagonal entry being:
\begin{equation}\label{line-1}
\left\{\nabla^2 {\mathcal L}(\M{u})\right\}_{ii} = n_{i}\exp(u_{i})/(1 + \exp(u_{i}))^2 \leq \frac{1}{4} n_{i},~~~ i = 1, \ldots, N.
\end{equation}
Hence, the largest eigenvalue of $\nabla^2 {\mathcal L}(\M{u})$, i.e., $\lambda_{\max}(\nabla^2 {\mathcal L}(\M{u})) \leq \frac14 \max_{i=1}^{N} n_{i}$, which justifies the choice of $\ell$, as above. 
For a fixed $L \geq \ell$, our  algorithm performs the following updates (for  all $k \geq 1$):
\begin{equation}\label{ist-algo}
\B\theta_{k+1} \in \argmin_{\B\theta} \; \frac{L}{2} \left\| \B\theta - \left(  \B\theta_{k} - \frac{1}{L} \nabla {\mathcal L}(\B\theta_{k}) \right) \right \|_{2}^2 + H(\B\theta).
\end{equation}
This leads to a decreasing sequence of objective values $\phi(\B\theta_{k+1}) \leq \phi(\B\theta_{k})$ for $k \geq 1$.
We now study the fate of the sequence $\B\theta_{k}$, depending upon the choice of $H(\B\theta)$.

\paragraph{The fused lasso penalty ($H_{\ell_{1}}(\B\theta)$)} We first consider the case where the regularizer and hence the function $\phi(\B\theta)$ is convex in $\B\theta$.
The sequence $\B\theta_{k}$ converges to a minimum of Problem~\eqref{eqn-opt-1} with the penalty function $H_{\ell_{1}}(\B\theta)$. More precisely, it can be shown that~\cite{fista-09}
\begin{equation}\label{conv-rate-0}
\phi(\B\theta_{k}) - \phi(\B\theta^*) \leq \frac{L}{2k} \| \B\theta_{k} - \B\theta^* \|_{2}^2,
\end{equation}
where, $\B\theta^*$ is an optimal solution to Problem~\eqref{eqn-opt-1}. Display~\eqref{conv-rate-0} states that the sequence 
$\phi(\B\theta_{k})$ converges to a minimizer of Problem~\eqref{eqn-opt-1} with a rate of $O(\tfrac{1}{k})$. In fact, this rate can be improved further under a minor additional assumption.
Let  $\mu_{k} = \min_{i=1, \ldots, N}~\left\{\nabla^2 {\mathcal L}(\B\theta_{k})\right\}_{ii}$, i.e., the smallest diagonal entry of the Hessian of ${\mathcal L}(\B\theta)$. As long as $\B\theta_{k}$'s are uniformly bounded and 
$\min_{i} \; n_{i} > 0$ then $\mu :=\min_k \mu_{k} > 0$. The rate of convergence is linear; and is given by:
\begin{equation}\label{conv-rate-1}
\phi(\B\theta_{k}) - \phi(\B\theta^*) \leq  \gamma^k \left(\phi(\B\theta_{0}) - \phi(\B\theta^*) \right),
\end{equation}
where, 
\begin{equation}
\gamma = \begin{cases}
\frac{L}{\mu} & \text{if $\frac{\mu}{L} \geq 2$}\\
(1 - \frac{\mu}{4L}) &\text{otherwise}.
\end{cases}
\end{equation}

Note that sub-problem~\eqref{ist-algo} requires solving a problem of the form:
\begin{equation}\label{dyn-prog-1}
\mini_{\M{u} \in \Re^{N} }  \;\; \frac12 \| \M{u} - \bar{\M{u}} \|_{2}^2 + \lambda' \sum_{i=1}^{N-1} | u_{i+1} - u_{i} |,
\end{equation}
which can be solved very efficiently via Dynamic Programming~\cite{Johnson2013fldp} 
with cost $O(N)$ for $\lambda' >0$ --  
one can solve instances with $N$ a few thousands in a few milliseconds. 

\paragraph{The $\ell_{0}$-segmentation penalty ($H_{\ell_{0}}(\B\theta)$)} The algorithm above, also applies to
the regularizer $H_{\ell_0}(\B\theta)$. In update~\eqref{ist-algo} we set $H(\theta)$ to $H_{\ell_0}(\B\theta)$. This requires solving
\begin{equation}\label{dyn-prog-0}
\mini_{\M{u} \in \Re^{N} }  \;\; \frac12 \| \M{u} - \bar{\M{u}} \|_{2}^2 + \lambda' \sum_{i=1}^{N-1}  \M{1} ( u_{i+1}  \neq u_{i}),
\end{equation}
which can also be computed efficiently using the dynamic programming algorithm of~\cite{Johnson2013fldp}. Describing the properties of this sequence 
$\B\theta_{k}, k \geq 1$ is subtle since the associated optimization problem~\eqref{eqn-opt-1}  is non-convex. Following~\cite{bertsimas2016best} it can be shown that 
the sequence $\phi(\B\theta_{k})$ is decreasing, bounded below\footnote{We will assume that the function $\phi(\B\theta)$ is bounded below and the minimum exists.} and it converges to 
$\phi^*$ (say), which may not be a global minimum. 
We say that $\tilde{\B\theta}$ is a first-order stationary point for Problem~\eqref{eqn-opt-1} if it satisfies the following 
fixed point equation:
$$\tilde{\B\theta} \in \argmin_{\B\theta} \; \frac{L}{2} \left\| \B\theta - \left(  \tilde{\B\theta} - \frac{1}{L} \nabla {\mathcal L}(\tilde{\B\theta}) \right) \right \|_{2}^2 + H_{\ell_0}(\B\theta).$$
 $\B\theta_{k}$ is said to be an $\epsilon$-accurate first order stationary point, if $ \|\B\theta_{k+1} - \B\theta_{k} \|^2_{2} \leq \epsilon$.
Following the convergence analysis in~\cite{bertsimas2016best}[Theorem 3.1], we obtain the following finite-time convergence rate of $\B\theta_{k}$ to a first order stationary point:
\begin{equation}\label{conv-rate-l0-1}
 \min\limits_{1 \leq k \leq K} \| \B\theta_{k+1} - \B\theta_{k} \|_{2}^2 \leq \frac{2(\phi(\B\theta_{1}) - \phi^*)}{K(L - \ell)}.
 \end{equation}
Note that this algorithm may not reach the global minimum of the $\ell_0$ version of Problem~\eqref{eqn-opt-1}. However, in practice, it reaches a high quality solution quite fast. 

\subsubsection{Estimated Signal}\label{sec:signals-1}

\begin{figure}[h!]
	\centering
	\begin{tabular}{cc}
	& {\sf \scriptsize $\ell_1$ regularization}\\
		\rotatebox{90}{{\graphFont ~~~~~Proportion}}
		&\includegraphics[scale = 0.5,clip = true, trim = 30 25 10 40]{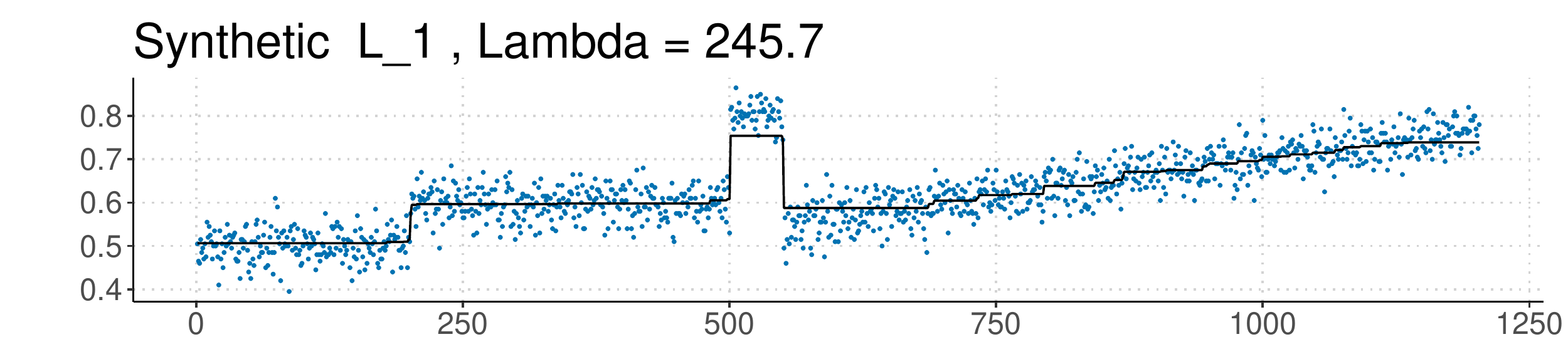}\\~\\
		& {\sf \scriptsize $\ell_0$ regularization}\\
		\rotatebox{90}{{\graphFont ~~~~~~~~Proportion}}
		&\includegraphics[scale = 0.5,clip = true, trim = 30 0 10 40]{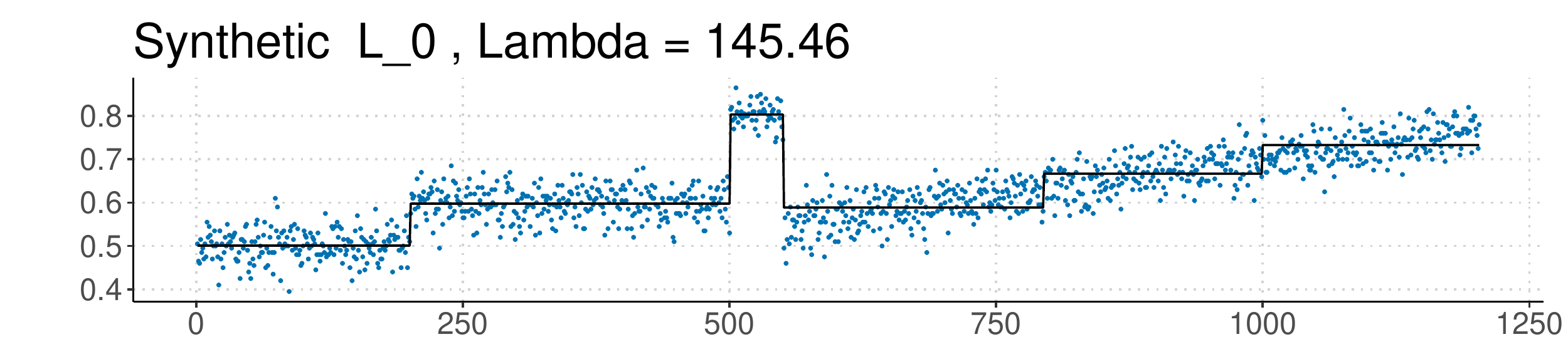}\\
	\end{tabular}
	\caption{\small{Estimators obtained from Problem~\eqref{eqn-opt-1} with $\ell_1$ (upper panel) and $\ell_0$ (lower panel) regularization. The data is synthetic and the underlying signal contains two sharp jumps and a gradual increasing trend. We use cross validation to select a value of $\lambda$. 
	$\ell_1$ penalty shrinks the estimated probability during a big burst ($501 \leq t \leq 550$) and gives more jumps during the gradual increase period ($551 \leq t \leq 1203$).
	} }
	\label{fig:synthL0L1}
\end{figure}

To gather some intuition about the behavior of the segmentation methods described above, we consider a synthetic example in Figure~\ref{fig:synthL0L1} -- here the underlying (true) signal is piecewise constant with three levels up to time point $t_0$, there is a right discontinuity at $t_0$ after which it becomes linear
\footnote{More specifically, data is generated by $y_t \stackrel{\text{ind}}{\sim} \text{Bin}(n_t=200, p_t), t = 1,..., 1203$, where $p_t = 0.5$ for $1 \leq t \leq 200$; $p_t = 0.6$ for $201 \leq t \leq 500$; $p_t = 0.8$ for $501 \leq t \leq 550$; $p_t = 0.55 + (t-550) / 3000$ for $551 \leq t \leq 1203$.}. The figure presents the signal estimates (for both the $\ell_{0}$ and $\ell_{1}$ penalties) at the cross-validated choices of the tuning parameter -- we use
$k$-fold cross validation~\cite{FHT-09-new} which is also used in the R package~{\texttt{genlasso}} (Since we want to ensure each fold is representative of the time series, instead of randomly assigning points to a fold, we systematically assign points by placing every $k$th point into the same fold).
For both segmentation schemes, the estimated signals $\{\hat{p}_{t}\}$ serve as good (overall) approximations of $\{p_{t}\}$ -- however, there are some subtle differences. 
First of all, the $\ell_{1}$-segmentation scheme leads to biased estimates and the bias becomes quite prominent in estimating the jump at the centre of the signal. This behavior is not present for the $\ell_{0}$-scheme.
In addition, the estimates for  the
linear component (at the right) also differ across the $\ell_{0}$ and $\ell_{1}$ schemes. The $\ell_{0}$ regularizer leads to a fewer number of segments (here three) compared to the $\ell_{1}$-penalty which has several smaller jumps.


\begin{figure}[h!]
		\centering
	\resizebox{1.01\textwidth}{0.3\textheight}{\begin{tabular}{ ccc}
		&{\graphFont $\ell_0$-segmentation}&{\graphFont $\ell_{1}$-segmentation}\\
\rotatebox{90}{~~~~~{\graphFont Proportion}}		&\includegraphics[width=0.45\textwidth,clip = true,trim = 30 25 0 60]{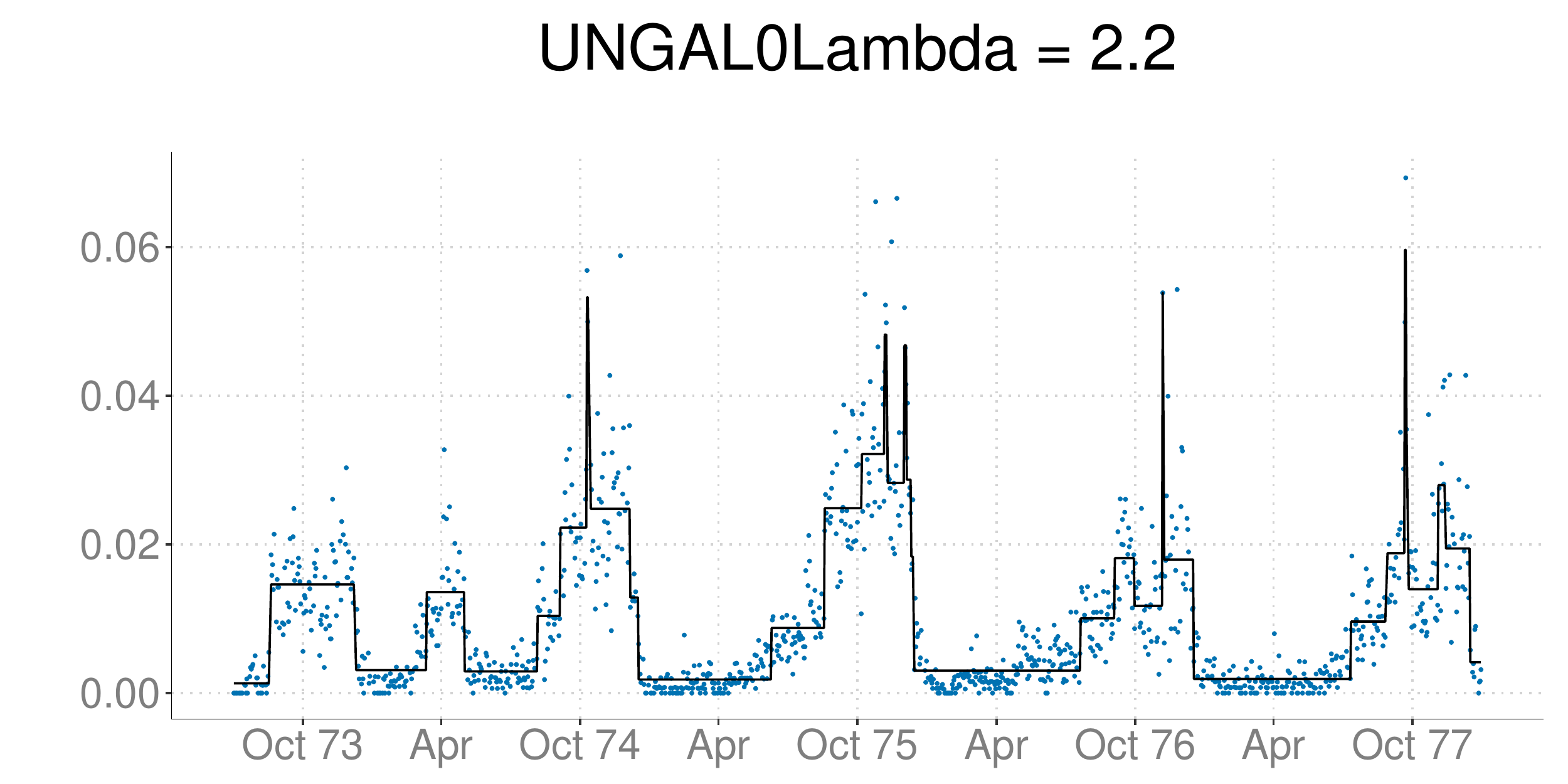}
		&\includegraphics[width=0.45\textwidth,clip = true,trim = 30 25 0 60]{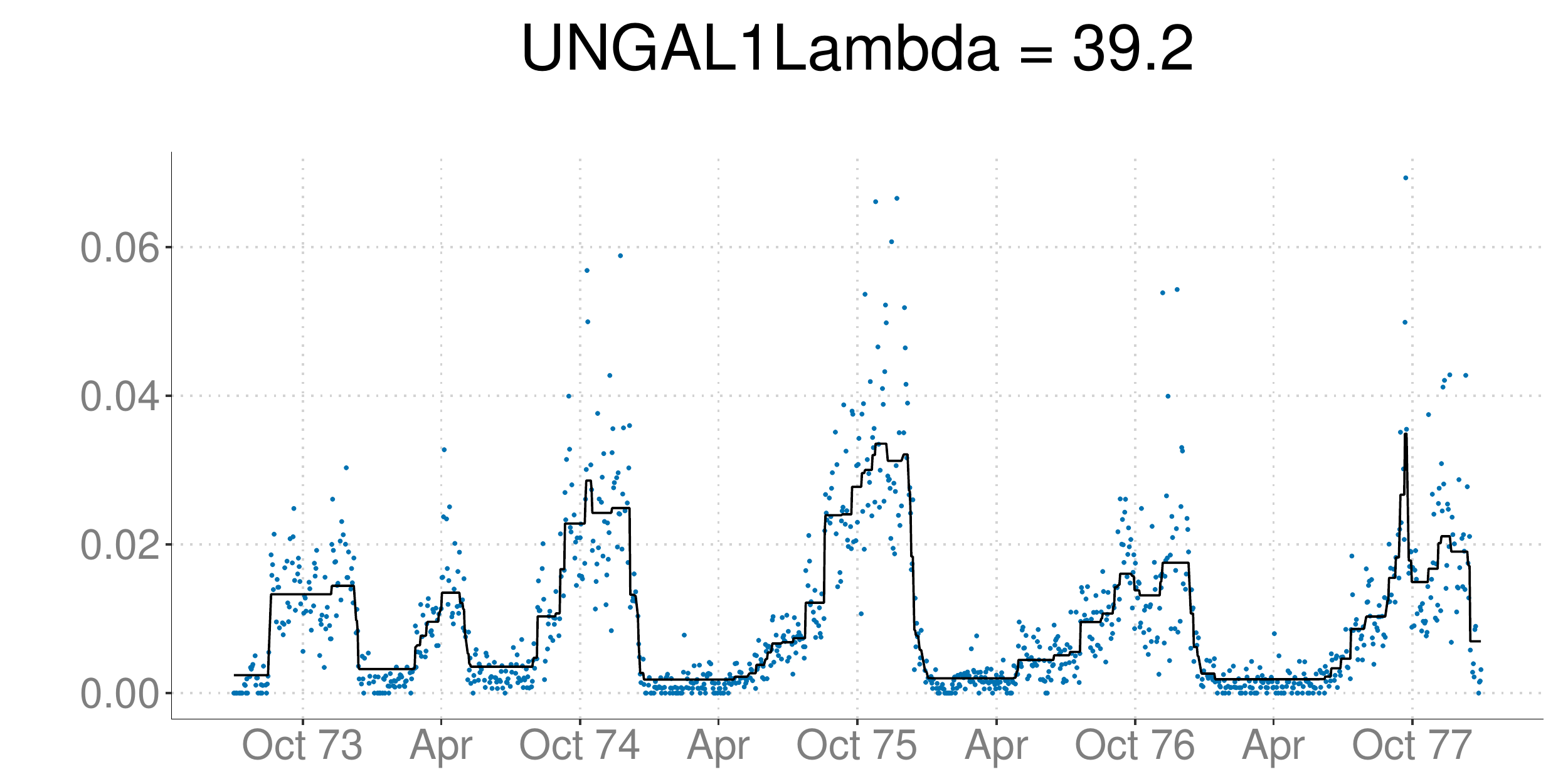}
		\\~\\
	\rotatebox{90}{~~~~~~{\graphFont Proportion}}	&\includegraphics[width=0.45\textwidth,clip = true,trim = 30 25 0 60]{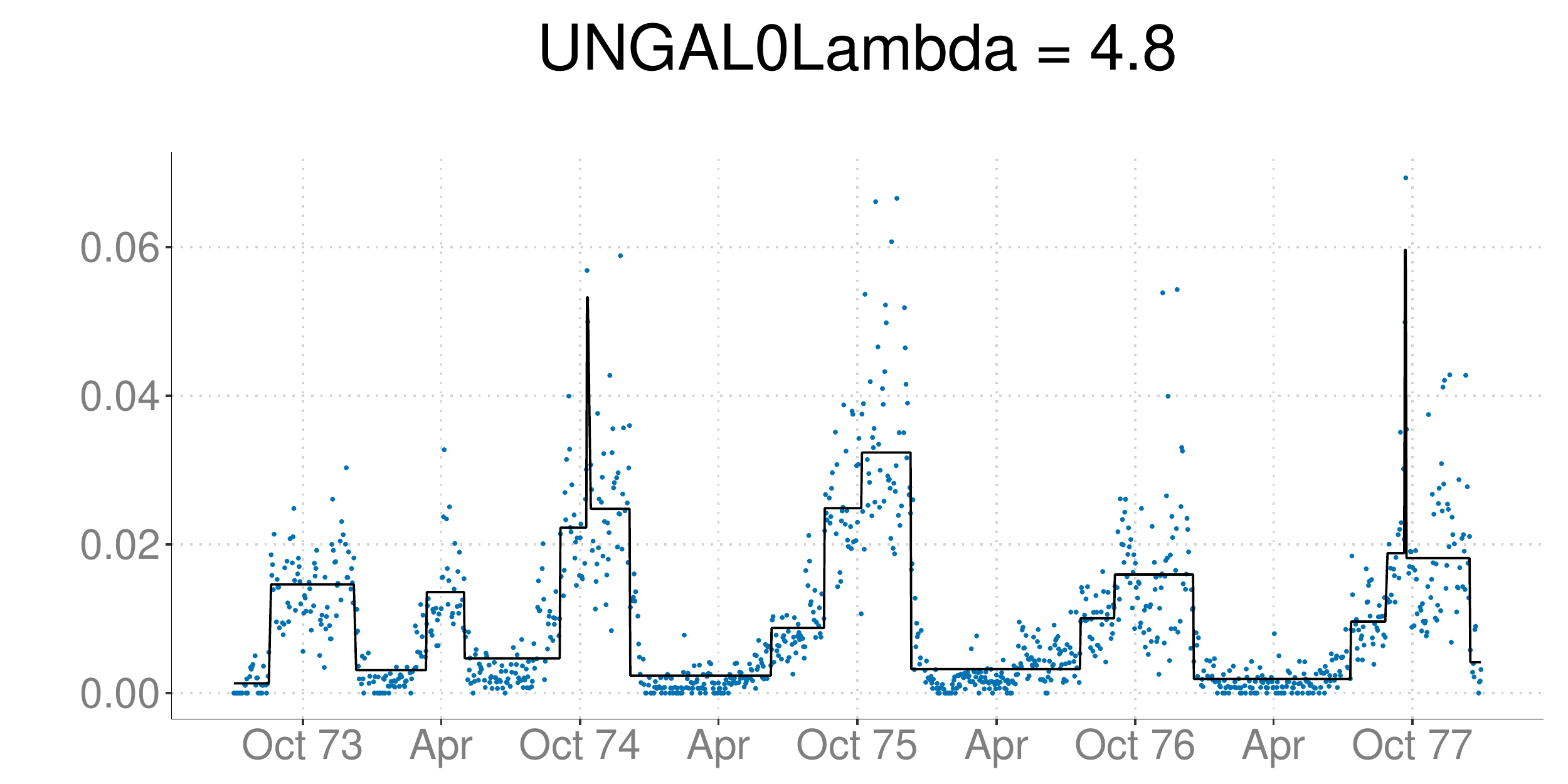}
		&\includegraphics[width=0.45\textwidth,clip = true,trim = 30 25 0 60]{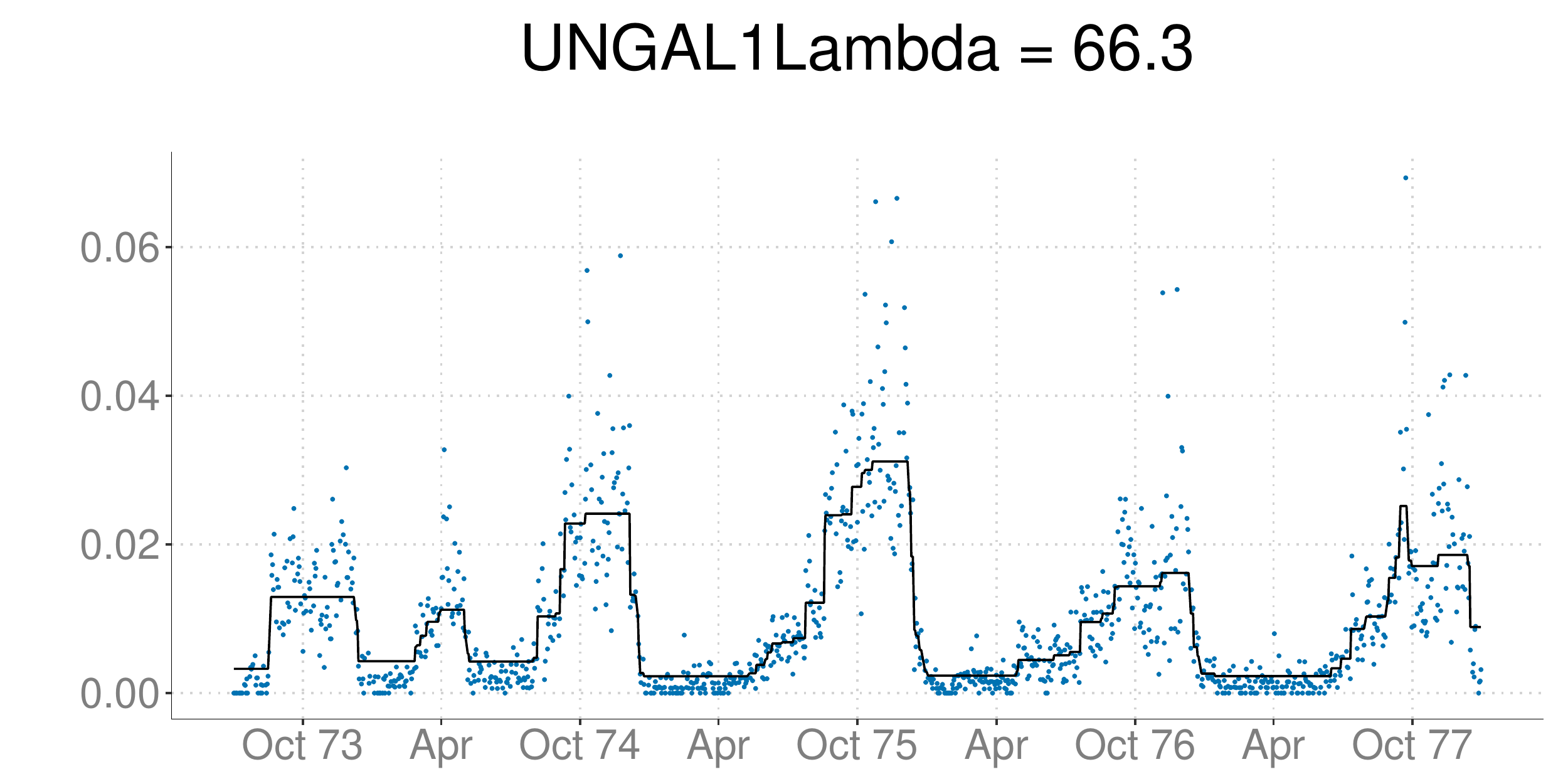}\\~\\
	\rotatebox{90}{~~~~~~{\graphFont Proportion}}	&\includegraphics[width=0.45\textwidth,clip = true,trim = 30 0 0 60]{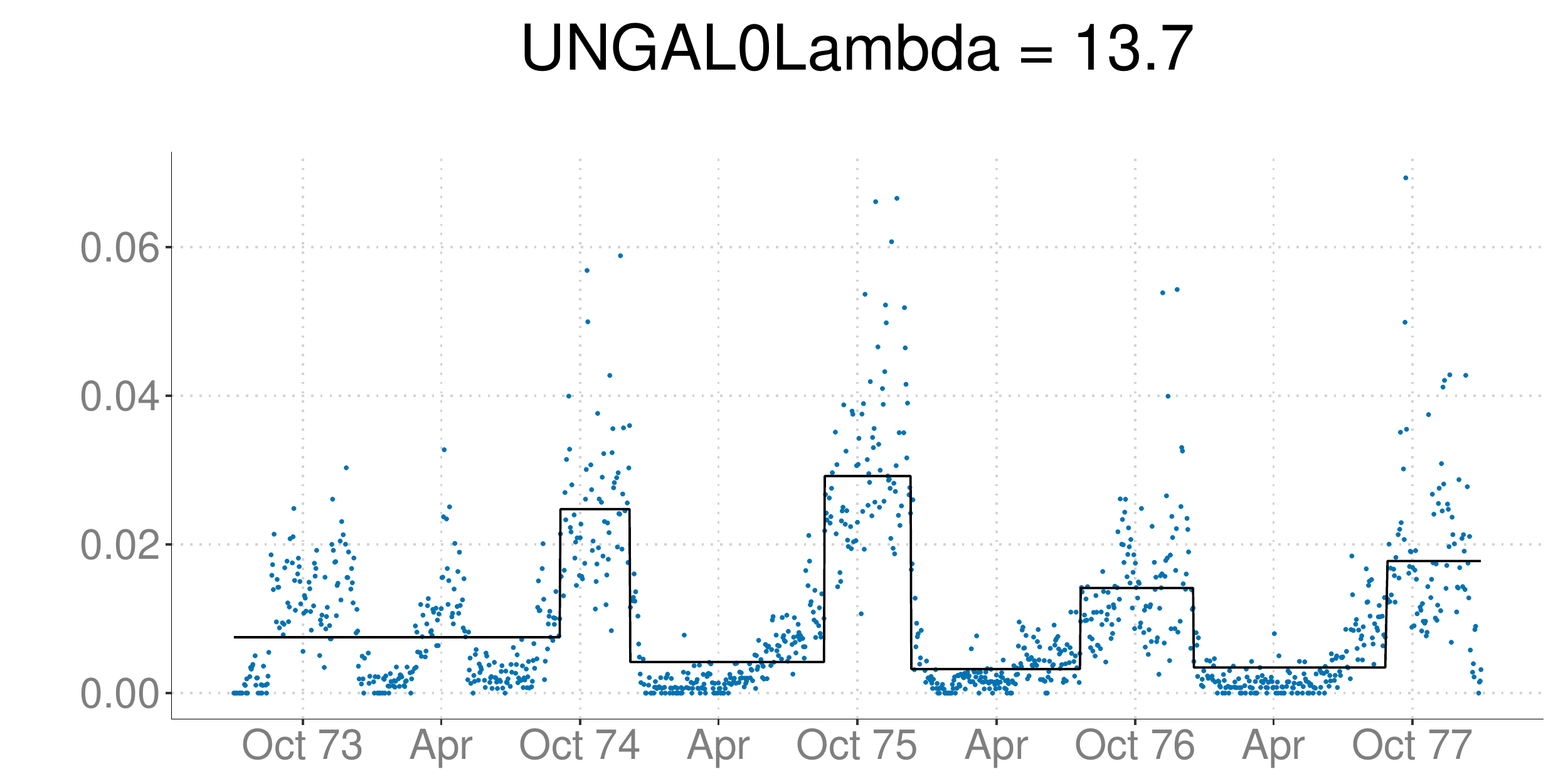}
		&\includegraphics[width=0.45\textwidth,clip = true,trim = 30 0 0 60]{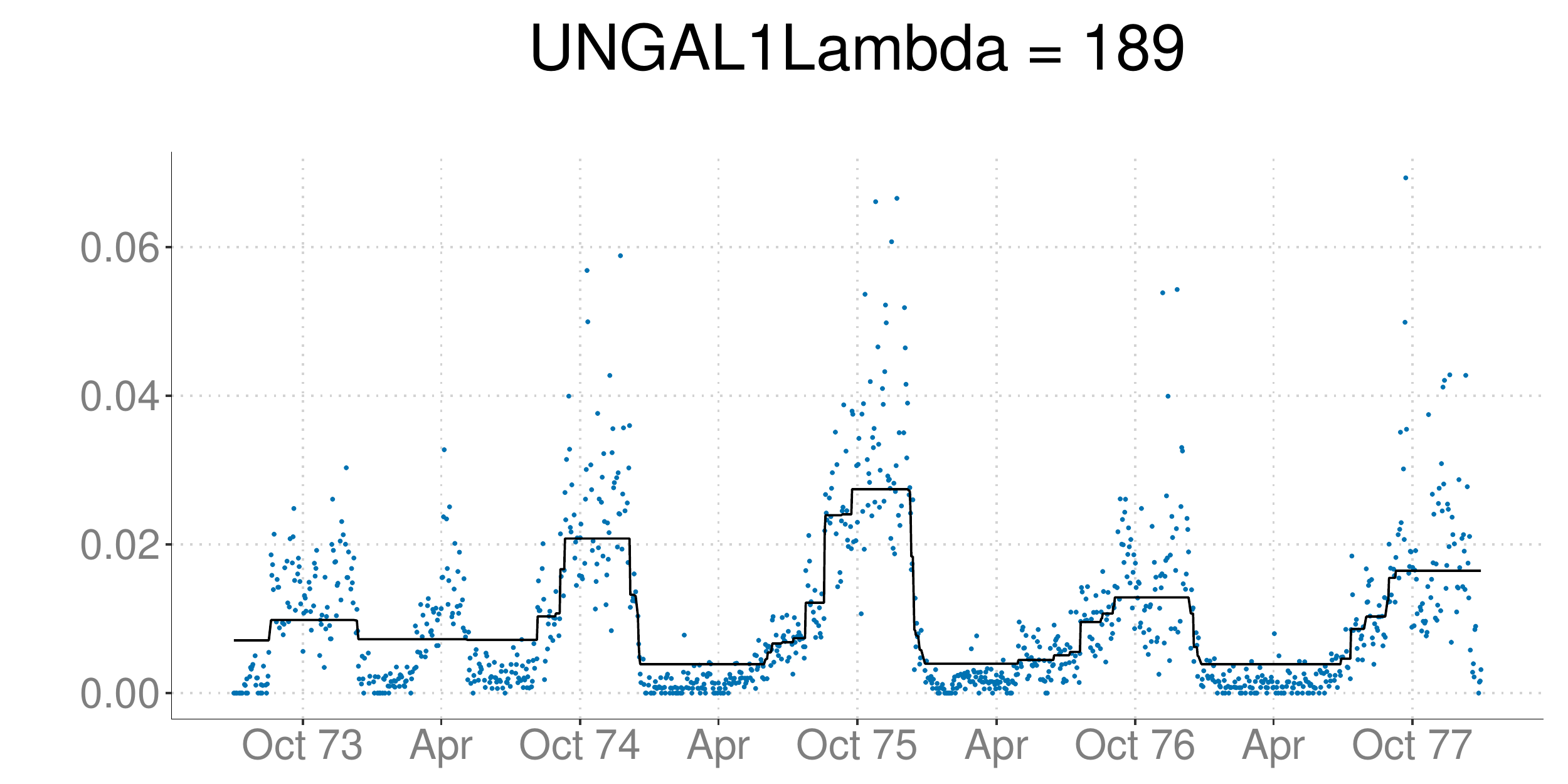}
			\end{tabular}}
		\caption{\small{Figure showing the raw proportions (in blue dots) for \TAGS UNGA (U.N. General Assembly) and the estimated proportions $\hat{p}_{t}, t \geq 1$, as obtained from the regularization framework in Problem~\eqref{eqn-opt-1}. The left panel shows the estimates obtained with the $\ell_{0}$-segmentation penalty  and the right panel shows the estimates with the $\ell_{1}$-segmentation penalty. 
				The middle rows correspond to the optimal $\lambda$ chosen by 10-fold cross-validation. It shows how, in between the cyclical jumps in UN-related communications relating to the regular Fall meetings of the General Assembly, there was also a jump in April-May 1974. This occurred when Algeria called a special session to demand UN support for a ``New International Economic Order." We show a few additional choices of the regularization parameter for each example (see text for details).
				The figure (bottom right panel)				 shows the large bias incurred by the $\ell_{1}$-penalization method in the estimation process -- this behavior is less pronounced with the $\ell_{0}$ penalty. The $\ell_{1}$-penalty also exhibits a stair-casing effect -- with many small jumps. Unlike the $\ell_{1}$-penalty, the $\ell_{0}$-penalty selects some segments that are \emph{very} short, these segments disappear upon increasing the penalty parameter. }
		}
	\label{fig:lamSeqUNGA}
\end{figure}

\begin{figure}[h!]
		\centering
	\resizebox{1.01\textwidth}{0.31\textheight}{\begin{tabular}{ ccc}
		&{\graphFont $\ell_0$-segmentation}&{\graphFont $\ell_{1}$-segmentation}\\
\rotatebox{90}{~~~~~~{\graphFont Proportion}}		&\includegraphics[width=0.45\textwidth,clip = true,trim = 30 25 0 60]{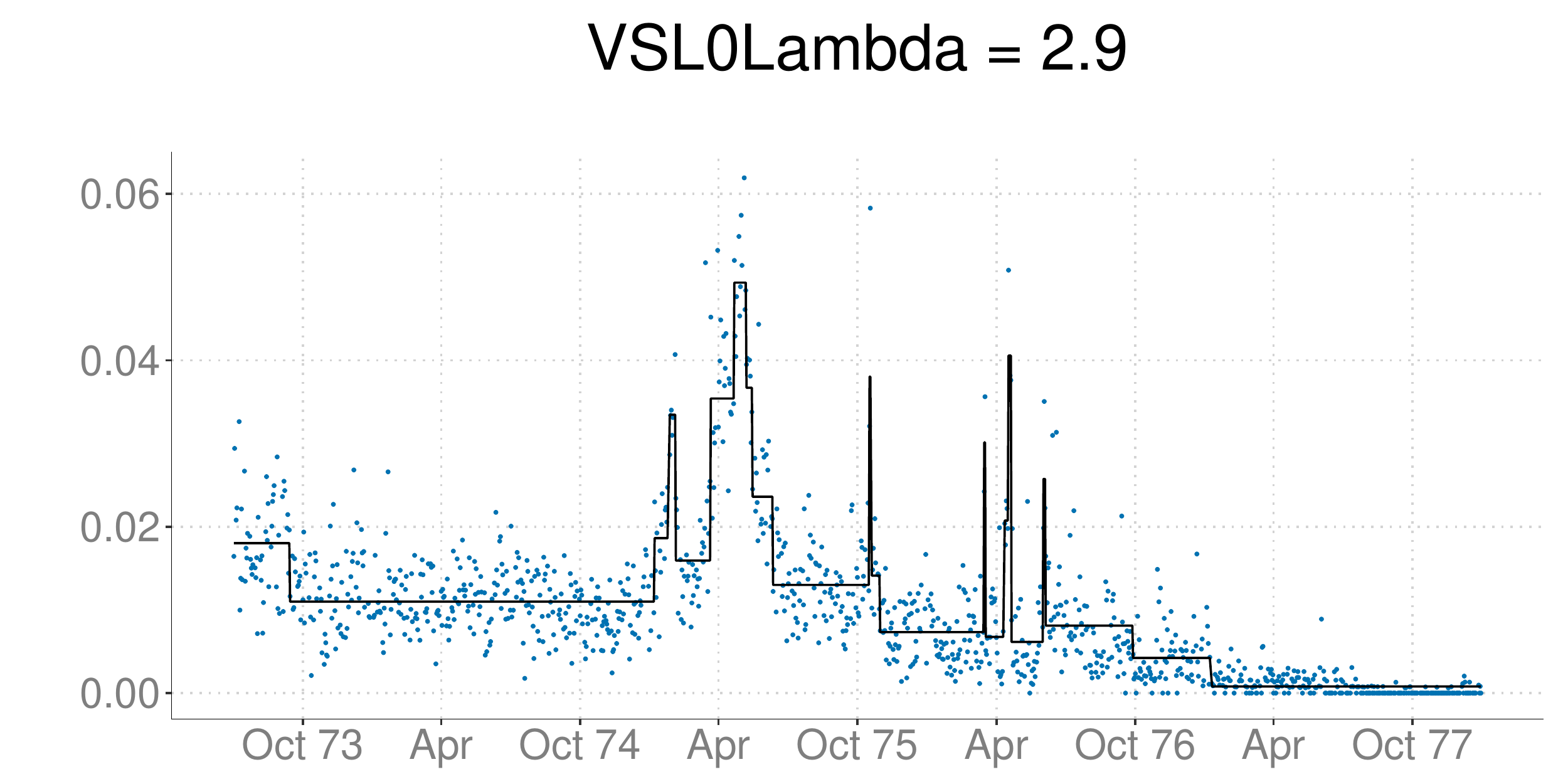}
		&\includegraphics[width=0.45\textwidth,clip = true,trim = 30 25 0 60]{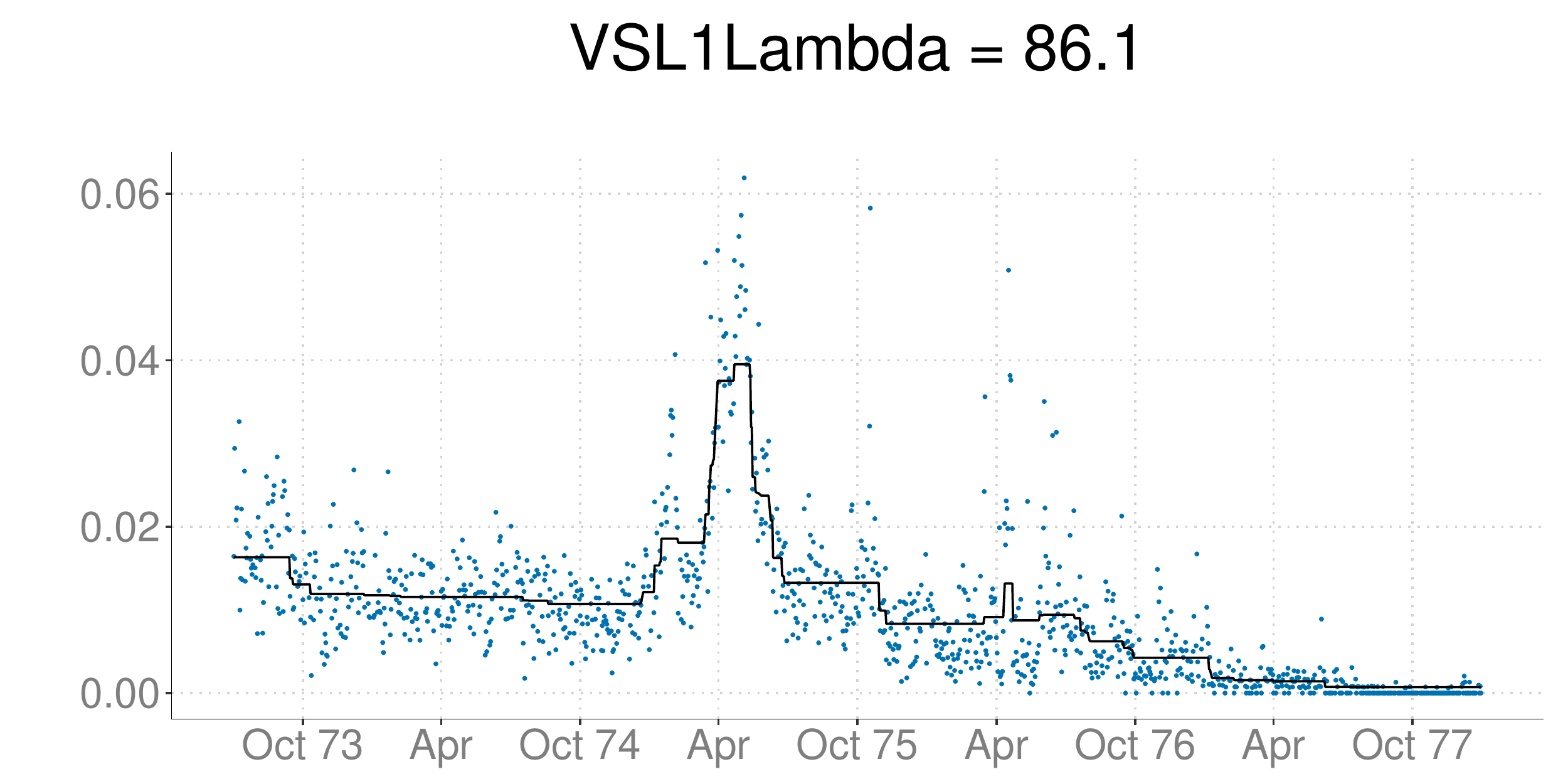}
		\\~\\
	\rotatebox{90}{~~~~~~{\graphFont Proportion}}	&\includegraphics[width=0.45\textwidth,clip = true,trim = 0 25 0 60]{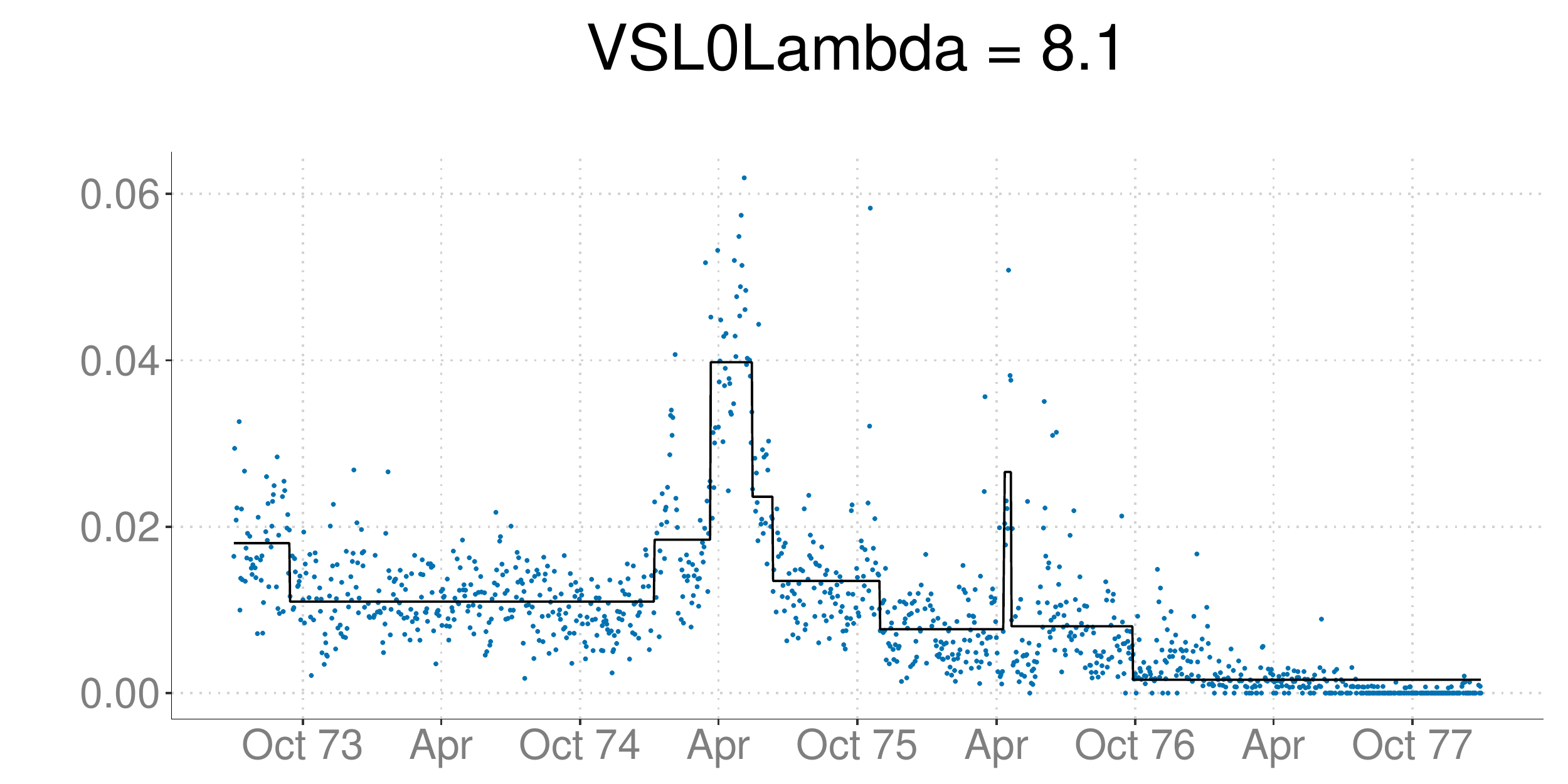}
		&\includegraphics[width=0.45\textwidth,clip = true,trim = 30 25 0 60]{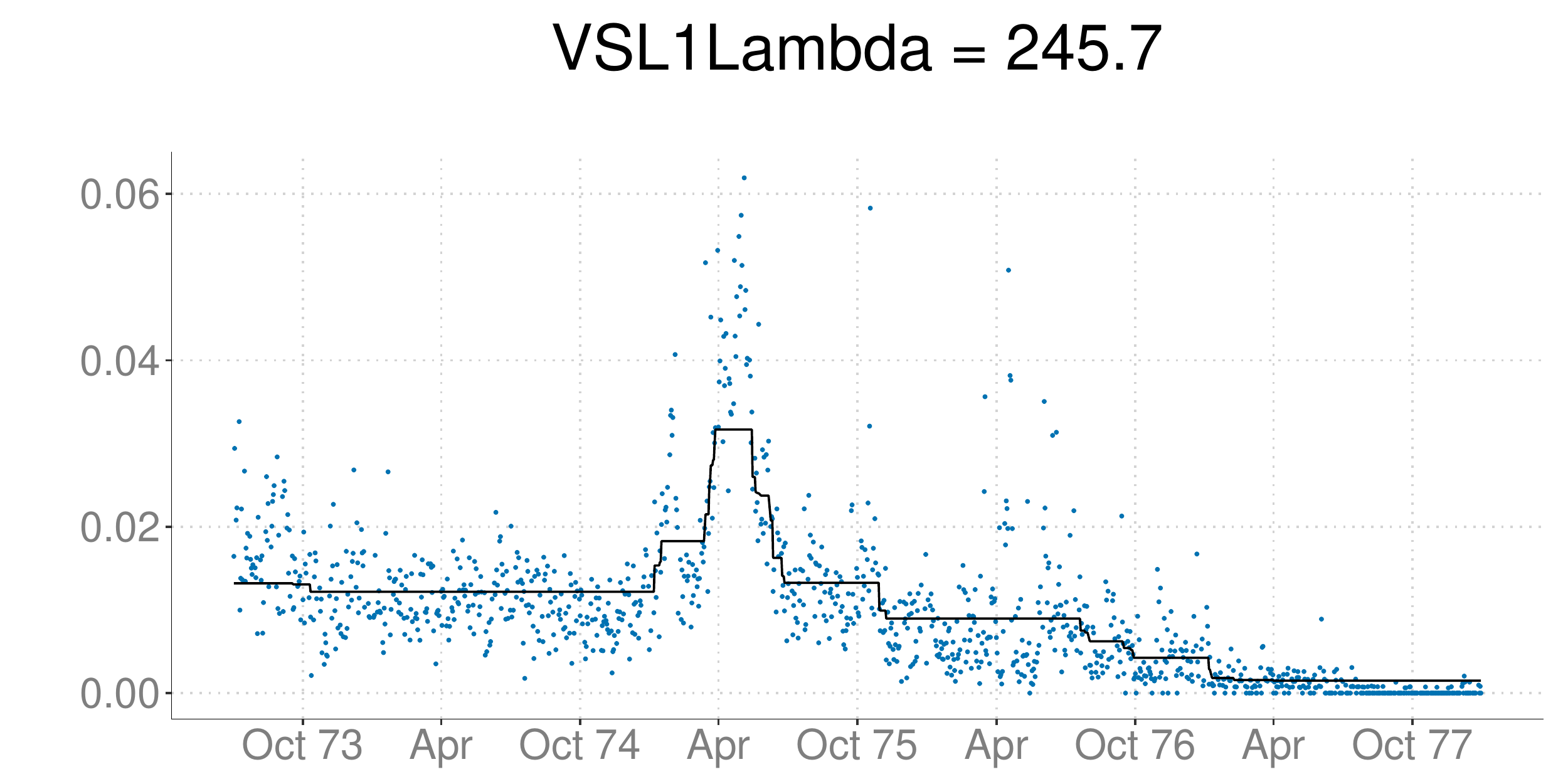}\\~\\
	\rotatebox{90}{~~~~~~{\graphFont Proportion}}	&\includegraphics[width=0.45\textwidth,clip = true,trim = 30 0 0 60]{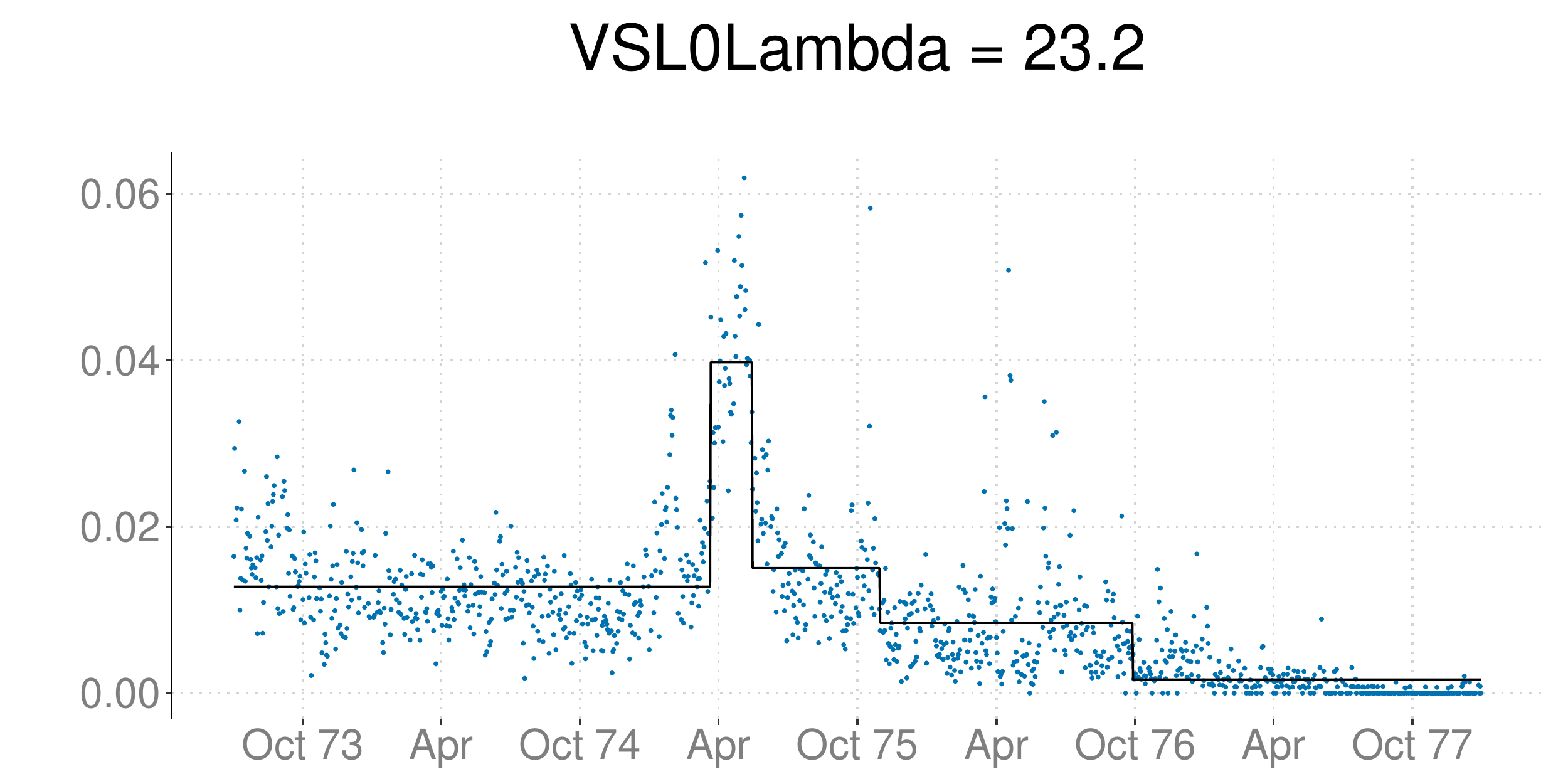}
		&\includegraphics[width=0.45\textwidth,clip = true,trim = 30 0 0 60]{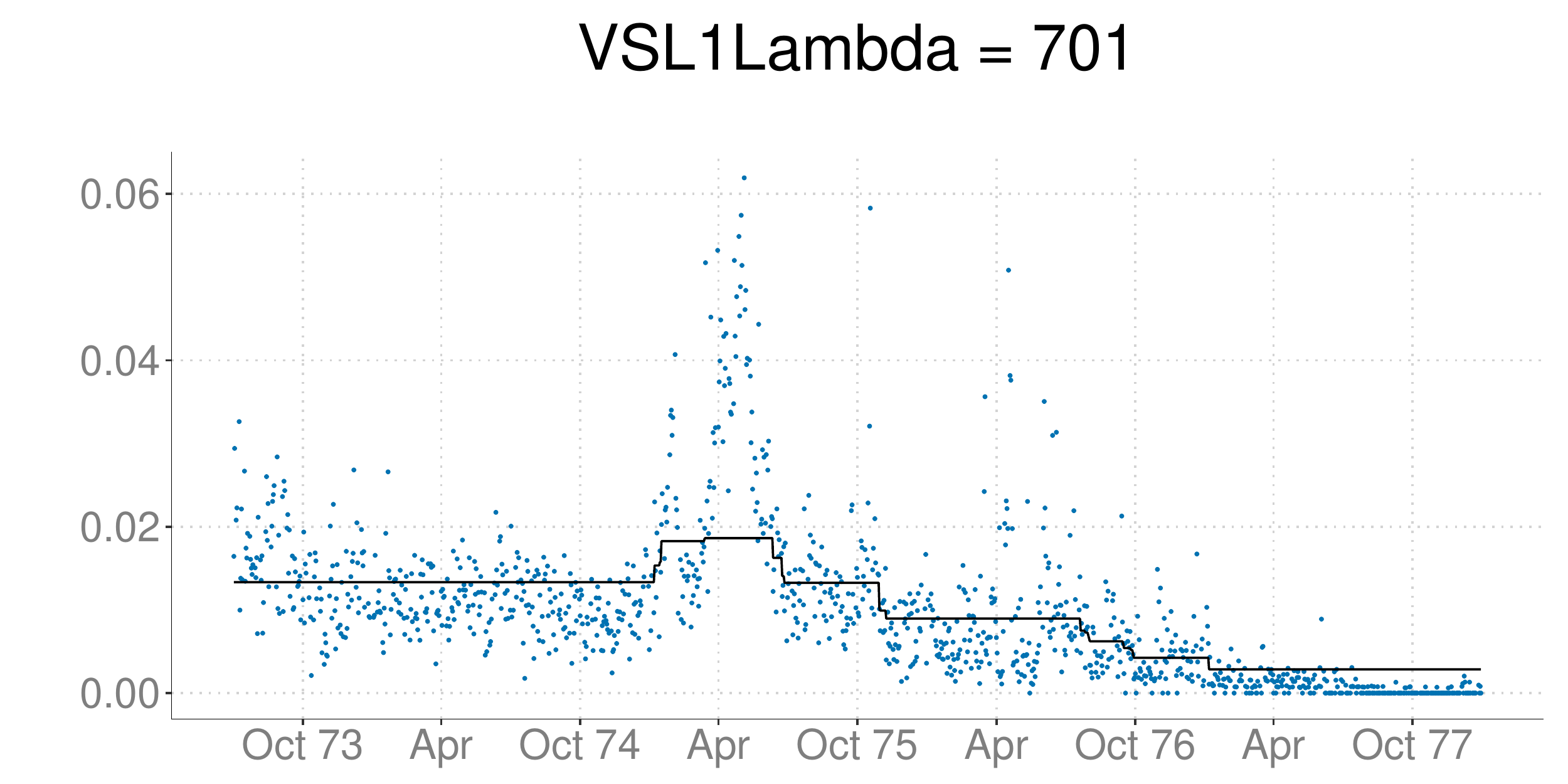}
			\end{tabular}}
		\caption{{\small Figure showing the raw proportions (in blue dots) for \TAGS VS (South Vietnam) and the estimated proportions $\hat{p}_{t}, t \geq 1$, as obtained from the regularization framework in Problem~\eqref{eqn-opt-1}. 
		The left panel shows the results for the $\ell_{0}$-segmentation penalty and the right panel the $\ell_1$-penalty.
		The middle rows correspond to the optimal $\lambda$ chosen by cross-validation, and we show a few additional choices of the regularization parameter for each example.  
		A practitioner might select one or another depending on whether they would want to identify smaller jumps that correspond, in this case, to the refugee crisis that followed the defeat of South Vietnam.}}

	\label{fig:lamSeqVS}
\end{figure}

 Figures~\ref{fig:lamSeqUNGA} and~\ref{fig:lamSeqVS} show the estimated signal proportions $\hat{p}_{t} = \exp(\hat{\theta}_t) / (1 + \exp(\hat{\theta}_t))$, where the $\hat{\theta}_{t}$'s are obtained from  Problem~\eqref{eqn-opt-1}. Both the penalty functions do a good job in estimating a piecewise constant version of the underlying signal -- the $\ell_{0}$ scheme leads to fewer jumps than its $\ell_{1}$ counterpart, for a comparable data-fidelity.  
The figures also show fitted signals for a few other values of $\lambda$ around the cross-validated choice at the center ($\lambda$ increases as one moves down the rows): we include the tuning parameter selected by the one-standard error rule~\cite{FHT-09-new} (see also the R package~{\texttt{genlasso}}).
 We can see that as $\lambda$ decreases, the algorithm captures a more granular structure of the data and estimates more jumps.

\subsection{A deeper investigation of Jumps} \label{deeper-invest-1}

\subsubsection{How intense is a Jump?}\label{signif-cha-point-1}

The procedure in Section~\ref{sec:model} provides an estimated piecewise constant signal $\{\hat{p}_t\}_{t \geq 1}$. In particular, this can be used to identify locations 
where the signal changes: $\hat{p}_{t+1} \neq \hat{p}_{t}$. Of course, this jump depends upon the choice of the tuning parameter, the penalty used and also the underlying signal. 
A jump estimated by the $\ell_{0}$ or $\ell_{1}$-segmentation procedure may reflect (a) a discontinuity in the signal -- in this case, the signal is well approximated by locally constant segments with pieces adapting to the data (b) a localized trend in the signal, as we saw in Figure~\ref{fig:synthL0L1} -- a jump here is a consequence of the slope of $t \mapsto p_{t}$ and not a
 discontinuity in the signal $t \mapsto p_{t}$.

Given an estimate of $\{\hat{p}_{t}\}$ a scholar accustomed to analyzing events through a close reading of historical documents may ask:

\begin{itemize}
\item Which of these jumps might be important or are indicative of a historical event of interest?
\item Can one obtain a rank ordering of the jumps based on their intensities?  
\end{itemize}

We formalize this question as follows: given an estimate of $\{\hat{p}_{t}\}$ and a set of candidate 
jumps, can we identify jumps that are \emph{strong enough}?
Ideally, we would like a simple measure that associates a score to the strength and size of a  jump selected by the estimation procedure -- this would help us select a smaller set of 
jumps that merit closer scrutiny. Towards this end, we use a sample splitting\footnote{Due to the large number of samples, sample splitting does not significantly reduce the size of the training dataset.} procedure~\cite{wasserman2009high}: a subsample of size 50\% 
of the data is used for estimating the location of the jumps and the remaining held out part of the data is used to associate a p-value score (the method is described below) to each jump identified in the first stage.  This method is simple, intuitive and provides a natural method to rank the jumps in a communication stream. Using this scheme, one can potentially reduce the number of jumps selected by the fitting procedure by screening out jumps with p-values larger than a  user-defined pre-specified threshold.

We describe our approach with reference to the $\ell_0$-penalization procedure, though the idea will also apply for the $\ell_{1}$-penalized estimate. 
Suppose a candidate location for the change point 
$\hat{t}$ is estimated based on the first part of the sample (used for estimating the signal). The test statistic is evaluated on the held out sample.
We take a neighborhood of size $2\Delta$ centered at $\hat{t}$, 
and denote the time points on the left of $\hat{t}$ as $L(\hat{t}, \Delta)$ and those on the right of $\hat{t}$ as $R(\hat{t}, \Delta)$.
Let us assume that $p_{t}$ for $t \in L(\hat{t}, \Delta)$ are all equal to $p(L, \hat{t})$; and $p_{t}$ for $t \in R(\hat{t}, \Delta)$ are all equal to $p(R, \hat{t})$.
We then test the null hypothesis ($H_0$) that the proportions on the left and right parts 
of $\hat{t}$ are equal: $p(L, \hat{t}) = p(R, \hat{t})$; versus the 
alternative ($H_1$) that $p(L, \hat{t}) \neq p(R, \hat{t})$.  We use the likelihood ratio test statistic for this purpose. 
To compute the null distribution, we used a two step procedure. We first identified segments of the time series
which did not overlap with any candidate change point location (i.e., parts of the series where the estimated signal was 
constant for stretches of size at least  $2\Delta+1$) of the time series. Based on the regions thus selected (i.e., the locally constant stretches of the signal), we simulated the null distribution 
of the test statistic by using a permutation test. These 2-sided p-values were consequently used as a measure of intensity of every jump.
 
\begin{figure}[h!]
	\centering
	\begin{tabular}{ccc}
		& {\sf \scriptsize No pruning} & {\sf \scriptsize Pruning with $\text{p-value} > 0.1$} \\
\rotatebox{90}{~~~~~~~{\graphFont Proportion}}		&\includegraphics[width=0.47\textwidth,clip = true, trim = 0 0 0 0]{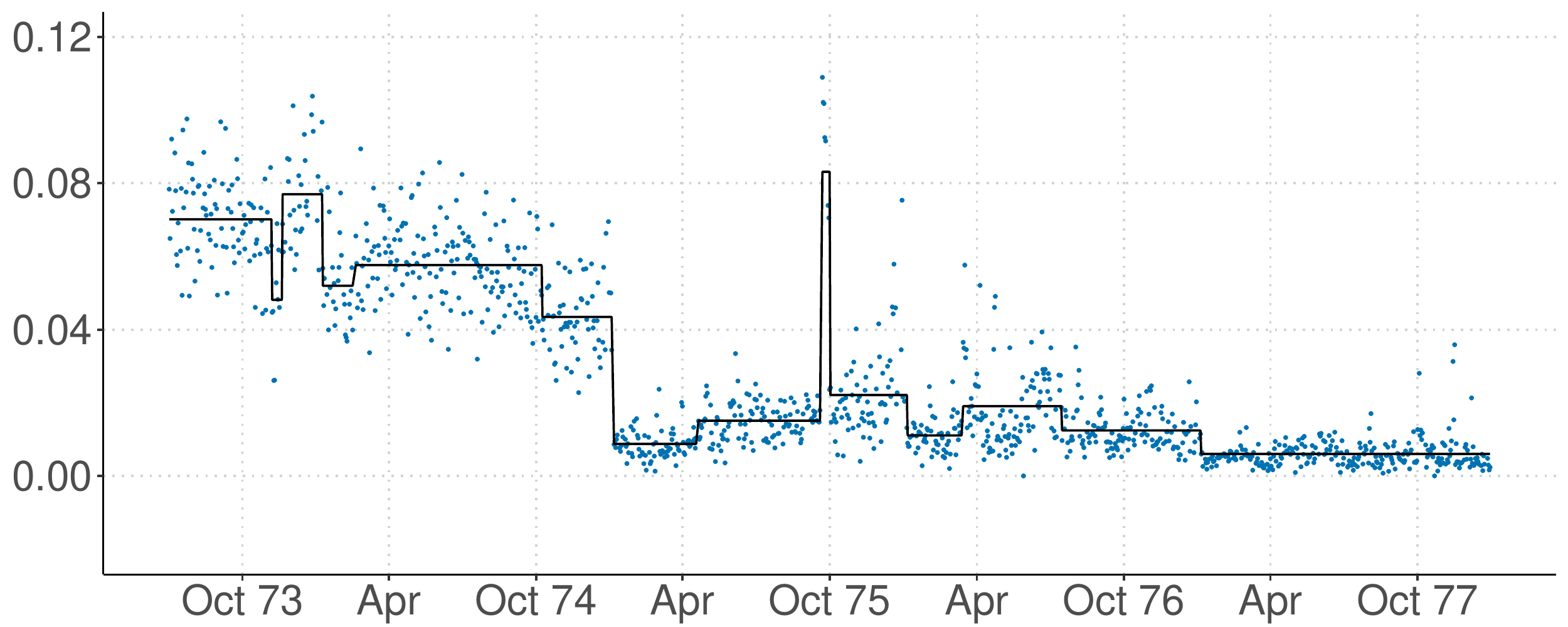}
		&\includegraphics[width=0.47\textwidth,clip = true, trim = 0 0 0 0]{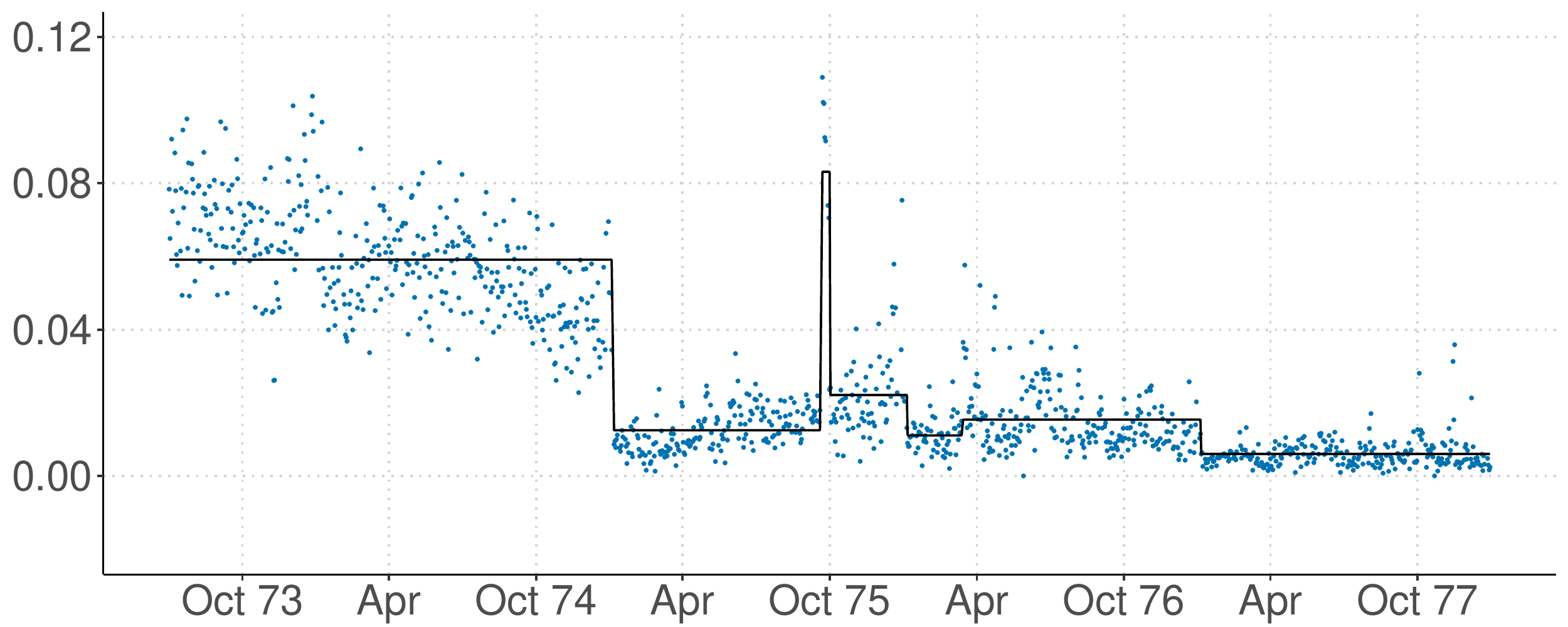}\\~\\
		& {\sf \scriptsize Pruning with $\text{p-value} > 0.01$} & {\sf \scriptsize Pruning with $\text{p-value} > 0.001$} \\
\rotatebox{90}{~~~~~~~{\graphFont Proportion}}		&\includegraphics[width=0.47\textwidth,clip = true, trim = 0 0 0 0]{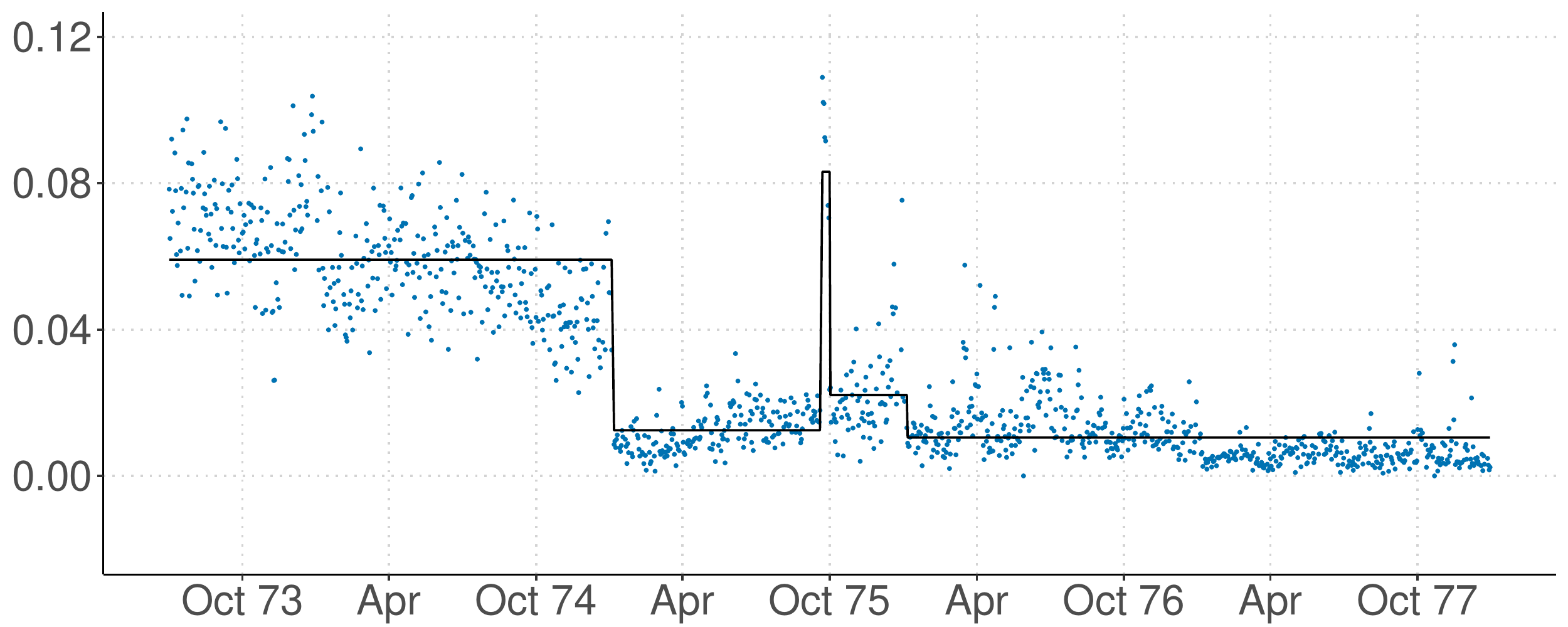}
		&\includegraphics[width=0.47\textwidth,clip = true, trim = 0 0 0 0]{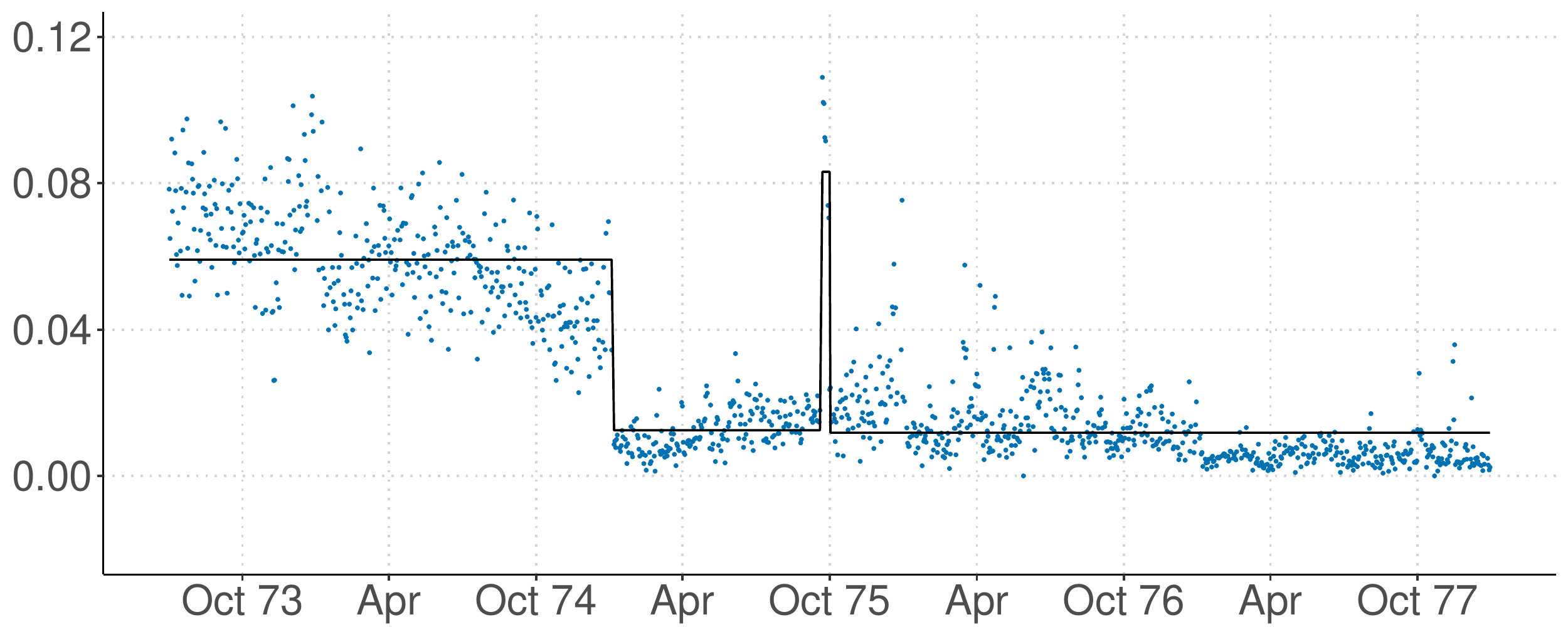}\\
	\end{tabular}
	\caption{\small{Figure showing the communication stream for \TAGS CVIS -- the estimated signal is obtained from the $\ell_{0}$-segmentation scheme (at the cross-validated choice of $\lambda$). We compute the p-values (based on sample splitting, as described in Section~\ref{signif-cha-point-1}) for every candidate jump location and prune the jump locations (and refit the signal with the new jump locations)
	based on different thresholds. We observe that a pruning rule based on p-values leads to a fairly robust 
	selection of intense jump locations, where each location 
	corresponds to a sharp change in local means. }
	}
	\label{fig:CVIS_prune}
\end{figure}

\begin{figure}[h!]
	\caption{ \small {\TAGS ENRG (energy) and US (for cables relating to the U.S.) with p-values associated with the estimated jumps (using the framework in Section~\ref{signif-cha-point-1}). The jumps for ENRG are much sharper and indicate rapid (though not instantaneous) changes in mean, starting with the 1973 OPEC oil embargo. This gives them low p-values. In contrast, the two jumps to the right of the signal for US are during less rapid changes in mean and thus have slightly larger p-values ($\sim 10^{-3}$). Figure~\ref{fig:L1-trend} presents more flexible fits of the underlying signal. (The notation $p=x$ is a shorthand for $\text{p-value}$ being equal to $x$.)} }
	\label{fig:US_ENRG}
	\centering
	\begin{tabular}{ccc}
	& {\sf \scriptsize \TAGS ENRG}& {\sf \scriptsize \TAGS US} \\
		\rotatebox{90}{{\graphFont ~~~~~~~~~~~~~Proportion}}
		&\includegraphics[height = 0.22\textheight, width = 0.47\textwidth, clip = true, trim = 36 10 10 0]{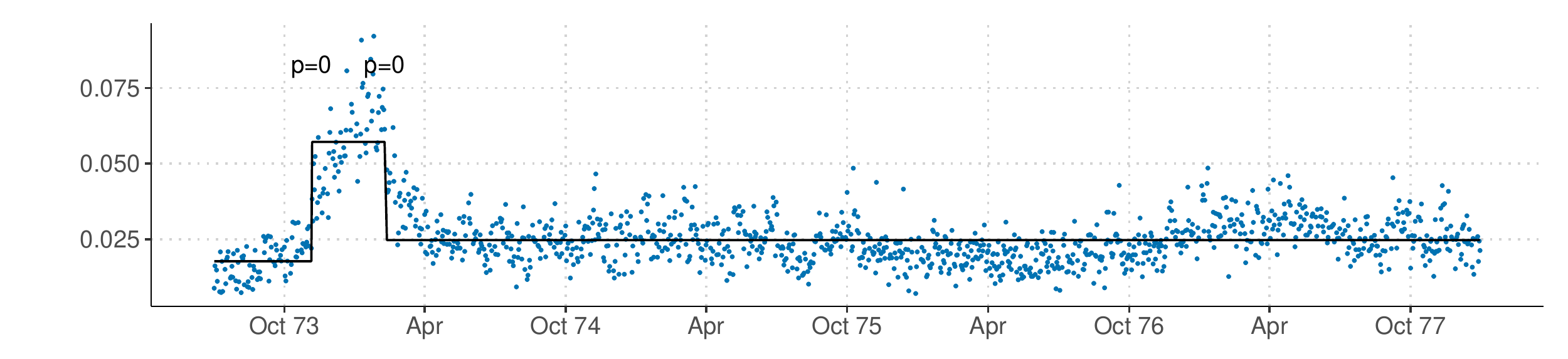} &\includegraphics[height = 0.22\textheight, width = 0.47\textwidth,clip = true, trim = 36 10 10 0]{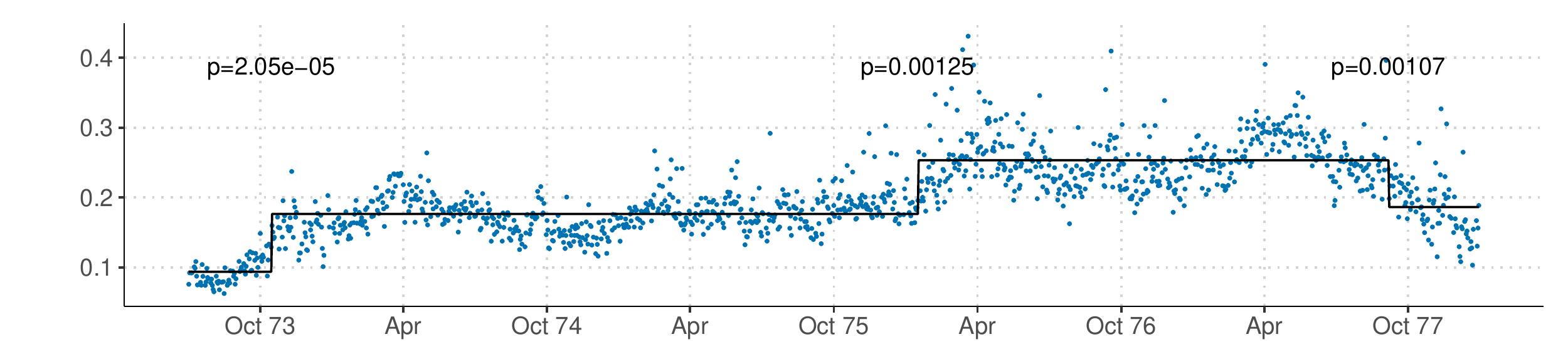}
	\end{tabular}
\end{figure}

\begin{figure}
	\centering
	\begin{tabular}{cc}
		\rotatebox{90}{{\graphFont ~~~~~~~~~~~~~~~~~~~Proportion}}
		&\includegraphics[height = 0.2\textheight, width = 0.85\textwidth,clip = true, trim = 30 0 0 0]{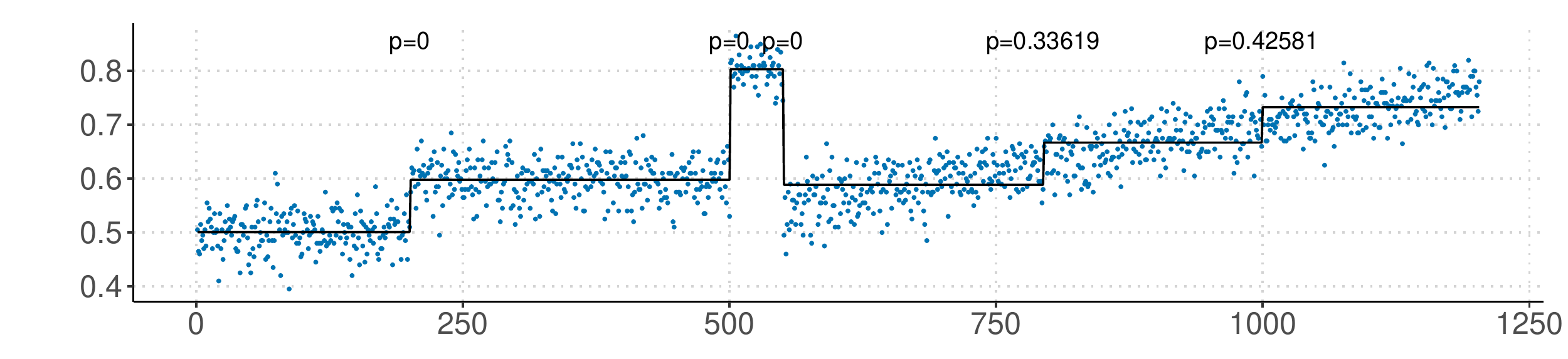}
	\end{tabular}
	\caption{\small {Synthetic data: data description is the same as Figure \ref{fig:synthL0L1}, which contains three real jumps and a linear increasing trend. The first three detected jumps (from left to right) 
	have small p-values (close to zero) -- they correctly correspond to the jumps in the underlying signal. The other two potential jumps have p-values $\sim 0.33$ and $\sim 0.42$ respectively -- these jumps are a consequence of the linear trend (we use the framework in Section~\ref{signif-cha-point-1}). (The notation $p=x$ is a shorthand for $\text{p-value}$ being equal to $x$.)
	}}
	\label{fig:synthEndRise}%
\end{figure}

\begin{figure}[h!]
	\caption{\small {Figure showing the estimates obtained from Problem~\eqref{eqn-opt-1} with the $\ell_{1}$-trend filtering regularizer (See Section~\ref{pwise-linear-segments}). The sharp spike in the CY (Cyprus) communication stream corresponds to an unanticipated event, when Greek forces launched a coup with the goal of annexing Cyprus. The first peak for the second stream (ENRG) corresponds to the 1973 energy crisis, after the OPEC oil ministers announced an embargo during the Yom Kippur War. The peak for VS, for South Vietnam, corresponds to the Fall of Saigon in 1975, which marked the end of the Vietnam War. SHUM, for communications related to human rights, shows the increasing attention the State Department gave to this subject, especially after the election of President Jimmy Carter. }} \label{fig:L1-trend}
	\centering
	\scalebox{0.99}{\begin{tabular}{lcc}
	& {\sf \scriptsize  CY} & {\sf \scriptsize  ENRG} \\
		\rotatebox{90}{{~~~~~~~~~\graphFont Proportion}}	&\includegraphics[width=0.45\textwidth,height=0.17\textheight,clip = true,trim = 35 10 20 50]{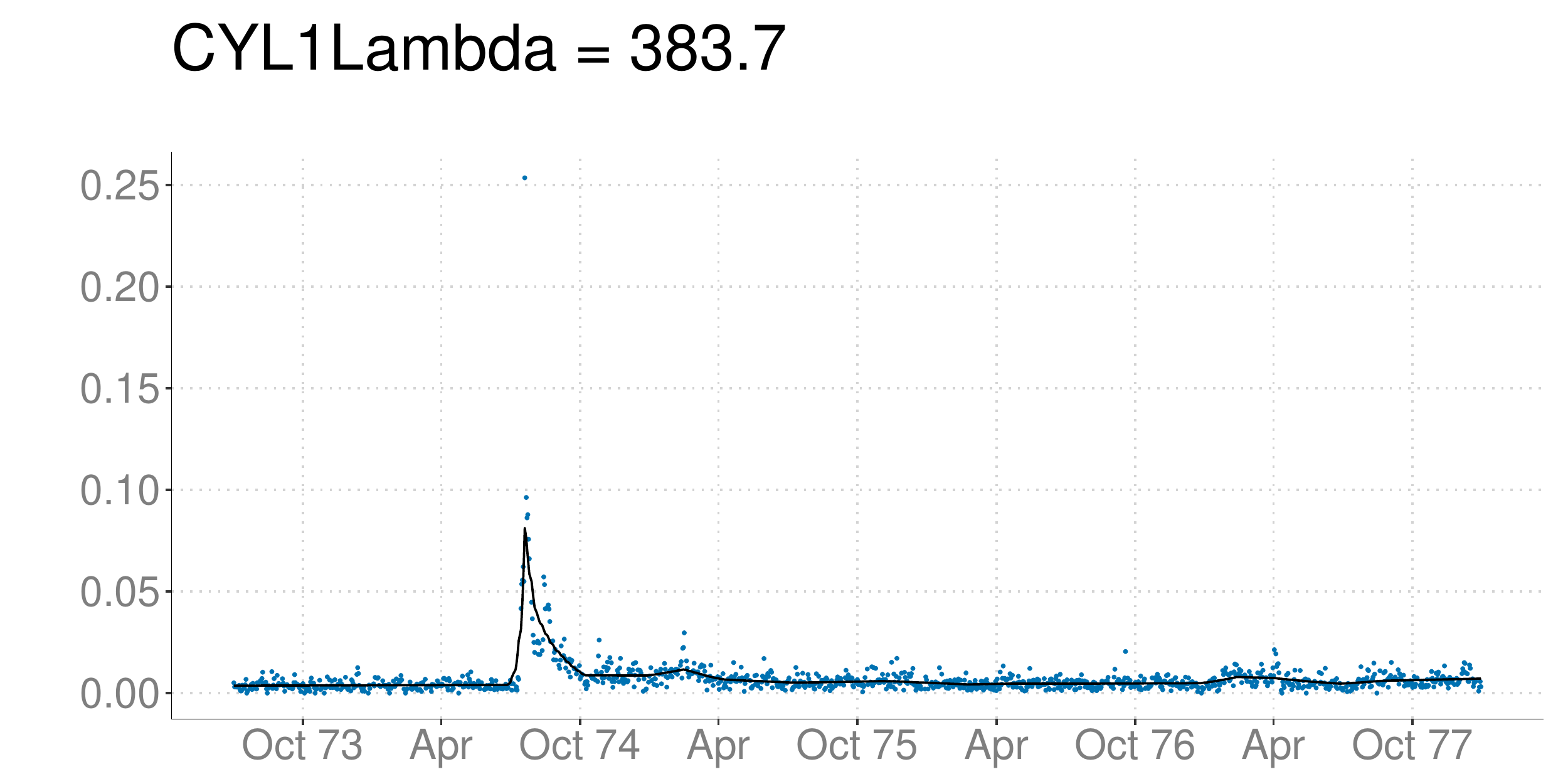}
		&\includegraphics[width=0.45\textwidth,height=0.17\textheight,clip = true,trim = 35 10 20 50]{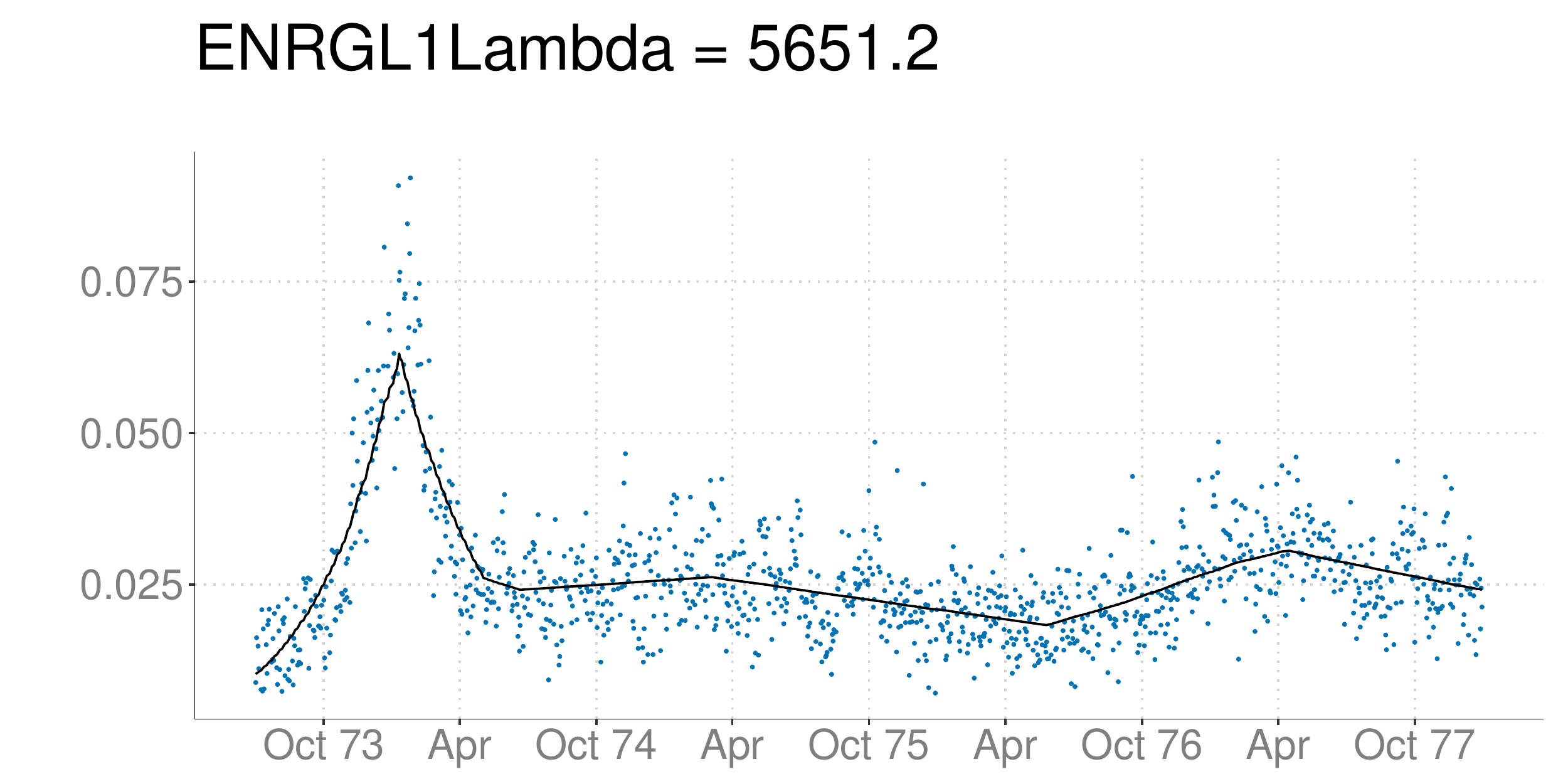}\\~\\
		& {\sf \scriptsize  VS} & {\sf \scriptsize  SHUM} \\
				\rotatebox{90}{{~~~~~~~~~\graphFont Proportion}}	&\includegraphics[width=0.45\textwidth,height=0.17\textheight,clip = true,trim = 35 10 20 50]{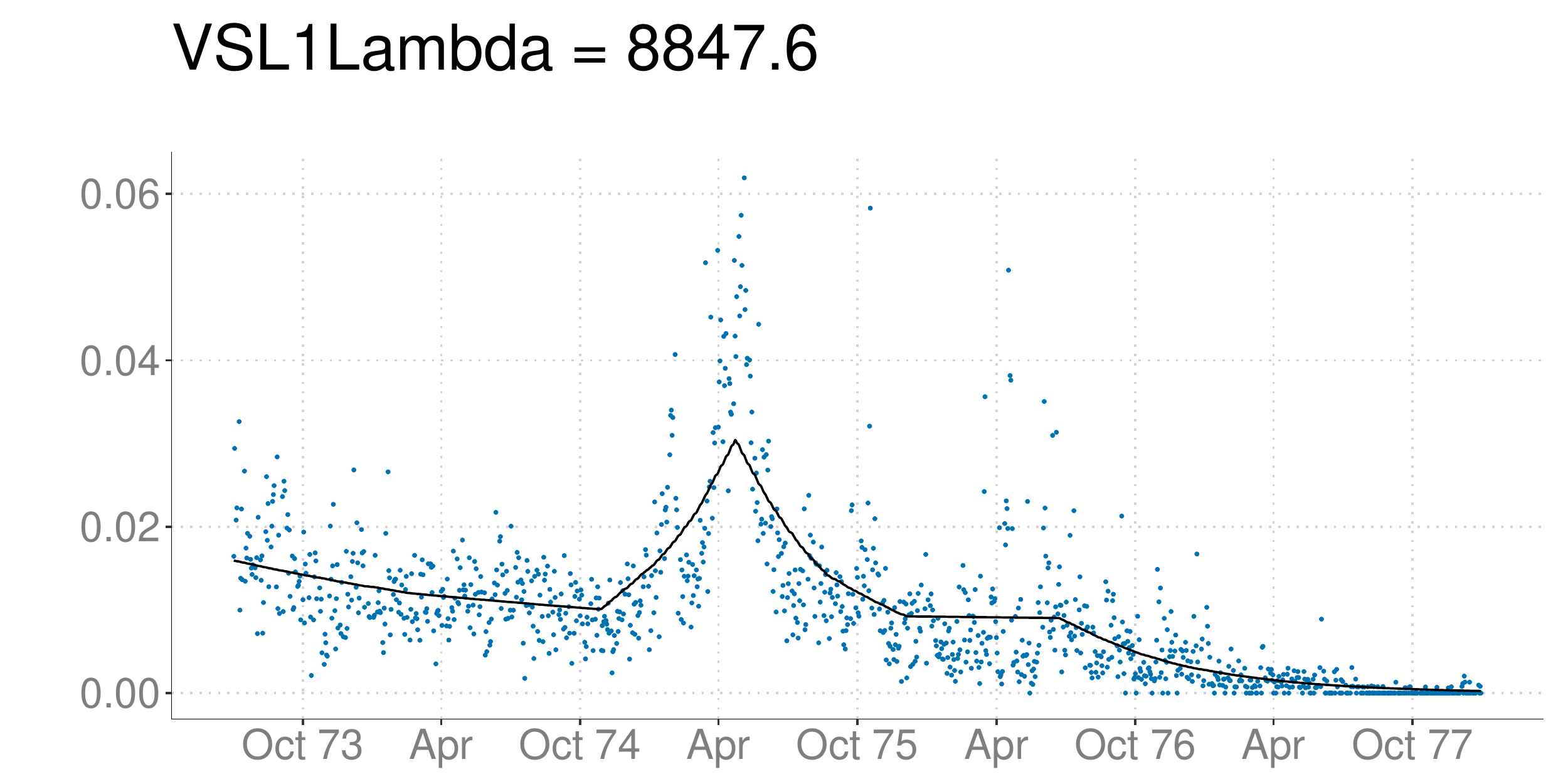}
		&\includegraphics[width=0.45\textwidth,height=0.17\textheight,clip = true,trim = 35 10 20 50]{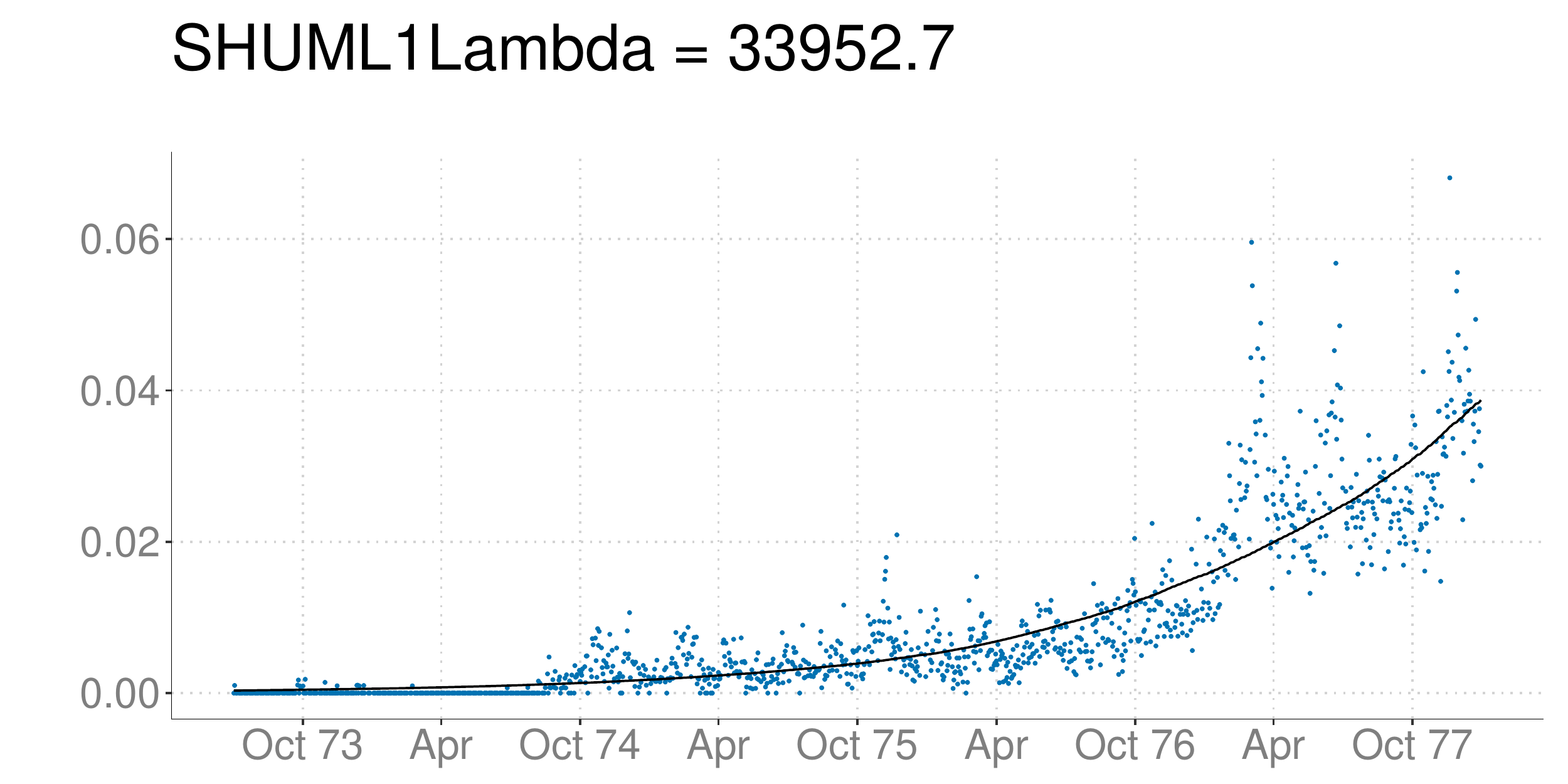}
\end{tabular}}
	
\end{figure}

 Note that a candidate jump estimated by the signal estimation procedure at the cross-validated choice of the tuning parameter need not necessarily correspond to a jump with a low p-value. A small 
p-value indicates that the intervals to the left 
and right of $\hat{t}$ have different proportions\footnote{A jump obtained from the estimated $\{\hat{p}_{t}\}$ may be due to a linear rise in the signal which need not correspond to a significant change in local proportions. Our experiments indicate that jumps in $\{\hat{p}_{t}\}$ that correspond to gradual linear rises in the signal, have higher p-values associated with them when compared to sudden or abrupt changes in $\{ \hat{p}_{t}\}$}.
Thus the p-values can be used to (a) devise a scoring mechanism to rank order multiple jumps observed in a series and/or (b) prune out redundant jumps and identify ones that exhibit a strong 
difference in proportions between the left and right intervals. 
Figure \ref{fig:CVIS_prune} shows the communication stream for \TAGS CVIS and the estimated signal obtained via $\ell_{0}$-segmentation. 
We also computed the p-values for each potential jump location as suggested via the $\ell_{0}$-segmentation fit.
We pruned them and refitted the model based on their p-value scores.
It helps to interpret these patterns alongside Figure~\ref{fig:L1-trend} (\TAGS CVIS) which presents more flexible fits of the underlying signals for this communication stream.
Figure~\ref{fig:US_ENRG} shows additional examples interpreting the p-values associated with 
the different jumps. Figure~\ref{fig:US_ENRG} suggests that the p-values are indicative of whether a jump is due to a shift in the piecewise constant level or 
a linear trend -- the p-values are larger when there is a linear trend rather than a sharp jump (as in a piecewise constant signal). Figure~\ref{fig:L1-trend}  shows a more flexible approximation of this communication stream which provides further insights into the patterns of \TAGS US series. 
To validate the intuition gathered above, we consider a synthetic example in Figure~\ref{fig:synthEndRise} (with the same data as in Figure \ref{fig:synthL0L1}) -- here we observe that the p-values tend to be larger for jumps in the right part of the signal -- 
these jumps in the piecewise constant segments result from estimating a linear trend (that appears at the right of the series) with piecewise constant segments. Note that the p-values associated with the first three jumps (at the left of the signal)
are quite small -- they correspond to jumps in the underlying piecewise constant signal.

In passing we note that it is also possible to perform multiple testing procedures~\cite{lehmann2006testing} to attach error rates to a family of jumps.
In particular, we can use Family Wise Error Rates or False Discovery Rate (FDR) control procedures to return a collection of candidate jumps with a certain 
prescribed control on the error rate~\cite{lehmann2006testing}. However, our goal here is to use the sample splitting and p-value framework to perform an exploratory analysis of the strengths of the 
different jumps -- we do not pursue an in-depth study of multiple testing in this work.


\subsection{From Jumps to Bursts}\label{sec:signals-post-1}
Section~\ref{sec:model} presents methods to detect several jumps in a signal, and also presents an in-depth investigation of how to associate scores (p-values) to each of the detected jumps using a sample splitting scheme.
We now explore if it is possible to summarize a single communication stream (corresponding to a specific  TAGS) with a score -- a measure that should ideally contain in it information pertaining to the strengths and stretches of the 
different jumps. Towards this end, following the terminology introduced by Kleinberg~\cite{kleinberg2003bursty} we formalize  the notion of a  ``burst''. The general approach pursued in this paper and the models used differ from~\cite{kleinberg2003bursty}. Informally speaking, a burst corresponds to a stretch of time where a communication stream depicts traffic which is 
larger than a baseline value. We present a  computational scheme to estimate such bursts for a \TAGS-specific communication 
stream.

\subsubsection{Computation of the strength of a Burst}\label{sec:burst-strength} As a starting point,  we consider an estimate of a baseline proportion $p_0$ (we discuss how to compute this below) that is specific to a communication stream. A ``burstiness period'' or simply burst corresponds to a time interval where the estimated signal lies above the baseline value $p_{0}$ and is given by 
${\mathcal T} = [t_{\text{start}}, t_{\text{end}}]$, where $\hat{p}_t > p_0, \forall t \in {\mathcal T}$. 
Following~\cite{kleinberg2003bursty}, 
we define the strength $S(\mathcal T)$ of the 
burst as the logarithm of the likelihood ratio (here, the numerator is the likelihood of the signal and the denominator is 
that evaluated at the baseline) given by:
$S(\mathcal T) = \sum_{t \in {\mathcal T} }\left( \log L(\hat{p}_t | n_t, y_t) - \log L(p_0 | n_t, y_t) \right),$
where $L(\hat{p}_t | n_t, y_t)$ denotes the likelihood at time $t$. 
Note that the baseline $p_0$ is specific to a communication stream and the score $S(\mathcal T)$ represents a deviation from this global baseline. $S({\mathcal T})$ is 
different than the magnitude of a jump given by $\hat{p}_{t+1} - \hat{p}_{t}$ -- it takes into account the deviation of $\hat{p}_t$ from the baseline $p_0$ as well as the duration of the burst  given by the length of ${\mathcal T}$.
A large value of $S(\mathcal T)$ means that a large part of the likelihood is explained by deviations from the baseline, and therefore,
 corresponds to a strong burst.  
 Note that each \TAGS can have multiple bursts leading to multiple intervals ${\mathcal T}$ -- each with an assigned strength $S({\mathcal T})$. 
 
 \medskip

\noindent {\bf{Choice of baseline:}} The baseline value $p_0$ should be representative of 
the behavior of the TAG-specific communication stream. The global proportion of a communication stream is a reasonable choice. 
We set $p_0$ to be one standard deviation larger than the global proportion
\[p_0 = \bar{p} + \sqrt{\frac{\bar{p}(1 - \bar{p})}{\bar{n}}}, ~~\text{where,}~~ \bar{p} = \frac{\sum_{t=1}^N y_t}{\sum_{t=1}^N n_t},~~~\bar{n} =  \frac{1}{N}\sum_{t=1}^N n_t.\]
A robust estimate like the median can also be used instead of the average. In our experiments we found that the top-ranked slots were relatively agnostic to the choice of 
the baseline $p_0$.


\subsubsection{Interpretation of  Bursts}\label{sec:burst-strength-interp}
Table~\ref{tab:top_burst2} presents the top thirty bursts, with the start and end dates, and the date with the highest value. A close study of the content of the cables shows that not all of these bursts correspond with what scholars would recognize as an even of historical importance. After all, the cable \TAGS that diplomats used do not necessarily correspond with diplomatic activity. For instance, the second biggest burst is made up of cables related to transportation (ETRN) a \TAGS that was commonly used, and overused, from when we begin to have records continuing until 1974, when diplomats' use of this \TAGS was largely discontinued. The biggest burst, for CVIS (visas), has a similar pattern (as shown in Figure \ref{fig:CVIS_prune}). But in this case, it appears to reflect a decision by archivists to stop preserving records related to visas \cite{langbart2007}. To the model, both of these look like bursts, but they simply reflect administrative procedures rather than historical events.

The bursts that follow, on the other hand, appear to correspond well to historical events. The next ten include the Carter administration's prioritization of human rights (SHUM), Anwar Sadat's surprise visit to Israel (PGOV), the Southeast Asian ``Boat People'' crisis (SREF), the U.S. withdrawal from the International Labor Organization (PORG), the conclusion of the Panama Canal Treaty (PDIP), the 1973 Yom Kippur War (XF, for Middle East), Portugal's withdrawal from Angola (AO), and the 1974 crisis over Cyprus (CY). All are included in each of the four standard reference works we consulted~\cite{brune2003chronological, flanders1993dictionary, jentleson1997encyclopedia, deconde2002encyclopedia}.

A systematic evaluation of hundreds of bursts for historical significance lies outside the scope of this paper. But the relative proportion of recognized historical events appears to diminish as one examines smaller bursts, like the ones ranked in the range 13-22. They include the denouement of the Vietnamese War (VM and VS), the OPEC oil embargo (ENRG), the Vladivostok summit (OVIP), and negotiations to end white rule in Rhodesia (RH). But there are also largely unrecognized events, like a 1975 UN General Assembly debate over the command of foreign military forces in South Korea, that would appear to merit closer scrutiny. The identification of such unstudied episodes, no less than rank-ordering well-known events, is valuable for historical scholarship.


\section{Beyond Piecewise Constant Segments}\label{pwise-linear-segments}
A major focus of Section~\ref{sec:methods} was on approximating a communication stream with a piecewise constant signal. This framework 
does help us answer some of the major data-driven questions of interest to a political scientist, based on a first order approximation of the communication streams. We now investigate more flexible signal approximations 
that provide us insights into the finer behavior of the signals. A natural extension of a  piecewise constant estimate $\{ \hat{p}_{t} \}$ is a piecewise linear estimate. However, there are subtleties in incorporating this 
structure into our framework, as we discuss below.

To settle ideas, let us consider the usual signal denoising problem with data: $\tilde{y}_{i} = \mu_{i} + \epsilon_{i}$, for $i=1, \ldots, N$ where, $\epsilon_{i} \stackrel{\text{iid}}{\sim}N(0,\sigma^2)$. We seek to estimate $\B\mu$ such that it is piecewise linear.
In this vein, it is common to use the $\ell_{1}$ trend-filtering approach~\cite{kim2009ell_1,tibshirani2014adaptive} with regularizer $H^{\text{tf}}_{\ell_1}(\B\mu) = \sum_{t} | \mu_{t+2} - 2\mu_{t+1} + \mu_{t} |$ to obtain a signal with piecewise linear segments:
$$\mini_{\B\mu} ~~ \frac12\sum_{i=1}^{N} (\tilde{y}_{i} - \mu_{i})^2 + \lambda H^{\text{tf}}_{\ell_1}(\B\mu).$$
  The penalty function $H^{\text{tf}}_{\ell_1}(\B\mu)$ encodes the
$\ell_{1}$-norm on the discrete second order derivative of the signal $\{\mu_{t}\}$ assuming that the time points are all equally spaced.
 $H^{\text{tf}}_{\ell_1}(\B\mu)$ can be interpreted as a convexification of its $\ell_0$ version: $H^{\text{tf}}_{\ell_0}(\B\mu) = \sum_{t} \M{1} ( \mu_{t+2} - 2\mu_{t+1} + \mu_{t} \neq 0)$ that counts the number of different piecewise linear segments.

Our situation is different from the denoising example outlined above.
Since we are working under the modeling assumption: $(y_{t}|n_t,p_t) \sim \text{Bin}(n_{t}, p_{t})$ with $p_{t} = \exp(\theta_{t})/(1+ \exp(\theta_{t}))$, imposing a trend filtering penalty on 
$p_{t}$ so as to maintain piecewise linearity will lead to a non-convex optimization problem due to the nonlinear dependence of $p_{t}$ on $\theta_{t}$. Thus, instead of enforcing the sequence 
$t \mapsto p_{t}$ to be piecewise linear, we allow the latent parameters $t \mapsto \theta_{t}$ to be piecewise linear -- this leads to a computationally tractable estimation framework. 
Encouraging $t \mapsto \theta_{t}$ to be piecewise linear enables $t \mapsto p_{t}$ to be more flexible than a piecewise constant signal.
Towards this end, we propose a simple adaption of the estimation criterion in Problem~\eqref{eqn-opt-1} by setting the regularizer
$H(\B\theta)= H^{\text{tf}}_{\ell_1}(\B\theta)$. 
 Figure~\ref{fig:L1-trend} shows the results of estimates obtained from some communication streams using the $\ell_{1}$-trend filtering penalty. 
If the time points are not equally spaced, then this penalty can be modified appropriately -- see for example~\cite{kim2009ell_1} and also Section~\ref{admm-details-irreg}.


\paragraph{Computation} The proximal gradient-stylized algorithm update~\eqref{ist-algo} can be adapted to the setting described above with $H(\B\theta) = H^{\text{tf}}_{\ell_1}(\B\theta)$. We use 
the specialized interior point solver\footnote{We use the R-package wrapper available from~\url{https://github.com/hadley/l1tf}} of~\cite{kim2009ell_1} -- this works quite nicely for the problem sizes encountered in this paper.
The resulting Problem~\eqref{eqn-opt-1} is convex and the sequence~\eqref{ist-algo} leads to the optimum of the problem.
The convergence rates outlined in~\eqref{conv-rate-0} and~\eqref{conv-rate-1} will also apply to this problem. 
If we set $H(\B\theta) = H^{\text{tf}}_{\ell_0}(\B\theta)$, the resulting Problem~\eqref{eqn-opt-1} becomes a challenging nonconvex optimization problem -- in this case, there is no analogue of the 
highly efficient dynamic programming implementation that was available for $H_{\ell_0}(\B\theta)$. Thus, for the case where we desire $t \mapsto \theta_{t}$ to be piecewise linear, 
we confine our study to the choice of the convex $\ell_{1}$-trend filtering regularizer.

\subsection*{Acknowledgements} 
R. Mazumder was supported by ONR (grant \# N000141512342) and NSF-IIS (grant \# 1718258) for support. The authors will like to thank: David Blei, David Madigan, Shawn Simpson for helpful comments and encouragement.
 The authors will also like to thank the workshop participants of 
``Famine and Feast--International Historical Research in the Digital Age'' (London, UK; 2015) for comments on the work. Preliminary results from this article appeared in a media article on {\texttt{buzzfeed.com}}
\footnote{Article~\url{www.buzzfeed.com/josephbernstein/can-a-computer-algorithm-do-the-job-of-a-historian?}}.

\appendix{

\section{Appendix}\label{sec:append}

\begin{table}[h!]
\begin{tabular}{|c|c|} \hline
	Significance level & number of TAGS for $\mathcal{T}$ test~\eqref{scan-stat-1}\\ \hline
	$<$ 0.1 & 914\\
	$<$ 0.01 & 768\\
	$<$ 0.001 & 622\\
	$<$ 0.0001 & 509\\
	$<$ 0.00001 & 391\\ \hline
\end{tabular}\caption{Table showing how many \TAGS-specific cables survive at different significance levels, with 0.00001 being the smallest detectible level. This is out of the first 1000 TAGS, which roughly corresponded to the TAGS with more than 50 total cables.}\label{tab:cutoffs-prelim}
\end{table}

\subsection{Problem~\eqref{ist-algo} with irregularly spaced time points}\label{admm-details-irreg}
If the time points are irregularly spaced then the penalty functions need to be adjusted accordingly.
The fused lasso penalty function becomes: $H(\B\theta) = \sum_{i} | \theta_{t+1} - \theta_{t}|/\Delta_{t}$ where, $\Delta_{t}$ denotes the time difference between time point $t$ and the next time point, indexed by $t+1$. 
For this choice, the associated proximal map i.e., Problem~\eqref{ist-algo} needs to be modified to:
\begin{equation}\label{time-depend-1}
\mini_{\M{u} \in \Re^{N} }  \;\; \frac12 \| \M{u} - \bar{\M{u}} \|_{2}^2 + \lambda' \| D \M{u} \|_{1},
\end{equation}
where, $\|D \M{u}\|_{1} = \sum_{i} | u_{i+1} - u_{i}|/\Delta_{t}$.
More generally, if we consider the $\ell_{1}$-trend filtering example with varying time intervals, we get an instance of Problem~\eqref{time-depend-1} with
$$\|D\M{u}\|_{1}:=\sum_{t} \left | \left(\frac{u_{t+2} - u_{t+1}}{\Delta_{t+1}}  -   \frac{u_{t+1} - u_{t}}{\Delta_{t}} \right) \frac{1}{\Delta_{t}}\right |. $$

We use the Alternating Direction Method of Multipliers (ADMM) procedure~\cite{boyd-admm1} which performs the following decomposition:
$\B\alpha = D \M{u}$ and obtains the Augmented Lagrangian:
$$ {\mathcal L}( \M{u},  \B\alpha; \B\nu) =  \frac12 \| \M{u} - \bar{\M{u}} \|_{2}^2 + \lambda' \| \B\alpha \|_{1} + \langle \B\alpha - D \M{u},  \B\nu \rangle + \frac{\rho}{2} \|\B\alpha - D \M{u} \|_{2}^2,$$
for some choice of $\rho >0$.
The usual ADMM approach performs the following sequence of updates:
\begin{equation}\label{admm-update1}
\begin{aligned}
\M{u} \leftarrow& \argmin_{\M{u}} ~ {\mathcal L}( \M{u},  \B\alpha; \B\nu) \\
\B{\alpha} \leftarrow& \argmin_{\B\alpha} ~ {\mathcal L}( \M{u},  \B\alpha; \B\nu) \\
\B\nu \leftarrow & \B\nu + \rho(\B\alpha - D \M{u}),
\end{aligned}
\end{equation}
where, in the update wrt $\M{u}$ other variables remain fixed, and the same applies for the update wrt $\B\alpha$.
We refer the reader to~\cite{boyd-admm1} for details pertaining to the convergence of this algorithm and choices of $\rho$.
We note that the update wrt $\M{u}$ in display~\eqref{admm-update1} can be solved quite easily via solving a system of linear equations:
$$ \M{u} \leftarrow (\rho D'D + \M{I})^{-1} ( \bar{\M{u}} + D'\M{u} + \rho \M{D}'\B\alpha).$$
Note that $(\rho D'D + \M{I})$ is a bidiagonal matrix when $D$ corresponds to the weighted fused lasso penalty and 
a tridiagonal matrix when it corresponds to the weighted trend filtering penalty.
The inverses in each of these cases can be computed with cost $O(2N)$ and $O(3N)$ (respectively) ~\cite{BV2004,kim2009ell_1} -- furthermore the inverse 
can be computed once (at the onset) as the matrix does not change across iterations.
The update wrt $\B\alpha$ in display~\eqref{admm-update1} requires a solving the following problem:
$$\B\alpha \leftarrow \argmin_{\B\alpha}~~\frac{\rho}{2} \| \B\alpha - \M{z} \|_{2}^2  + \lambda' \| \B\alpha \|_{1},$$
where, $\M{z} = (D\M{u} -\B\nu/\rho)$. A solution to the above problem is given by the familiar soft-thresholding~\cite{FHT-09-new} operation where, 
$\alpha_{i} =  \sgn(z_{i}) \max \{ |z_{i}| - \lambda'/\rho,0 \}$. The sequence of updates in~\eqref{admm-update1} are performed till some form of convergence criterion is met~\cite{boyd-admm1}.


\begin{table}[h!]
\small
\centering
\caption{\small {Top 30 bursts identified using $\ell_0$ segmentation algorithm, using the method in Section~\ref{sec:signals-post-1} to compute burst strengths. For interpretations regarding the bursts please see the discussion in 
Section~\ref{sec:burst-strength-interp}.}}
\label{tab:top_burst2}
 \resizebox{1.05\textwidth}{0.42\textheight}{\begin{tabular}{|  ll ll ll l | }
  \hline
& TAGS & meaning & start & end & peak & Burst Strength \\ 
  \hline
1 & ETRN & Economic Affairs-Transportation & 1973-07-02 & 1974-08-09 & 1973-09-28 & 5146.05 \\ 
  2 & CVIS & Consular Affairs-Visas & 1973-07-02 & 1975-01-02 & 1974-06-28 & 4839.35 \\ 
  3 & SHUM & Social Affairs-Human Rights & 1977-01-19 & 1977-12-30 & 1977-11-18 & 2872.02 \\ 
  4 & US & United States & 1976-01-28 & 1977-09-16 & 1976-04-15 & 2516.03 \\ 
  5 & PGOV & Political Affairs-Government & 1977-06-03 & 1977-12-30 & 1977-11-18 & 2484.57 \\ 
  6 & SREF & Social Affairs-Refugees & 1975-04-22 & 1976-07-20 & 1976-06-02 & 1662.58 \\ 
  7 & SOPN & Social Affairs-Public Opinion and Information & 1976-11-26 & 1977-12-30 & 1977-08-26 & 1597.14 \\ 
  8 & PORG & Political Affairs-Policy Relations With International Organizations & 1977-06-15 & 1977-12-30 & 1977-11-11 & 1547.35 \\ 
  9 & PDIP & Political Affairs-Diplomatic and Consular Representation & 1977-05-24 & 1977-12-30 & 1977-09-02 & 1462.93 \\ 
  10 & XF & Middle East & 1973-10-09 & 1973-12-19 & 1973-10-16 & 1453.76 \\ 
  11 & AO & Angola & 1975-11-08 & 1976-02-23 & 1975-11-10 & 1439.58 \\ 
  12 & CY & Cyprus & 1974-07-15 & 1974-07-29 & 1974-07-20 & 1378.79 \\ 
  13 & VM & Vietnam & 1977-10-11 & 1977-12-30 & 1977-10-12 & 1365.45 \\ 
  14 & PDEV & Political Affairs-National Development & 1977-06-13 & 1977-12-30 & 1977-08-31 & 1344.70 \\ 
  15 & VS & Vietnam (South) & 1973-07-02 & 1975-06-06 & 1975-04-25 & 1150.46 \\ 
  16 & UNGA & UN General Assembly & 1975-08-19 & 1975-12-13 & 1975-11-07 & 1044.29 \\ 
  17 & CARR & Consular Affairs-Americans Arrested Abroad & 1977-06-01 & 1977-12-30 & 1977-06-28 & 951.98 \\ 
  18 & MCAP & Political Affairs-Military Capabilities & 1973-07-02 & 1974-08-15 & 1974-07-03 & 903.26 \\ 
  19 & ENRG & Economic Affairs-Energy & 1973-11-08 & 1974-02-21 & 1974-01-25 & 760.64 \\ 
  20 & PBOR & Political Affairs-Boundary and Sovereignity Claims & 1977-07-01 & 1977-12-30 & 1977-11-09 & 685.75 \\ 
  21 & OVIP & Operations-VIP Travel Arrangements & 1974-10-09 & 1974-11-09 & 1974-10-31 & 607.17 \\ 
  22 & RH & Rhodesia & 1976-09-01 & 1977-12-30 & 1977-08-31 & 569.89 \\ 
  23 & AEMR & Administration-Emergency and Evacuation & 1975-03-28 & 1975-05-12 & 1975-04-28 & 524.59 \\ 
  24 & MPLA & Popular Movement for the Liberation of Angola & 1975-11-07 & 1976-02-24 & 1976-02-18 & 507.98 \\ 
  25 & MSG & Marine Security Guards & 1976-09-02 & 1977-12-30 & 1977-11-28 & 507.27 \\ 
  26 & OREP & Operations-Congressional Travel & 1976-10-27 & 1976-11-18 & 1976-11-02 & 481.08 \\ 
  27 & PRG & Provisional Revolutionary Government of South Vietnam & 1975-01-16 & 1975-02-06 & 1975-02-03 & 470.51 \\ 
  28 & MNUC & Military and Defense Affairs-Military Nuclear Applications & 1977-03-11 & 1977-12-30 & 1977-08-22 & 421.53 \\ 
  29 & UNGA & UN General Assembly & 1974-09-05 & 1974-12-05 & 1974-10-10 & 417.20 \\ 
  30 & CB & Cambodia (Khmer Republic) & 1973-07-02 & 1975-05-21 & 1975-04-16 & 370.14 \\ 
   \hline
\end{tabular}}
\end{table}

\begin{figure}[]
	\caption{\small {Communication streams with different significance scores in the spirit of Section~\ref{sec:interest-1}. Top row shows the series where, like FI (for Finland), the p-values are larger than 0.1; second row has p-values in 0.1-0.001 (such as for OSCI, scientific grants). The third and fourth rows show series that seemed to have a high degree of intense activity: all p-values were smaller than 0.001. This includes, for example, SREF (for refugees) and SF (for South Africa.) As we will see, a strong deviation from the null model under the global testing framework does not necessarily imply a communication stream with a significant change point.
	}} 
	\centering
	\scalebox{0.95}{\begin{tabular}{llccc}
	\setlength{\tabcolsep}{.1pt} \\
	&& {\sf \scriptsize  SA} & {\sf \scriptsize  TQ} &{\sf \scriptsize  FI}  \\
	\rotatebox{90}{{~~~~~~~~~~\graphFont $\text{p-value}> 0.1$}}&\rotatebox{90}{{~~~~~~~~~\graphFont Proportion}}	&\includegraphics[width=0.3\textwidth,clip = true,trim = 35 25 20 50]{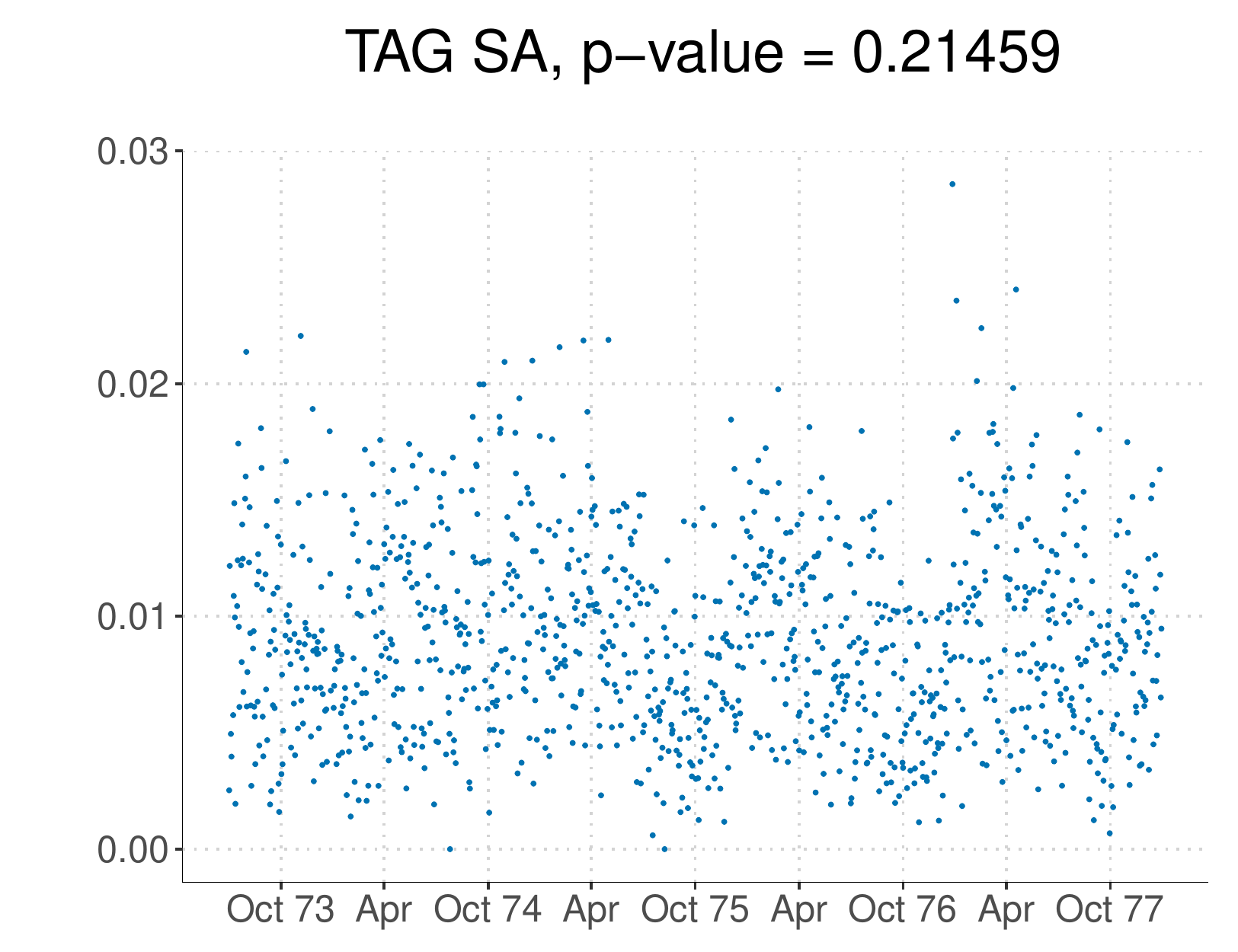}
		&\includegraphics[width=0.3\textwidth,height=0.17\textheight,clip = true,trim = 35 25 20 50]{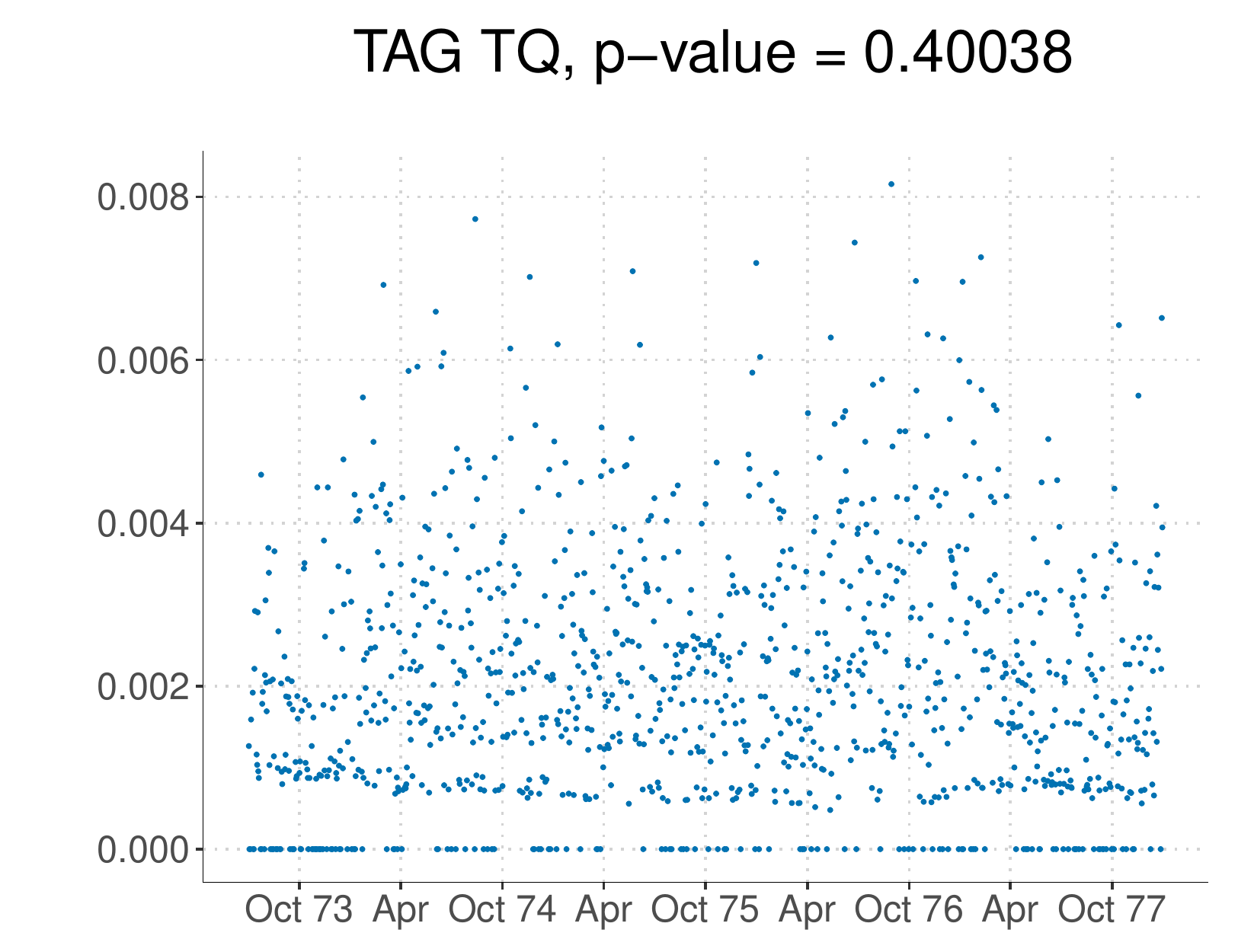}
		&\includegraphics[width=0.3\textwidth,height=0.17\textheight,clip = true,trim = 35 25 20 50]{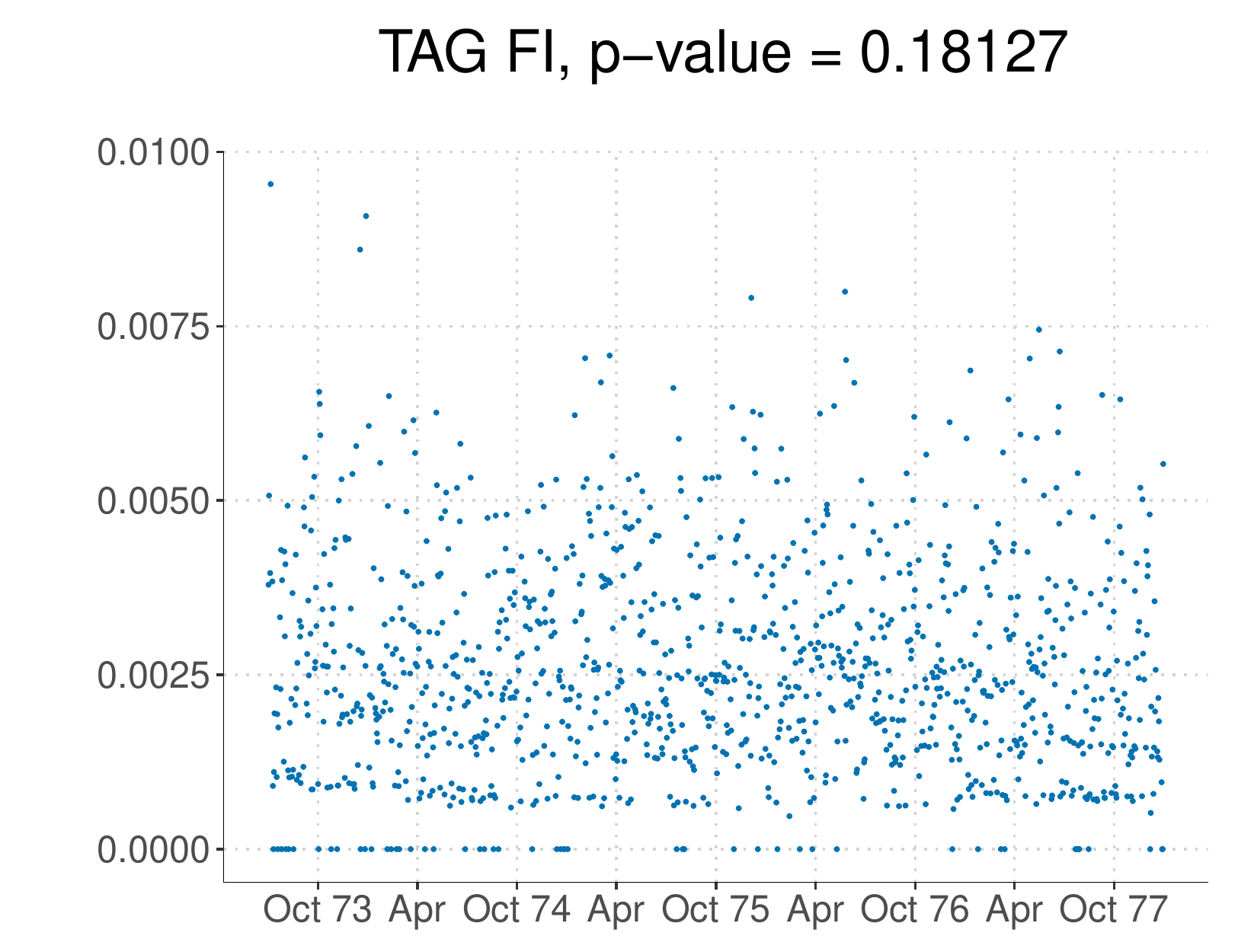}\\~\\
			&& {\sf \scriptsize  SW} & {\sf \scriptsize  AR} &{\sf \scriptsize  OSCI}  \\
	\rotatebox{90}{{~~~~~~~\graphFont $0.001 < \text{p-value} < 0.1$}}&\rotatebox{90}{{~~~~~~~~~\graphFont Proportion}}		&\includegraphics[width=0.3\textwidth,height=0.17\textheight,clip = true,trim = 35 25 20 50]{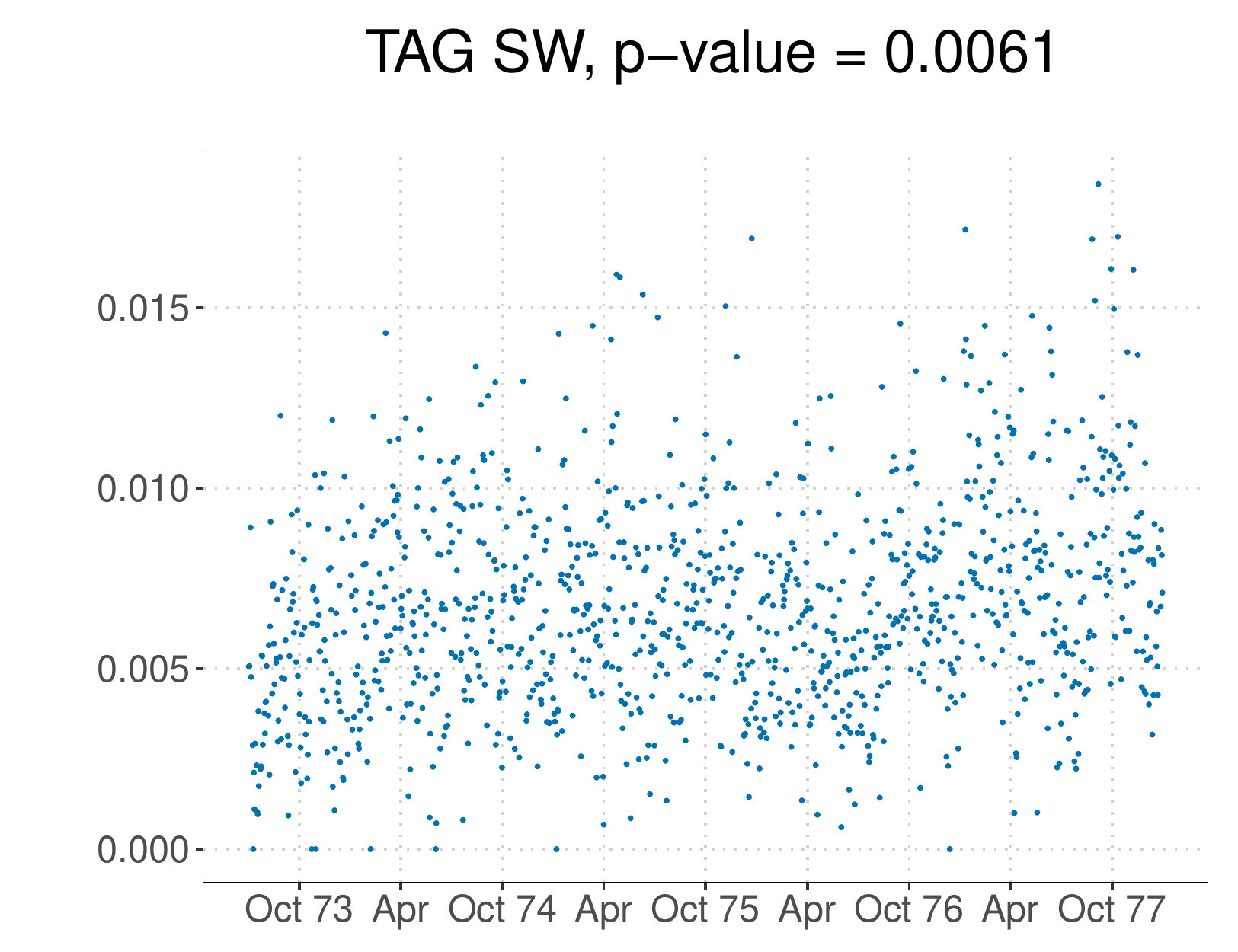}
		&\includegraphics[width=0.3\textwidth,height=0.17\textheight,clip = true,trim = 35 25 20 50]{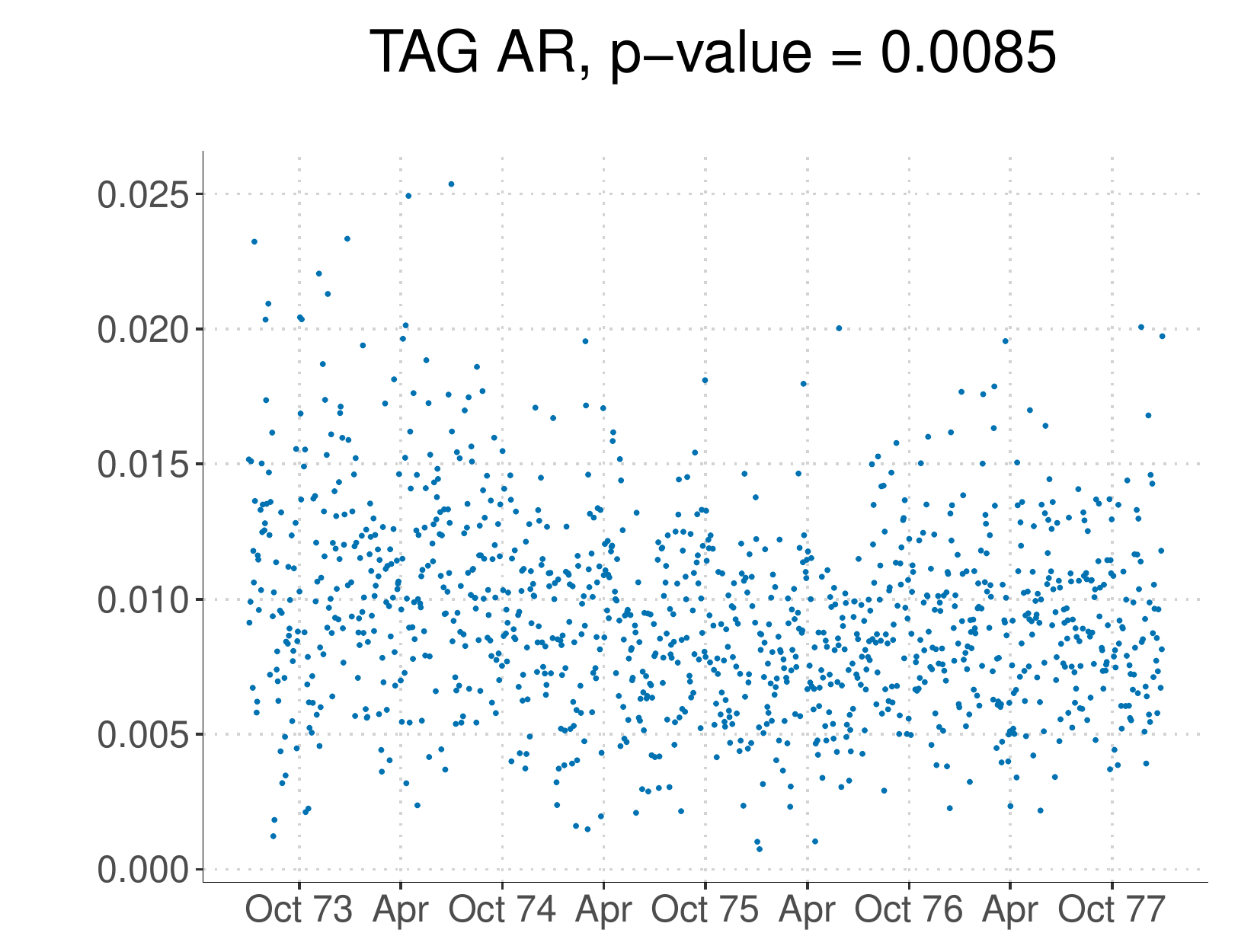}
		&\includegraphics[width=0.3\textwidth,height=0.17\textheight,clip = true,trim = 35 25 20 50]{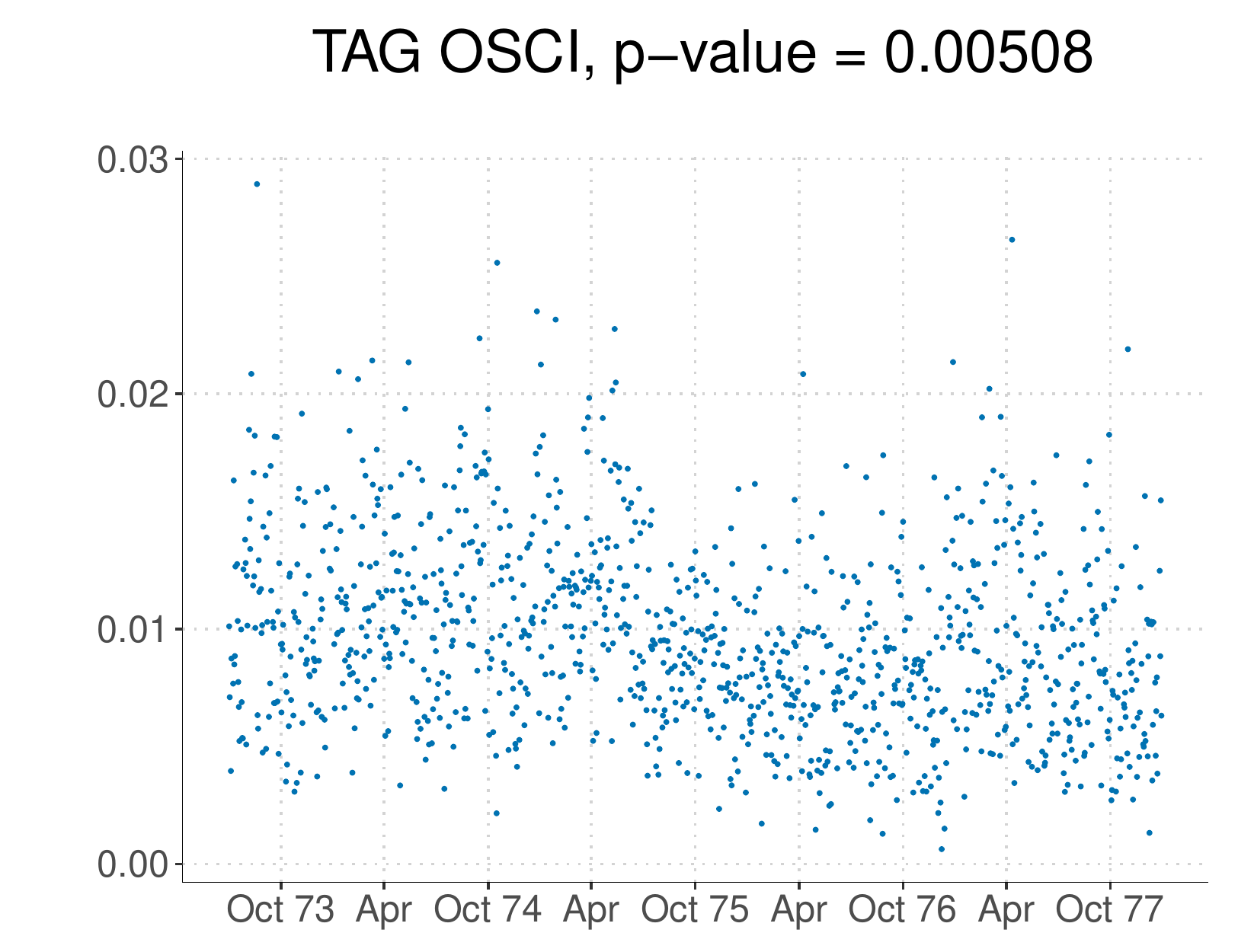}\\ ~\\
					&& {\sf \scriptsize  PFOR} & {\sf \scriptsize  SREF} &{\sf \scriptsize  OEXC}  \\
	\rotatebox{90}{{~~~\graphFont $\text{p-value} < 0.001$; rapid change}}&\rotatebox{90}{{~~~~~~~~~\graphFont Proportion}}		&\includegraphics[width=0.3\textwidth,height=0.17\textheight,clip = true,trim = 35 25 20 50]{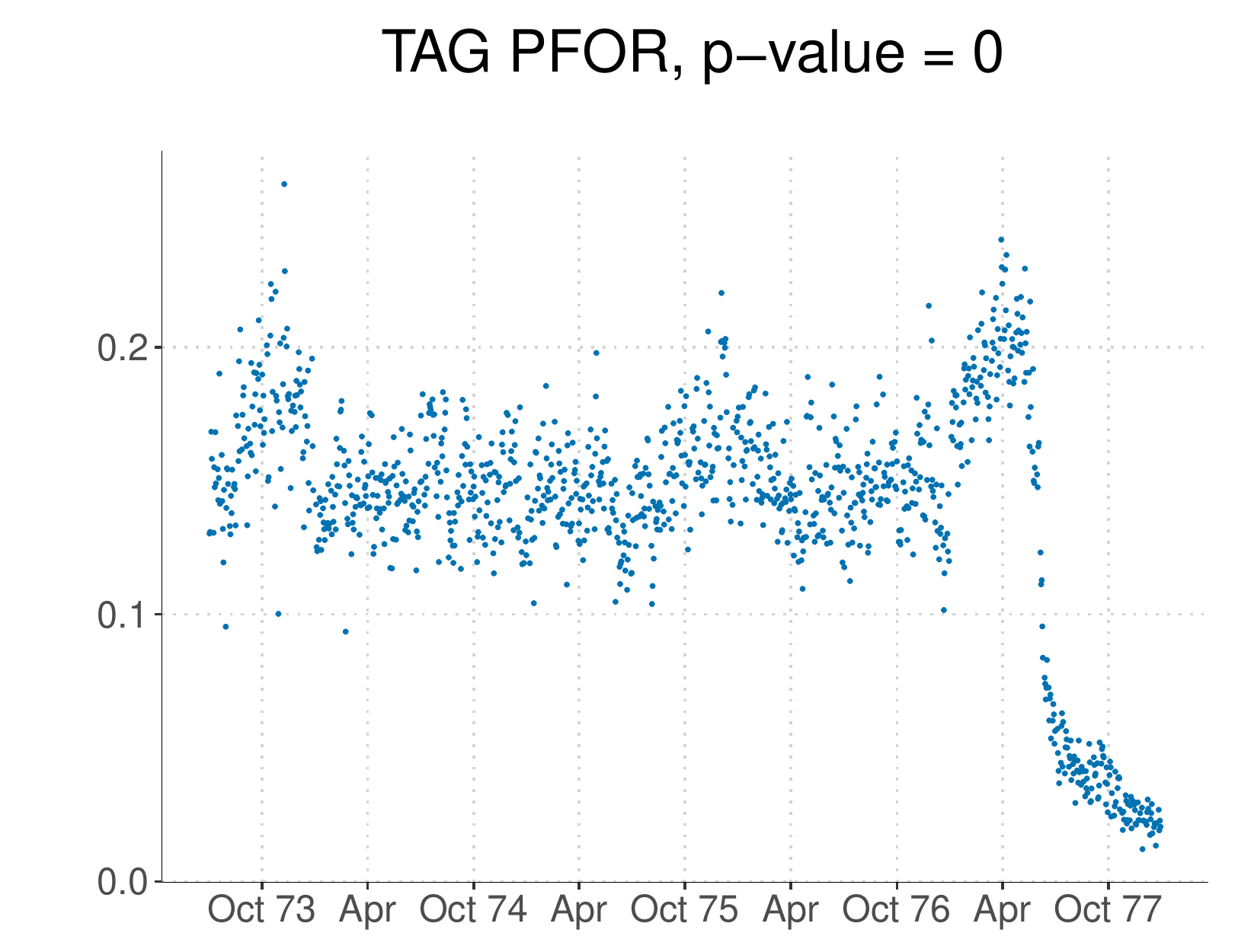}
		&\includegraphics[width=0.3\textwidth,height=0.17\textheight,clip = true,trim = 35 25 20 50]{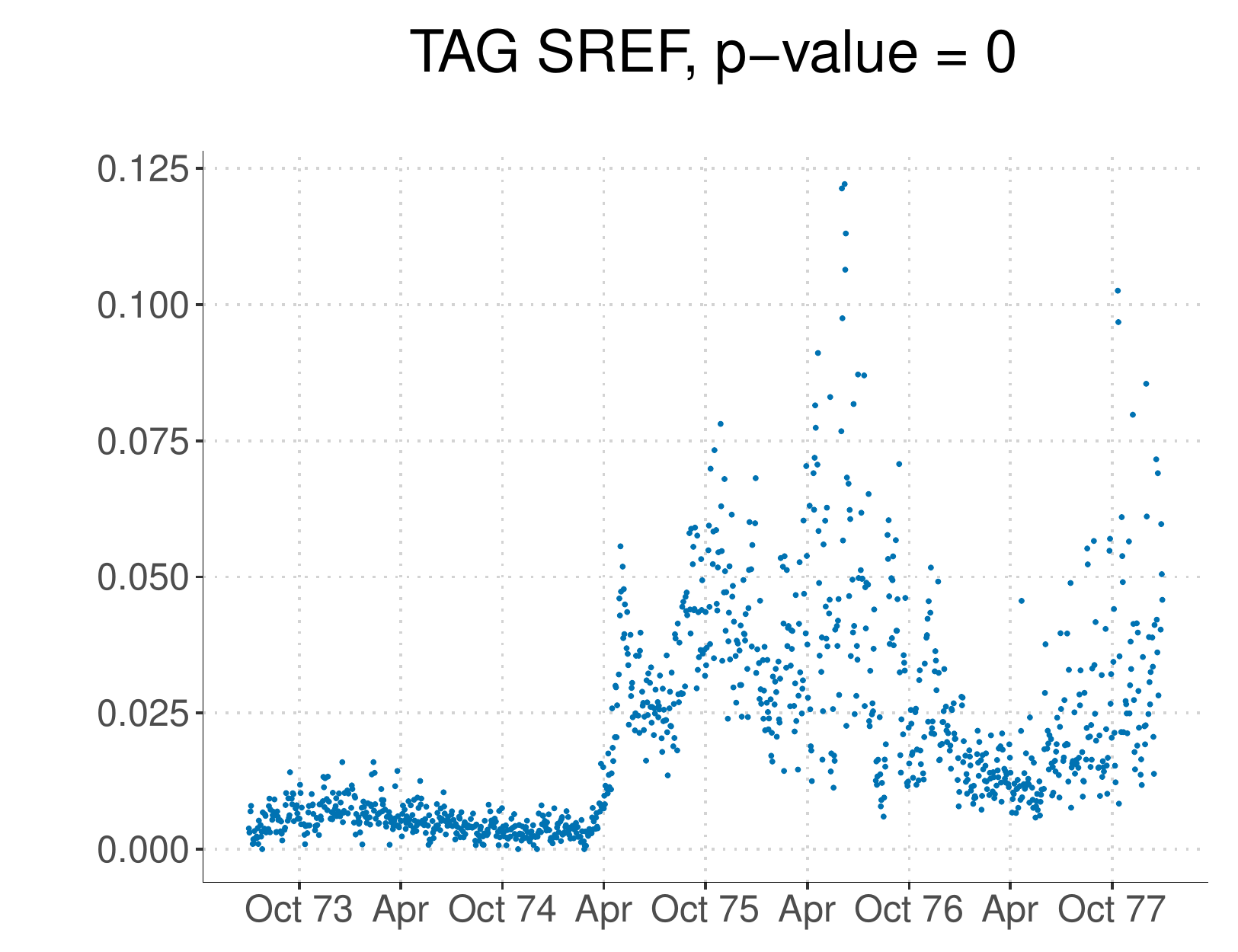}
		&\includegraphics[width=0.3\textwidth,height=0.17\textheight,clip = true,trim = 35 25 20 50]{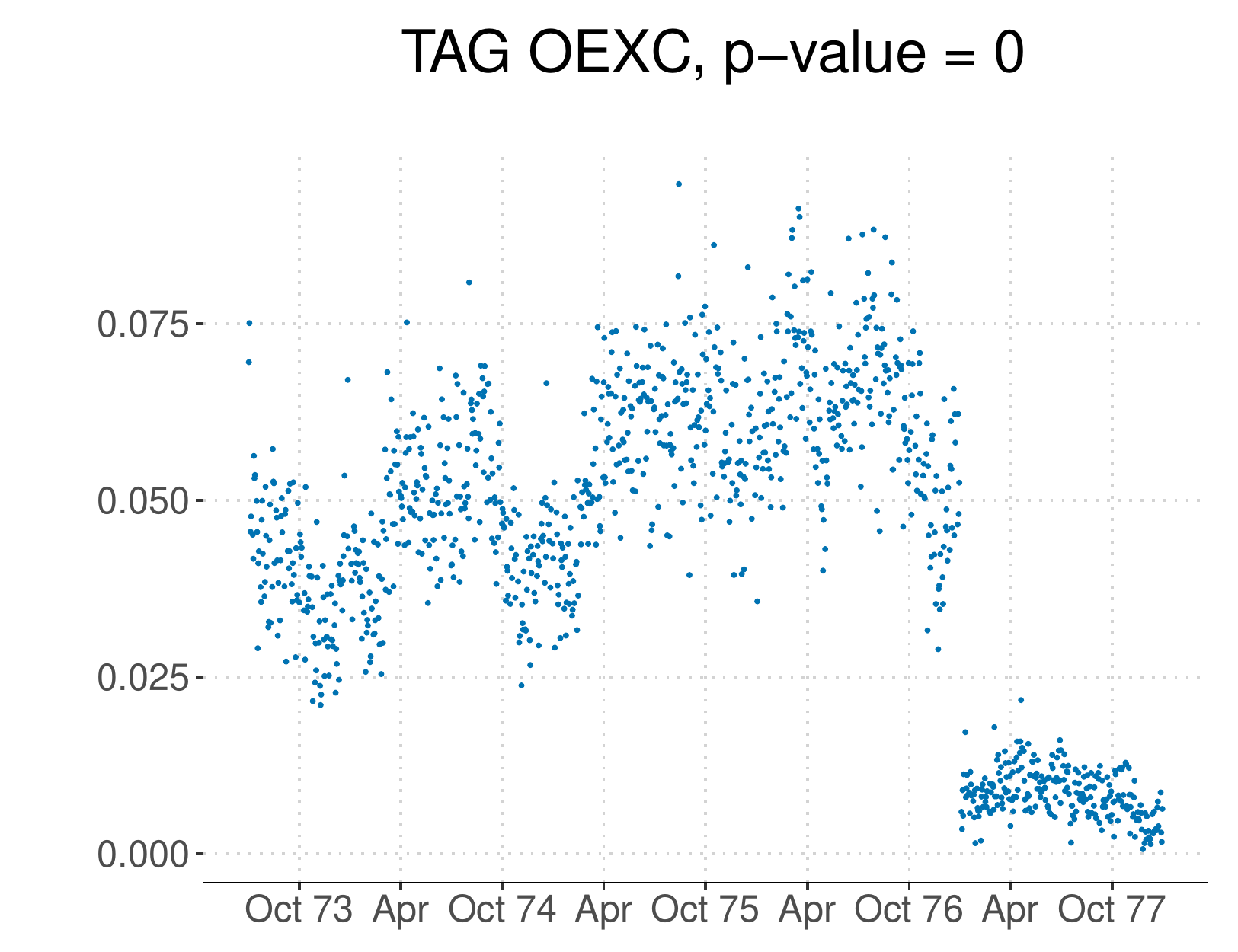}\\ ~\\
								&& {\sf \scriptsize  PINT} & {\sf \scriptsize  OCON} &{\sf \scriptsize  SF}  \\
	\rotatebox{90}{{~~~\graphFont $\text{p-value} < 0.001$; gradual change}}&\rotatebox{90}{{~~~~~~~~~\graphFont Proportion}}		&\includegraphics[width=0.3\textwidth,height=0.17\textheight,clip = true,trim = 35 0 20 50]{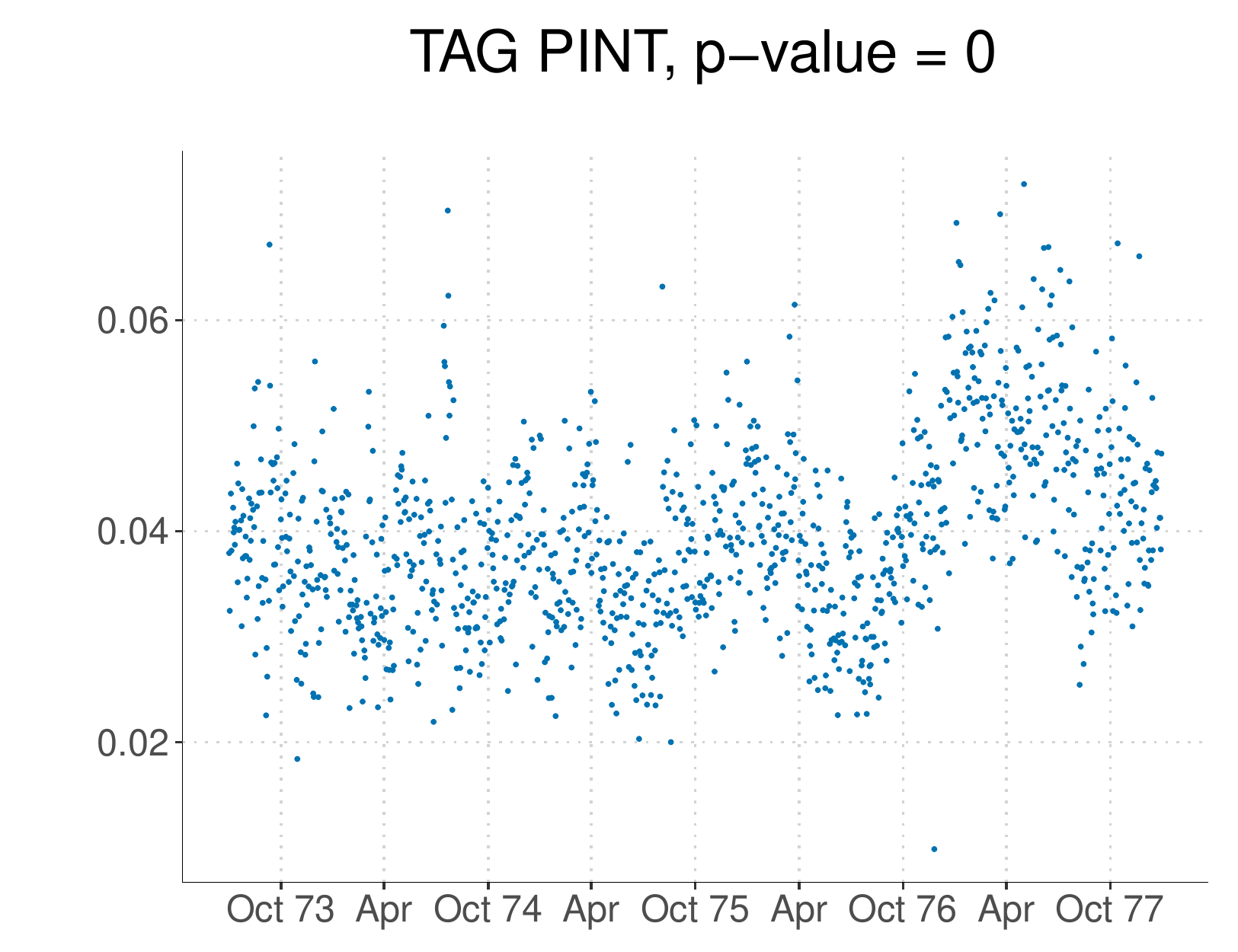}
		&\includegraphics[width=0.3\textwidth,height=0.17\textheight,clip = true,trim = 35 0 20 50]{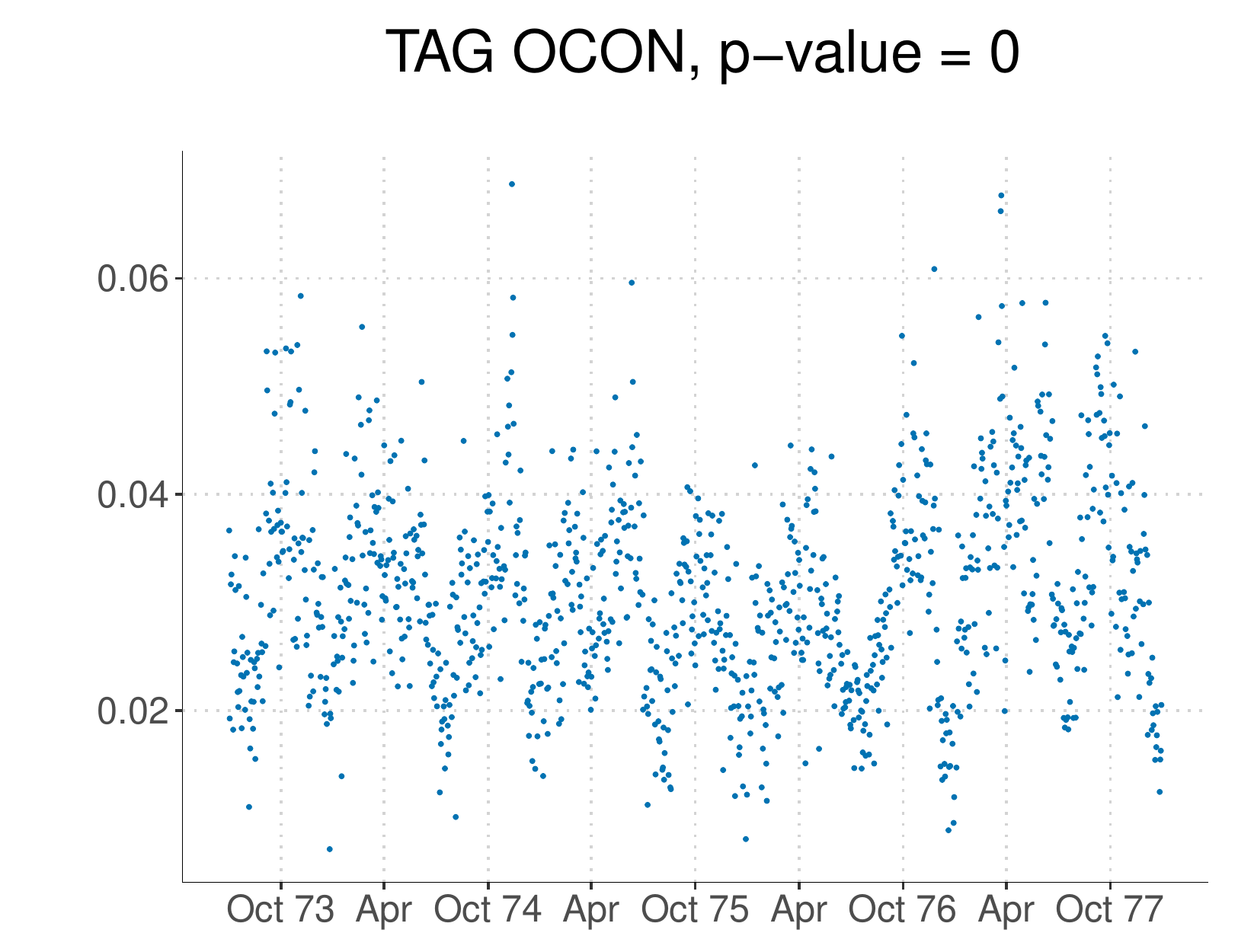}
		&\includegraphics[width=0.3\textwidth,height=0.17\textheight,clip = true,trim = 35 0 20 50]{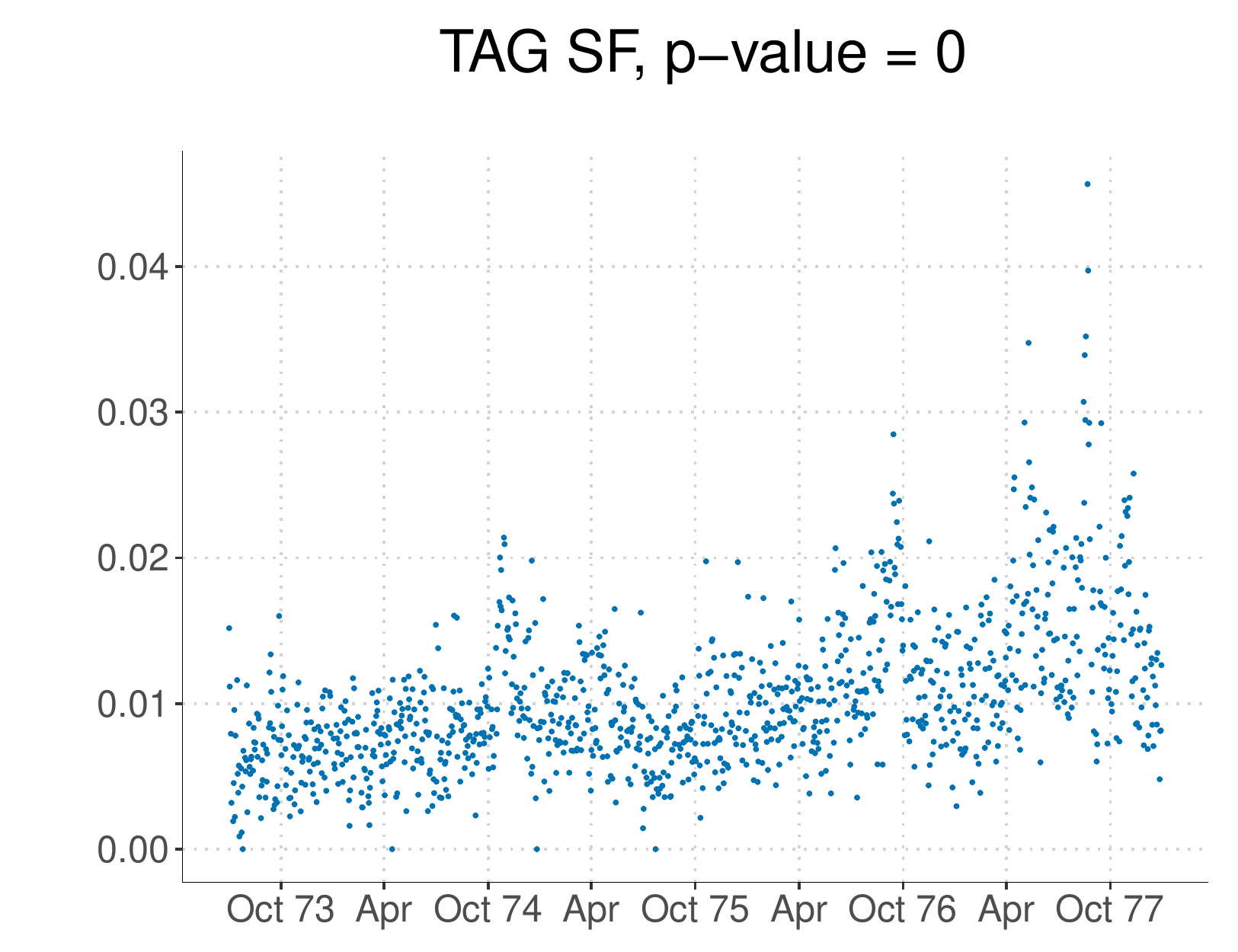}\\
	\end{tabular}}
	\label{fig:all_sig}
\end{figure}

}

\clearpage

\bibliographystyle{plain}

{\small{\bibliography{project_ref,rahul_dbm3}}}


\end{document}